\documentclass[aps,prd,a4paper,superscriptaddress,nofootinbib,10pt, one column]{revtex4-1}
\usepackage{amsmath,amssymb,bm,latexsym,soul,nicefrac,graphicx,subfigure,multirow,dcolumn,aas_macros,xcolor,hyperref}
\usepackage{CJK}
\usepackage{geometry}
\usepackage{acronym}
\usepackage{graphicx,url,bm,makecell,verbatim}

\newcommand{\TRC}{MOE Key Laboratory of TianQin Mission, TianQin Research Center for Gravitational Physics, Frontiers Science Center for TianQin, CNSA Research Center for Gravitational Waves, Sun Yat-Sen University (Zhuhai Campus), Zhuhai 519082, People's Republic of China}
\newcommand{\SPA}{School of Physics and Astronomy, Sun Yat-Sen University (Zhuhai Campus), Zhuhai 519082, People's Republic of China}
\renewcommand\thetable{\Roman{table}}

\begin{document}

\title{Constraining the cosmological parameters using gravitational wave observations of massive black hole binaries and statistical redshift information}

\author{Liang-Gui Zhu}
\email{zhulg3@mail2.sysu.edu.cn}
\affiliation{\SPA}
\affiliation{\TRC}

\author{Yi-Ming Hu}
\email{huyiming@sysu.edu.cn}
\affiliation{\SPA}
\affiliation{\TRC}

\author{Hai-Tian Wang}
\email{wanght@pmo.ac.cn}
\affiliation{Purple Mountain Observatory, Chinese Academy of Sciences, Nanjing 210023, People's Republic of China}
\affiliation{School of Astronomy and Space Science, University of Science and Technology of China, Hefei, Anhui 230026, People's Republic of China}
\affiliation{\TRC}

\author{Jian-dong Zhang}
\email{zhangjd9@sysu.edu.cn}
\affiliation{\SPA}
\affiliation{\TRC}

\author{Xiao-Dong Li}
\email{lixiaod25@sysu.edu.cn}
\affiliation{\SPA}
\affiliation{CSST Science Center for the Guangdong-Hong Kong-Macau Greater Bay Area, Sun Yat-Sen University (Zhuhai Campus), Zhuhai 519082, People's Republic of China}

\author{Martin Hendry}
\email{Martin.Hendry@glasgow.ac.uk}
\affiliation{SUPA, School of Physics and Astronomy, University of Glasgow, Glasgow G12 8QQ, United Kingdom}

\author{Jianwei Mei}
\email{meijw@sysu.edu.cn}
\affiliation{\SPA}
\affiliation{\TRC}


\begin{abstract}
  Space-borne gravitational wave detectors like TianQin are expected to detect gravitational wave signals emitted by the mergers of massive black hole binaries. 
  Luminosity distance information can be obtained from gravitational wave observations, and one can perform cosmological inference if redshift information can also be extracted, which would be straightforward if an electromagnetic counterpart exists.
  In this paper, we concentrate on the conservative scenario where the electromagnetic counterparts are not available, and comprehensively study if cosmological parameters can be inferred through a statistical approach, utilizing the non-uniform distribution of galaxies as well as the black hole mass-host galaxy bulge luminosity relationship.  
  By adopting different massive black hole binary merger models, and assuming different detector configurations, we conclude that the statistical inference of cosmological parameters is indeed possible. 
  TianQin is expected to constrain the Hubble constant to a relative error of about 4\%-7\%, depending on the underlying model.
  The multidetector network of TianQin and LISA can significantly improve the precision of cosmological parameters.
  In the most favorable model, it is possible to achieve a level of 1.7\% with a network of TianQin and LISA.
  We find that without electromagnetic counterparts, constraints on all other parameters need a larger number of events or more precise sky localization of gravitational wave sources, which can be achieved by the multidetector network or under a favorable model for massive black hole mergers. 
  However, in the optimistic case, where electromagnetic counterparts are available, one can obtain useful constraints on all cosmological parameters in the Lambda cold dark matter cosmology, regardless of the population model. 
  Moreover, we can also constrain the equation of state of the dark energy without the electromagnetic counterparts, and it is even possible to study the evolution of equation of state of the dark energy when the electromagnetic counterparts are observed.

\end{abstract}

\maketitle

\acrodef{GW}{gravitational wave}
\acrodef{EM}{electromagnetic}
\acrodef{LCDM}[$\Lambda$CDM]{Lambda cold dark matter}
\acrodef{CPL}{Chevallier-Polarski-Linder}
\acrodef{SNIa}[SN Ia]{type Ia supernova}
\acrodefplural{SNIa}[SNe Ia]{type Ia supernovae}
\acrodef{CMB}{cosmic microwave background}
\acrodef{BNS}{binary neutron star}
\acrodef{StBBH}{stellar-mass binary black hole}
\acrodef{EMRI}{extreme mass ratio inspiral}
\acrodef{MBH}{massive black hole}
\acrodef{MBHB}{massive black hole binary}
\acrodef{SNR}{signal-to-noise ratio}
\acrodef{FIM}{Fisher information matrix}
\acrodef{MDPL}{MultiDark Planck}
\acrodef{MCMC}{Markov chain Monte Carlo}
\acrodef{EoS}{equation of state}

\section{Introduction}    \label{s1}

The first direct \ac{GW} detections by Advanced LIGO and Advanced Virgo \cite{2016PhRvL.116f1102A, 2016PhRvL.116x1103A, 2017PhRvL.118v1101A, 2017ApJ...851L..35A, 2017PhRvL.119n1101A, 2017PhRvL.119p1101A, GWTC-1:2018, LIGOScientific:2020ibl} have opened an era of \ac{GW} astronomy and the detected \ac{GW} events have provided powerful tests for astrophysics and fundamental physics \cite{LIGOScientific:2018jsj, 2016ApJ...818L..22A, 2016PhRvL.116v1101A}. 
The first direct \ac{GW} detection  from a \ac{BNS} merger, GW170817\cite{2017PhRvL.119p1101A} and its \ac{EM} counterpart identification \cite{2017ApJ...848L..13A, 2017ApJ...848L..12A} provided a ``cosmic distance ladder''-free determination of the Hubble constant $H_0$ \cite{2017Natur.551...85A, Hotokezaka:2018dfi}. Aside from the \ac{GW} signals with \ac{EM} counterpart, \ac{GW} signals without \ac{EM} counterpart have also provided effective measurements of $H_0$ \cite{Fishbach:2018gjp, 2019ApJ...876L...7S, Abbott:2019yzh, Palmese:2020aof}, and \acp{GW} have become standard sirens for cosmological investigations, as first proposed over thirty years ago \cite{1986Natur.323..310S}. 
There is an especially noticeable tension between local (or so-called late Universe) measurements of the Hubble constant and cosmological (or so-called early Universe) measurements of $H_0$ \cite{2016ApJ...826...56R, Riess:2019cxk, Freedman:2019jwv, Aghanim:2018eyx, Freedman:2017yms, Verde:2019ivm, Riess:2020sih, Riess:2020fzl}. The independent determinations of $H_0$ from \ac{GW} detections offer an effective way to clarify this ``Hubble tension'' \cite{Feeney:2018mkj}. 

The key to the success of the standard siren method involves obtaining redshift information for \ac{GW} sources.  One needs to either (1) identify the \ac{EM} counterpart and the host galaxy of \ac{GW} sources, or (2) obtain the statistical redshift distribution of candidate host galaxies, based on our knowledge of galaxy clustering \cite{1986Natur.323..310S}.  Novel methods have also been developed to obtain the redshift information of \ac{GW} sources, using such effects/information as \cite{2018SSPMA..48g9805Z}: 
(3) the strong gravitational lensing of GWs \cite{Sereno:2010dr, Sereno:2011ty, Liao:2017ioi}; 
(4) the mass distribution function of compact binary \ac{GW} sources, such as \cite{Taylor:2011fs, Taylor:2012db} for \ac{BNS} and \cite{Farr:2019twy, You:2020wju} for \ac{StBBH}, dependent on our understanding of the history of binary mergers; 
(5) the redshift probability distribution of compact binary mergers based on their intrinsic merger rates \cite{Ding:2018zrk}, requiring a large number of \ac{GW} events and also dependent on our understanding of the history of binary mergers; 
(6) the phase correction of \ac{BNS} merger GWs due to tidal effects \cite{DelPozzo:2015bna, Messenger:2011gi, Messenger:2013fya}, requiring detectors with higher sensitivity such as third-generation ground-based \ac{GW} detectors; 
(7) the evolution of \ac{GW} phase with cosmological expansion \cite{Seto:2001qf, Nishizawa:2010xx, Nishizawa:2011eq}, requiring detectors with both higher sensitivity and operating in the decihertz band. 

The potential of the second-generation ground-based \ac{GW} detector network, including Advanced LIGO, Advanced Virgo, KAGRA \cite{KAGRA:2018plz} and LIGO-India \cite{LIGO-India:2013}, to measure $H_0$ has been studied in detail, e.g., the measurements of $H_0$ from \ac{BNS} merger \ac{GW} detections \cite{Fishbach:2018gjp, Taylor:2011fs, Nissanke:2013fka, Mortlock:2018azx, Chen:2017rfc}, measuring $H_0$ with neutron star black hole mergers \cite{Vitale:2018wlg} and the constraints on $H_0$ via \ac{StBBH} mergers \ac{GW} detections \cite{Chen:2017rfc, DelPozzo:2012zz, Nair:2018ign, Farr:2019twy, Gray:2019ksv}. 

Third-generation ground-based \ac{GW} detectors such as the Einstein Telescope (ET) \cite{Punturo:2010zz} and Cosmic Explorer (CE) \cite{Evans:2016mbw} are expected to detect thousands of \ac{GW} events with a redshift concentration at $z \approx 2$ and a horizon of $z \approx 10$ \cite{Sathyaprakash:2009xt, Taylor:2012db, deSouza:2019ype, Ding:2018zrk}. 
The high-redshift \ac{GW} events detected by ET or CE can provide precise measurements of $H_0$ and other cosmological parameters, such as the density of dark matter $\Omega_M$, and of dark energy $\Omega_{\Lambda}$ \cite{Nair:2018ign, Taylor:2012db, Ding:2018zrk, You:2020wju, Zhao:2010sz, DelPozzo:2015bna, Cai:2016sby, Zhang:2018byx, Du:2018tia, Yu:2020vyy}. 
In addition, a large number of high-redshift \ac{GW} events would allow us to reconstruct the dark energy \ac{EoS} and expansion dynamics by nonparametric methods \cite{Seikel:2012uu, Cai:2016sby, Yang:2019vni}. 

Space-borne \ac{GW} detectors like TianQin will open for exploration the millihertz to Hertz band of the \ac{GW} spectrum. TianQin is a constellation of three satellites orbiting around the Earth, using drag-free control to lower noise and measure \ac{GW} effects through laser interferometry \cite{Luo:2015,Mei:2020lrl}.
TianQin is expected to observe multiple different types of \ac{GW} source, including \ac{StBBH} inspirals\cite{Liu:2020eko}, \acp{EMRI} \cite{Fan:2020zhy}, Galactic compact binaries \cite{Huang:2020rjf} and \ac{MBHB} mergers \cite{Wang:2019}. 
Such detections are also expected to put stringent constraints on deviations from general relativity or testing specific gravity theories \cite{Shi:2019hqa,Bao:2019kgt}.  Studies have revealed the availability of stable orbits that fulfils the requirement for \ac{GW} detections \cite{Zhang:2020paq,Ye:2020tze,Tan:2020xbm,Ye:2019txh}.  Moreover, the technology demonstration satellite, TianQin-1, has met the design requirements \cite{Luo:2020bls}.

Studies have been performed on the ability of \acp{GW} observations to constrain cosmological models with the Laser Interferometer Space Antenna (LISA) \cite{Holz:2005df, Dalal:2006qt, LISA:2017pwj, Arun:2008xf, Babak:2010ej, 2016JCAP...04..002T, Caprini:2016qxs, Cai:2017yww, Wang:2020dkc}.
\acp{GW} can usefully constrain cosmology even when no \ac{EM} information is available, according to studies carried out on the measurement of $H_0$ with \ac{MBHB} mergers \cite{Petiteau:2011we, Wang:2020dkc}, \ac{StBBH} inspirals \cite{Kyutoku:2016zxn, DelPozzo:2017kme}, and \acp{EMRI} \cite{macleod2008precision, Laghi:2021pqk}.
Next generation missions like the Decihertz Interferometer Gravitational Wave Observatory (DECIGO) \cite{Kawamura:2006up} and the Big Bang Observer (BBO) \cite{Crowder:2005nr} can also serve as powerful cosmological probes \cite{Nishizawa:2010xx, Nishizawa:2011eq, Namikawa:2015prh}. 

In this paper, we focus on the ability of space-borne \ac{GW} detectors to constrain the cosmological parameters using \ac{MBHB} merger \ac{GW} signals. 
Some studies have suggested that \ac{MBHB} mergers may have observable electromagnetic signatures \cite{Armitage:2002uu, Palenzuela:2010nf, Gold:PRD2014, Farris:mnrasl2014}, and some thus assume the availability of an \ac{EM} counterpart when performing \ac{GW} cosmology studies \cite{Holz:2005df, Arun:2008xf, 2016JCAP...04..002T, Cai:2017yww, Wang:2019tto, Wang:2021srv}. 
In order to be conservative, however, we set as our default assumption that no \ac{EM} counterparts are detectable for our \ac{GW} sources, so that we rely on the luminosity distance information from the \ac{GW} detections combined with statistical information about their host galaxy redshifts to constrain the cosmological parameters.
Our method can be simply described as follows: \ac{GW} detection provides a localisation error cone for the three-dimensional (3D) position of the \ac{GW} source; one can use the redshift distribution of the galaxies within the cone as the proxy for the \ac{GW} redshift.  However, there could be a lot of galaxies within the localisation error volume that will cause contamination and limit the effectiveness of the dark standard siren method.
We therefore study how using the relationship between the central \acp{MBH} and their host galaxies \cite{Graham:2007uq, Bentz:2008rt, Kormendy:2013dxa, Jiang:2011bt} can better pinpoint the host and improve the cosmological estimation. 
We study this under the assumption that both TianQin \cite{Luo:2015} and LISA \cite{LISA:2017pwj, Robson:2019} will be operating in the 2030s, with overlapping operation time and the scope for joint detections.

The remainder of this paper is organized as follows. 
In Sec. \ref{analysis-method}, we present the appropriate cosmology theory and introduce the astrophysical background needed for our analysis. 
In Sec. \ref{data-simulation}, we present the simulation method used to generate our observations and describe the characteristics of our simulated data. 
In Sec. \ref{cosmo_constraint-results}, we show the results of our constraints on the cosmological parameters. 
In Sec. \ref{discussion-part}, we discuss several key issues in our simulations. Finally, in Sec. \ref{conclusion-outlook}, we summarize our findings.

\section{Methodology}    \label{analysis-method}
\subsection{The cosmological models}    \label{LCDM-model}

Throughout this paper, we consider a Universe in which the spacetime can be described by the Friedmann-Lema\^itre-Robertson-Walker metric, and adopt the \ac{LCDM} model as our fiducial model, with the dark energy \ac{EoS} being described by a constant $\omega \equiv p_{\Lambda} / \rho_{\Lambda} = -1 $.
In this model the expansion of the Universe can be characterized by the Hubble parameter $H(z)\equiv \dot a/a$,  with $a$ being the scale factor, the current value of which is $a_0$, and with redshift $z$ defined as $z + 1 \equiv a_0/a$.
The Hubble parameter can therefore be expressed as 
\begin{equation}  \label{H_z}
H(z) = H_0 \sqrt{\Omega_M(1+z)^3 + \Omega_K(1+z)^2 + \Omega_{\Lambda}},
\end{equation}
where the Hubble constant $H_0 \equiv H(z=0)$ describes the current expansion rate of the Universe, and $\Omega_M$, $\Omega_K$ and $\Omega_{\Lambda}$ are respectively the current dimensionless fractional densities for the total matter, curvature and dark energy with respect to the critical density.
They satisfy the relationship  $\Omega_M + \Omega_K + \Omega_{\Lambda} \equiv 1$. 

We will also consider the possibility that the dark energy component has dynamical properties, by adopting the \ac{CPL} parametrization \cite{Chevallier:2000qy, Linder:2002et} 
\begin{equation}  \label{DE_EoS}
\omega(z) = w_0 + w_a \frac{z}{1+z}.  
\end{equation}
This yields the equation
\begin{equation}  \label{H_z_DE}
H(z) = H_0 \sqrt{\Omega_M(1+z)^3 + \Omega_K(1+z)^2 + \Omega_{\Lambda} \exp \Big( -\frac{3 w_a z}{1+z}\Big)(1+z)^{3(1+w_0+w_a)}}. 
\end{equation}

Estimates of the cosmological parameters can be inferred by fitting the relationship between observed luminosity distances $D_L$ and redshifts $z$.
This relationship encodes information about the expansion history of universe, and is given by
\begin{equation} \label{DL_z}
\renewcommand\arraystretch{1.5}
D_L = \frac{c(1+z)}{H_0} \left\{  \begin{array}{ll}
\frac{1}{\sqrt{\Omega_K}} \sinh[\sqrt{\Omega_K} \int_{0}^{z} \frac{H_0}{H(z')} d z']   &  \textrm{for $\Omega_K > 0$} \\
 \int_{0}^{z} \frac{H_0}{H(z')} d z'   &  \textrm{for $\Omega_K = 0$} \\
\frac{1}{\sqrt{|\Omega_K|}} \sin [\sqrt{|\Omega_K|} \int_{0}^{z} \frac{H_0}{H(z')} d z']   &  \textrm{for $\Omega_K < 0$}
\end{array} \right.
\end{equation}
where $c$ is speed of light in vacuum. 

We adopt cosmological parameters derived from various recent observations, like Planck \cite{Aghanim:2018eyx}, $H_0 = 67.8 $ km/s/Mpc, $\Omega_M = 0.307$, $\Omega_{\Lambda} = 0.693$ and we adopt $w_0 = -1$, $w_a = 0$, respectively, consistent with the observed galaxy distribution \cite{Klypin2016}.

\subsection{Standard sirens}    \label{standard-siren}

For the inspiral of a compact binary system with component masses $m_1$ and $m_2$, the frequency domain \ac{GW} waveform can be expressed as \cite{Sathyaprakash:2009xs}
\begin{equation} \label{h_f}
\widetilde{h}(f) = \frac{1}{D_L} \sqrt{\frac{5}{24}} \frac{(G {\mathcal{M}_z})^{5/6}}{\pi^{2/3} c^{3/2}} f^{-7/6} \exp  \big( -i\Phi(f; \mathcal{M}_z, \eta) \big)
\end{equation}
where $G$ is the gravitational constant, $\mathcal{M}=\eta^{3/5} M$ is the chirp mass, $\eta = m_1 m_2/M^2$ is the symmetric mass ratio, $M = m_1 + m_2$ is the total mass, and $\Phi(f; \mathcal{M}_z, \eta)$ is phase of the waveform depending on the parameters $\mathcal{M}$ and $\eta$. 
The chirp mass $\mathcal{M}$ largely determines the overall evolution of the \ac{GW} waveform, but the parameter directly measured from the data is actually the redshifted chirp mass $\mathcal{M}_z \equiv \mathcal{M}(1+z)$. 

One can see from Eq. (\ref{h_f}) that the luminosity distance $D_L$ of the binary has a direct impact on the measured waveform amplitude.
Therefore, \ac{GW} observations of compact binary coalescences can be used to estimate their corresponding luminosity distance directly.
If the redshifts of such mergers can be inferred through other observational or theoretical channels, one can use information from both luminosity distance and redshift to constrain the cosmological parameters \cite{1986Natur.323..310S}. 
Such a one-stop measurement of luminosity distance makes the inspiral signal of compact binary systems desirable objects for cosmological studies, since they are largely immune to systematic errors caused by intermediate calibration stages like those required by type Ia supernovae; consequently, they have been coined as ``standard sirens" .

There is a catch, however.
Both $D_L$ and $z$ are needed in order to use Eq. (\ref{DL_z}) to infer cosmological parameters.
Moreover, as noted above, the redshift $z$ is deeply intertwined with the chirp mass $\mathcal{M}$ and can not be solely determined by \ac{GW} observations of the inspiral.
The measurement of redshift information thus relies on extra information, like the identification of the host galaxy or the direct observation of an \ac{EM} counterpart.

\subsection{Bayesian framework}    \label{bayes}

We adopt a Bayesian framework to infer cosmological parameters through the \ac{GW} observations of massive black hole binary mergers.
Consider a set of data composed of $N$ \ac{GW} observations, $D \equiv \{ D_1, D_2, ..., D_i, ...,D_N \}$, as well as an \ac{EM} data set $S$ derived from \ac{EM} observations.
Then one can derive the posterior probability distribution of the cosmological parameters $\vec{\Omega}$ as
\begin{equation} \label{bayes_formula}
p(\vec{\Omega}|D, S, I) = \frac{p_0(\vec{\Omega}|I) p(D, S|\vec{\Omega}, I)}{p(D, S|I)} = \frac{p_0(\vec{\Omega}|I) \prod_i p(D_i, S|\vec{\Omega}, I)}{p(D, S|I)},
\end{equation}
where $I$ indicates all relevant background information. 
Since the normalisation factor, also known as the Bayesian evidence, $p(D, S|I)$, is irrelevant in the calculation, the posterior can be written as
\begin{equation} \label{likeli0}
p(\vec{\Omega}|D, S, I) \varpropto p_0(\vec{\Omega}|I) \prod_i p(D_i, S|\vec{\Omega}, I).
\end{equation}

We can classify parameters into three categories: the common parameters (which affect both \ac{GW} waveforms and \ac{EM} observations); \ac{GW}-only parameters $\vec{\theta}'$, and \ac{EM}-only parameters $\vec{\phi}'$.
Throughout this paper, we identify the common parameters as the luminosity distance $D_L$, the redshift $z$, the longitude $\alpha$, the latitude $\delta$, the total mass $M$ of the compact binary, and the bulge luminosity $L_{\rm bulge}$ of the host galaxy.
We can express the likelihood as
\begin{equation} \label{likeli1}
p(D_i, S|\vec{\Omega}, I) = \frac{\int p(D_i, S, D_L, z, \alpha, \delta, M_z, L_{\rm bulge}, \vec{\theta}', \vec{\phi}' | \vec{\Omega}, I) d D_L d z d \alpha d\delta d M_z d L_{\rm bulge} d \vec{\theta}' d \vec{\phi}'}{\beta(\vec{\Omega}|I)},
\end{equation}
where $M_z = M(1+z)$ is the redshifted total mass of the \ac{GW} source.  Notice that we introduce a correction term $\beta(\vec{\Omega} | I)$ to eliminate the effect of selection biases \cite{2017Natur.551...85A, Chen:2017rfc, Mandel:2018mve}. 
The integrand in the numerator of Eq. (\ref{likeli1}) can be factorized as 
\begin{align} \label{likeli3}
& p(D_i, S, D_L, z, \alpha, \delta, M_z, L_{\rm bulge}, \vec{\theta}', \vec{\phi}' | \vec{\Omega}, I)   \nonumber \\
= \!~& p(D_i | D_L, \alpha, \delta, M_z, \vec{\theta}', I) p(S | z, \alpha, \delta, L_{\rm bulge}, \vec{\phi}', I) p_0(D_L | z, \vec{\Omega}, I) p_0(M_z | z, L_{\rm bulge}, \vec{\Omega}, I)  \nonumber \\ 
 &\!\! \times p_0(z, \alpha, \delta, L_{\rm bulge} | \vec{\Omega}, I) p_0(\vec{\theta}'|\vec{\Omega}, I) p_0(\vec{\phi}'|\vec{\Omega}, I) .
\end{align}
For details on the derivation of Eq. (\ref{likeli3}), please refer to Appendix \ref{derivation_Eq9}. 
The general mathematical treatment of the likelihood $p(D_i|D_L,\alpha,\delta,M_z,\vec{\theta}',I)$ can be given by \cite{Finn:1992} 
\begin{equation}  \label{likeli_GW}
p(D_i|D_L,\alpha,\delta,M_z,\vec{\theta}',I) \varpropto \exp \Big[-\frac{1}{2} \big \langle D_i - h(D_L,\alpha,\delta,M_z,\vec{\theta}')|D_i - h(D_L,\alpha,\delta,M_z,\vec{\theta}')\big \rangle  \Big],
\end{equation}
where $\langle \cdot|\cdot \rangle$ is the inner product as defined in Eq. (\ref{inner_product}). 
When the host galaxy of the \ac{GW} signal cannot be uniquely identified, one can set $p(S|z,\alpha,\delta,L_{\rm bulge}, \vec{\phi}',I)$ as a constant \cite{Chen:2017rfc}. 
We assume that $p_0(D_L|z,\vec{\Omega},I) \equiv \delta \big( D_L - \hat D_L(z, \vec{\Omega}) \big)$ depends on the cosmological model 
and $p_0(M_z|z,L_{\rm bulge},\vec{\Omega},I) \equiv \delta  \big( M_z - (1+z) \hat M(L_{\rm bulge})  \big)$ is based on the relation of massive black hole mass with galactic luminosity  \cite{Bentz:2008rt, Kormendy:2013dxa}, where the total mass $\hat M(L_{\rm bulge})$ is a function of the galactic bulge luminosity $L_{\rm bulge}$. 

The \ac{EM} measurements of parameters like position of the galaxies $(\alpha, \delta)$ are much more precise compared with the measurements from \ac{GW} observations. 
The prior in Eq. (\ref{likeli3}) can therefore be approximated as \cite{Chen:2017rfc, Petiteau:2011we, DelPozzo:2012zz}
\begin{equation} \label{prior_EM}
p_0(z,\alpha,\delta,L_{\rm bulge} | \vec{\Omega},I) = \frac{1}{N_{\rm gal}} \sum_{j=1}^{N_{\rm gal}} \mathcal{N}(z | {\bar z^j}, \sigma_z^j) \delta(\alpha - \alpha^j) \delta(\delta - \delta^j) \mathcal{N}(L_{\rm bulge} | \bar{L}_{\rm bulge}^j, \sigma_{L_{\rm bulge}}^j),
\end{equation}
where $N_{\rm gal}$ is the number of galaxies in our \ac{EM} catalog, and $\mathcal{N}(x|\bar x, \sigma_x)$ is a Gaussian distribution on $x$, with expectation $\bar{x}$ and standard deviation $\sigma_x$, here $x=\{z, L_{\rm bulge}\}$. 
However, if the luminosity function $\Phi(L)$ is also regarded as redshift-dependent (as in \cite{Cohen:2001ag, Marchesini:2006vg, Bouwens:2014fua} etc.), then Eq. (\ref{prior_EM}) needs to be supplemented, as shown in Eq. (\ref{catalog_debias}).

Marginalizing over the parameters $D_L$, $M_z$, $\vec{\theta}'$, and $\vec{\phi}'$, Eq. (\ref{likeli1}) becomes
\begin{equation} \label{likeli4}
p(D_i, S|\vec{\Omega}, I) \varpropto \frac{\int p \big( D_i | \hat D_L(z, \vec{\Omega}), \alpha, \delta, (1+z)\hat M(L_{\rm bulge}), I \big) p_0(z, \alpha, \delta, L_{\rm bulge} | \vec{\Omega}, I) d z d \alpha d \delta d L_{\rm bulge}}{\beta(\vec{\Omega}|I)}.
\end{equation}

We next examine the correction term $\beta(\vec{\Omega}|I)$.
Bias of the inferred cosmological parameters might arise from two sources: the \ac{GW} data and the \ac{EM} information. 
The \ac{EM} information contains the following bias: (1) the 3D error volume is cone like, so this will in general lead to a bias towards higher redshift due to the larger associated volume; and (2) the incompleteness of the catalog will introduce a Malmquist bias, where brighter galaxies are disproportionately recorded and weighted \cite{1922MeLuF.100....1M}. 
To obtain cosmological constraints from ``dark'' standard sirens depends on the non-uniform distribution of galaxies, due to large scale structures (LSS) or smaller-scale clustering of galaxies, so the correction term should exclude the influence of LSS and galaxy clustering information as far as possible.
We account for the aforementioned two biases by counting the detectable galaxies over the whole sky, and considering the redshift evolution of this galaxy count. 
We assume that the EM observations are isotropic within the survey region, and define the prior for the redshift distribution of galaxy catalog as
\begin{equation} \label{p_noLSS}
p_\textrm{c}(z|\vec{\Omega},I)  \varpropto \frac{1}{2\Delta z} \int_{(z-\Delta z)}^{(z+\Delta z)} \int \!\!\!\! \int_{4\pi} \int p_0(z',\alpha,\delta,L_{\rm bulge} | \vec{\Omega},I) d L_{\rm bulge} d \alpha d \delta d z',
\end{equation}
where $\Delta z$ is chosen to be much larger than the typical redshift scale of the LSS. 
The correction term after marginalization is then
\begin{equation} \label{N_term}
\beta(\vec{\Omega}|I) \approx \int p(D_i|\hat D_L(z,\vec{\Omega}),I) p_\textrm{c}(z|\vec{\Omega},I) d z.
\end{equation}
Notice that the \ac{GW} selection effect is accounted for by integrating over only events detectable by \ac{GW} detectors. 
In addition, if the catalog of survey galaxies contains two or more sky areas with different observation depths, then $p_\textrm{c}(z|\vec{\Omega},I)$ and $\beta(\vec{\Omega}|I)$ terms need to be calculated separately for each sky area.

For our analysis, the incompleteness of a galaxy catalog can introduce two effects.
First, it could bring Malmquist bias so that more distant galaxies are disproportionately weighted.
Such a systematic bias is caused by the selection effect that a catalog tends to be more complete for brighter galaxies.
The correction term $\beta(\vec{\Omega}|I)$ accounts for this bias.
On the other hand, a less complete catalog means there is a higher chance for the host galaxy cluster to be missed.
 In this case, an additional bias could be introduced to the analysis.
In Sec. \ref{consistency-check} we demonstrate that such a large deviation can be identified through  application of a consistency check.
Meanwhile, in \cite{Gray:2019ksv} the authors take a step further than Eq. (\ref{N_term}), so that more distant events are down-weighted and a consistency check is considered unnecessary.

\subsection{Parameter estimation of the \ac{GW} sources}    \label{GWparas-estimate}

\begin{figure}[htbp]
\centering
\includegraphics[height=6.cm, width=9.cm]{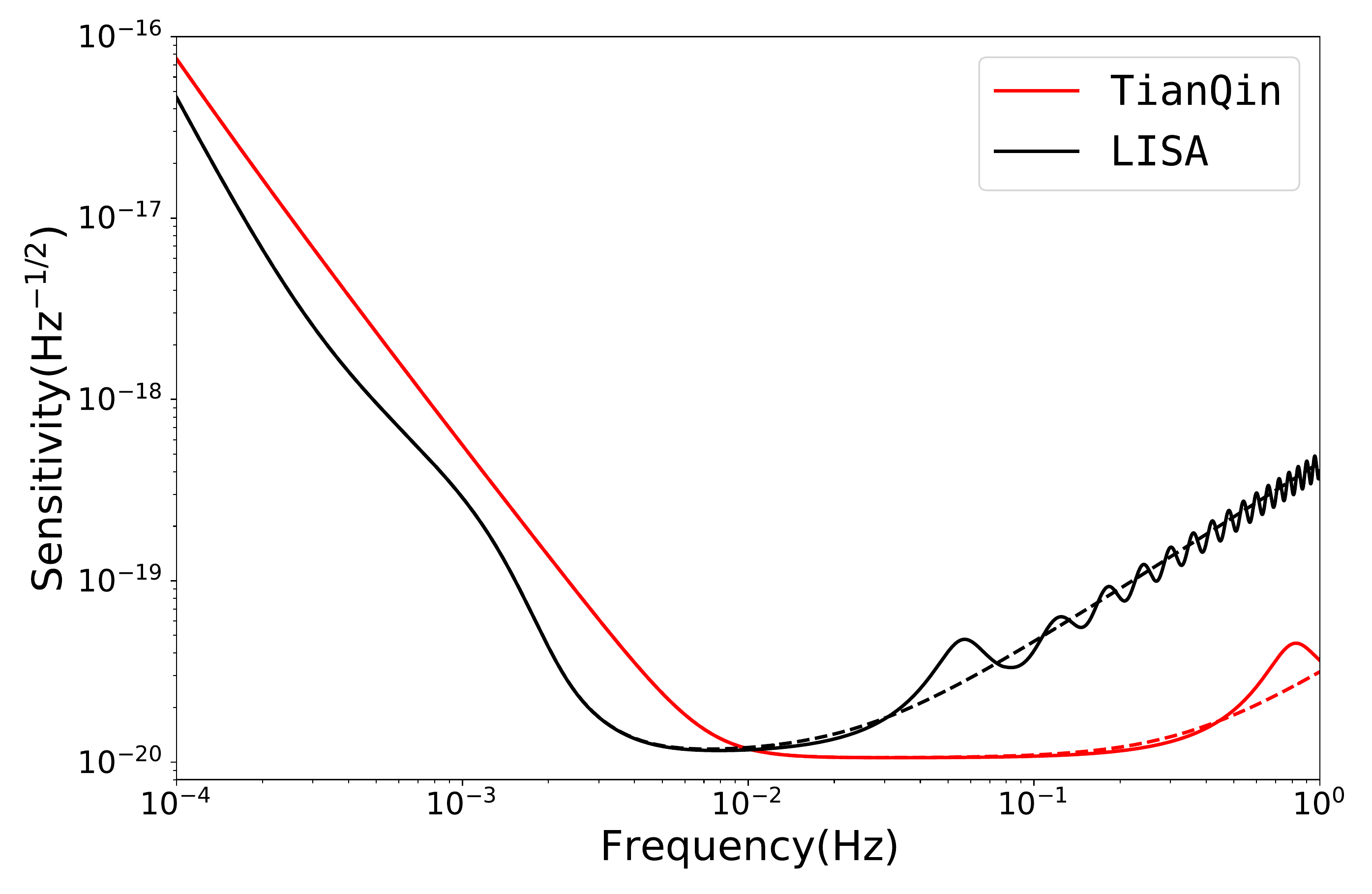}
  \caption{Anticipated averaged sensitivity curves of TianQin (red line) and LISA (black line). Base on Eqs. (\ref{S_N-TQ}) and (\ref{S_N-LISA}).  }
\label{sensi_cur}
\end{figure}

We define the sensitivity curve $S_n(f)$  in terms the expected power spectral density  $S_N(f)$ according to the relation,
\begin{equation} \label{reation-SN_Sn}
  S_n (f) = \frac{S_N (f)}{\bar{\mathcal{T}}(f)}, 
\end{equation}
where $\bar{\mathcal{T}}(f)$ is the sky and polarization averaged response function of the detector,
\begin{equation} \label{sky_averaged}
  \bar{\mathcal{T}}(f) \approx \frac{1}{1 + 0.6 {(2 \pi f L / c )}^2}, 
\end{equation}
and
\begin{align} 
S_N (f) &=  \frac{1}{L^2} \bigg[ \frac{4 S_a}{(2\pi f)^4}\left(1 + \frac{10^{-4}\textrm{Hz}}{f}\right) + S_x \bigg] + S_c(f), && \rm{for \ TianQin},  \label{S_N-TQ}  \\
S_N (f) &=  \frac{1}{L^2} \bigg[ \frac{4 S_a}{(2\pi f)^4} \bigg(1 + \Big( \frac{4 \times 10^{-4}\textrm{Hz}}{f} \Big)^2 \bigg) + S_x \bigg] + S_c(f), && \rm{for \ LISA}.  \label{S_N-LISA}
\end{align}
Here $L$ is the arm length, $S_a$ is the acceleration noise, $S_x$ is the positional noise, and $S_c(f)$ is the Galactic foreground noise. 
For TianQin, $S_a^{1/2}=1 \times 10^{-15} ~\textrm{m} \ \textrm{s}^{-2}  \textrm{Hz}^{-1/2}$, $S_x^{1/2}=1 \times 10^{-12} ~\textrm{m} \ \textrm{Hz}^{-1/2}$, and the arm-length $L=\sqrt{3} \times 10^8 ~\textrm{m}$ \cite{Luo:2015,Wang:2019}; 
for LISA, we adopt $S_a^{1/2}=3 \times 10^{-15} ~\textrm{m} \ \textrm{s}^{-2} \textrm{Hz}^{-1/2}$, $S_x^{1/2}=1.5 \times 10^{-11} ~\textrm{m} \ \textrm{Hz}^{-1/2}$, and arm length $L= 2.5 \times 10^9 ~\textrm{m}$ \cite{Robson:2019}. 
In addition, we added the Galactic foreground on top of the sensitivity curve assuming an observation time of 4 years according to \cite{Robson:2019}. 
On the other hand, no foreground is added for TianQin and it has been suggested that the anticipated foreground will be below TianQin's sensitivity \cite{Huang:2020rjf,Liang:2021bde}. 
The sky- and polarization-averaged sensitivity curves of TianQin and LISA are shown in Fig. \ref{sensi_cur}. 

A simultaneous observation from multiple detectors can improve the sky localisation of the \ac{GW} source.
Specifically, the long baseline between different \ac{GW} detectors makes it possible to use the time delay information to perform sky localisation \cite{2017PhRvL.119n1101A, 2017PhRvL.119p1101A, GWTC-1:2018, 2017Natur.551...85A, Wang:2020dkc, Ruan:2020smc}. 

Suppose that the $k$th detector record the \ac{GW} strain $h^k(t)$, the strain takes the form
\begin{equation} \label{h_t}
h^k (t) =  \frac{1+\cos^2 \iota}{2} A(t) \cos \!\big[\Phi(t) + \Phi_D^k(t) + \Phi_P^k(t)\big] F_+^k(t) + \cos \iota ~A(t) \sin \!\big[\Phi(t) + \Phi_D^k(t) + \Phi_P^k(t) \big] F_{\times}^k(t)
\end{equation}
where $\iota$ is the source inclination angle, $A(t)$ and $\Phi(t)$ are the amplitude and phase of the \ac{GW} respectively, $F_{+,\times}^k(t)$ are the response 
functions of the $k$th detector, $\Phi_D^k(t)$ is the Doppler frequency modulation due to the Doppler effect of the solar orbital revolution of the detector, and $\Phi_P^k(t)$ is the phase modulation. 
The expressions for the phase modulation and the response function are closely related to the orbit of the detectors. We follow \cite{Wang:2019, Hu:2018} for TianQin, and \cite{Cutler:1998, 2003PhRvD..67b2001C} for LISA, adopting the low-frequency limit.  

For a circularised binary black hole merger event, the \ac{GW} signal can be generally described by a set of parameters including the binary component masses $m_1$ and $m_2$ (which can be reexpressed as chirp mass $\mathcal{M}$ and symmetric mass ratio $\eta$), the luminosity distance $D_L$, the longitude $\alpha$ and latitude $\delta$ of the source sky position, the coalescence phase $\phi_c$, the coalescence time $t_c$, and the direction of the orbital angular momentum of the \ac{GW} source $(\alpha_L, \delta_L)$ (which can be reexpressed as an inclination angle $\iota$ and polarization angle $\psi$). More parameters would be needed if the spins of the black holes are included.

Let us consider a detector network including $N$ independent detectors, for which the frequency domain \ac{GW} signal $\widetilde{\boldsymbol{h}}(f)$ can be written as
\begin{equation} \label{h_t-vector}
\widetilde{\boldsymbol{h}}(f) = \Big[ \widetilde{h}^1(f),\widetilde{h}^2(f),\cdots, \widetilde{h}^k(f),\cdots, \widetilde{h}^N(f) \Big]^{\rm T}
\end{equation}
where $\widetilde{h}^k(f)$ denotes the Fourier transform of $h^k(t)$. 
The \ac{SNR} $\rho$ of a \ac{GW} signal  $\widetilde{\boldsymbol{h}}(f)$ can be defined as \cite{Finn:1992, Cutler:1994}
\begin{equation} \label{SNR}
\rho = \langle \widetilde{\boldsymbol{h}}(f) | \widetilde{\boldsymbol{h}}(f) \rangle ^{1/2},
\end{equation}
where the inner product symbol $\langle \cdot | \cdot \rangle$ is defined as
\begin{equation} \label{inner_product}
\langle \widetilde{\boldsymbol{h}}(f) | \widetilde{\boldsymbol{h}}(f) \rangle = \sum_k \langle \widetilde{h}^k(f)|\widetilde{h}^k(f) \rangle \equiv \sum_k 4 \mathfrak{Re} \int_{0}^{\infty} \frac{\widetilde{h}^{k*}(f) \widetilde{h}^k(f)}{S_n^k(f)} d f,
\end{equation}
where $*$ represents complex conjugate, $\mathfrak{Re}$ denotes the real component, and $S_n^{k} (f)$ are the sensitivity curve functions of the $k$th detector respectively when the response functions $F_{+,\times}(t)$ adopt the low-frequency approximation \cite{Wang:2019, Robson:2019}. 

For a \ac{GW} signal from a binary characterised by physical parameters $\boldsymbol{\theta}=(\mathcal{M}, \eta, D_L, \alpha, \delta, \cos \iota, \psi, \phi_c, t_c)$, the inverse of the corresponding \ac{FIM} sets the Cram\'{e}r-Rao lower bound for the covariance matrix \cite{Vallisneri_2008}.
The \ac{FIM} can be written as 
\begin{equation} \label{FIM}
\Gamma_{mn} \equiv \bigg \langle \frac{\partial \widetilde{\boldsymbol{h}}(f)}{\partial \theta_m} \bigg| \frac{\partial \widetilde{\boldsymbol{h}}(f)}{\partial \theta_n} \bigg \rangle,
\end{equation}
where $\theta_m$ indicates the $m$th parameter. 
With $\Sigma = \Gamma^{-1}$, we take the estimation uncertainty on parameter 
$\theta_m$ as $\Delta \theta_m = \sqrt{\Sigma_{mm}}$, with the sky localization error $\Delta \Omega$ given by the combination $\Delta \Omega = 2\pi |\sin \delta| \sqrt{\Sigma_{\alpha\alpha} \Sigma_{\delta\delta} - \Sigma_{\alpha\delta}^2}$. 

If the \ac{GW} signals are gravitationally lensed, or the peculiar velocity of its host galaxy is not properly accounted for, the inferred luminosity distance would be in error. Therefore, in addition to the measurement error from \ac{GW} observation, weak lensing and peculiar velocity can both contribute to the intrinsic uncertainty of the estimated $D_L$. 
We denote $\sigma_{D_L}^{\textrm{GW}}$ as the uncertainty arising from the \ac{GW} observation, and $\sigma_{D_L}^{\textrm{tot}}$ as the overall uncertainty including all sources of error. 
We adopt a fitting formula from \cite{Hirata:2010ba} (also see \cite{Cusin:2020ezb}) to estimate the weak lensing error $\sigma_{D_L}^{\textrm{lens}}$, given by
\begin{equation} \label{sigma_lens}
  \sigma_{D_L}^{\textrm{lens}}(z) = \frac{1}{2} D_L(z) \sigma_{\ln D_L^2}  = \frac{1}{2} D_L(z) \times C_l \bigg[ \frac{1 - (1+z)^{-\beta_l}}{\beta_l} \bigg]^{\alpha_l}, 
\end{equation}
where $C_l = 0.066$, $\beta_l = 0.25$ and $\alpha_l = 1.8$. 
For the error, $\sigma_{D_L}^{\rm pv}$, due to peculiar velocity we adopt the fitting formula from \cite{Kocsis:2005vv,Gordon:2007zw}, 
\begin{equation} \label{sigma_v}
\sigma_{D_L}^{\rm pv}(z) = D_L(z) \times \bigg[ 1 + \frac{c (1+z)^2}{H(z)D_L(z)} \bigg] \frac{\sqrt{\langle v^2 \rangle}}{c}, 
\end{equation}
with $\sqrt{\langle v^2 \rangle}=500 \ \textrm{km}/\textrm{s}$ as the root mean square peculiar velocity of the host galaxy with respect to the Hubble flow \cite{He:2019dhl}. 
We then define the total uncertainty
\begin{equation} \label{sigma_tot}
\sigma_{D_L}^{\textrm{tot}} = \sqrt{(\sigma_{D_L}^{\textrm{GW}})^2 + (\sigma_{D_L}^{\textrm{lens}})^2 + (\sigma_{D_L}^{\rm pv})^2 },
\end{equation}

Throughout this paper, we use $\sigma_{D_L}^{\textrm{tot}}$ for our likelihood calculation.
However, we remark that the above is a conservative estimate of the total uncertainty, since ``delensing'' methods like the use of weak lensing maps \cite{Shapiro:2010MNRAS}, observations of the foreground galaxies \cite{Jonsson:2006vc} or deep shear surveys \cite{Hilbert:2010am} could in principle be used to alleviate $\sigma_{D_L}^{\textrm{lens}}$, while peculiar velocity maps could also be used to reduce $\sigma_{D_L}^{\rm pv}$ \cite{2017Natur.551...85A,Howlett:2019mdh}. 
We do not apply such corrections in this paper; hence we can expect better results for \ac{GW} cosmological inference from realistic future detections. 

\section{Simulations}    \label{data-simulation}   

For the purpose of our paper, we require a catalog of massive black hole mergers that mimic our understanding of the real Universe. 
In this section, we describe the simulation of these catalogs, and how we use the 3D localisation information derived from a \ac{GW} detection, as well as the empirical $M_{\rm MBH}-L_{\rm bulge}$ relation, to allocate probabilities to candidate host galaxies.

\subsection{Massive black hole binary mergers and galaxy catalog}    \label{MBHB_population-survey_catalog}
Following previous studies, e.g., \cite{Klein:2016,Wang:2019}, we adopt the massive black hole binary merger populations from \cite{2012MNRAS.423.2533B}.
Both the ``light-seed'' scenario and the ``heavy-seed'' scenario are considered for the seeding models of massive black holes.
In the light-seed scenario, seed black holes are assumed to be the remnants of first generation (or population III) stars and the mass of the seed black hole is around 100 $M_{\odot}$ \cite{Madau:2001sc, Volonteri:2002vz}.
This model is later referred to as \emph{popIII}.
In the heavy-seed scenario, the MBH seeds are assumed to be born from the direct collapse of protogalactic disks that may be driven by bar instabilities \cite{Klein:2016}, with a mass of $\sim 10^5 ~M_\odot$. 
The critical Toomre parameter $Q$ that determines when the protogalactic disks become unstable is set to 3 \cite{Volonteri:2007ax}.
Two models are derived from this scenario, with \emph{Q3d} considering the time lag between the merger of \acp{MBH} and merger of galaxies, and \emph{Q3nod}, which does not consider such a time lag. 
In both scenarios, the seed BHs grow via accretion and mergers, eventually becoming the massive black holes, while the evolution of \acp{MBH} are deeply coupled to the evolution of their host galaxies \cite{Ferrarese:2000se}.

In addition to the simulated catalog of \ac{GW} mergers, we also require a simulated galaxy catalog, so that we can associate each \ac{MBHB} merger with a certain host galaxy and use multimessenger information to infer cosmological parameters.
TianQin has the ability of observing very distant mergers, but no existing galaxy survey project can extend to these distances so we choose to adopt the \ac{MDPL} cosmological simulation \cite{Klypin2016} from the Theoretical Astrophysical Observatory (TAO) \cite{TAO-web} for this purpose.
The \ac{MDPL} simulated a catalog of galaxies based on an $N$-body simulation which tracks the evolution of dark matter halos assuming a Planck cosmology \cite{Ade:2015xua}, with $3840^3$ particles and a box side length to $h^{-1}$ $\textrm{Gpc}$ (where $h \equiv \frac{H_0 }{100 \ \textrm{km/s/Mpc}}$), and the simulation is performed from $z=100$ to $z=0$.
Additional information such as galaxy evolution was obtained from the semi-analytic galaxy evolution model \cite{Croton:2016etl}, and the luminosity distribution from Conroy \emph{et al.} \cite{Conroy:2009gw}.

\subsection{\ac{GW} event catalog}  \label{GW_catalog}
We first simulated \ac{MBHB} mergers according to the three models (popIII, Q3d, and Q3nod), where only events with network \ac{SNR} $\rho > 8$ are later used.
Notice that although TianQin has the ability to observe very distant events, it is anticipated that complete galaxy catalogs would be extremely hard to obtain for galaxies beyond redshift $z=3$, therefore we apply a redshift cut beyond this point. 
The remaining parameters are then generated from simple distributions: all angle parameters are chosen uniformly in solid angle, $\alpha \in \textrm{U}[0, 2\pi]$, $\cos \delta \in \textrm{U}[-1, 1]$, $\alpha_L \in \textrm{U}[0, 2\pi]$, and $\cos \delta_L \in \textrm{U}[-1, 1]$, and we choose the merger time $t_c \in \textrm{U}[0, 5]$ years, merger phase $\phi_c \in \textrm{U}[0, 2\pi]$ and the spin $\chi_{1,2} \in \textrm{U}[-1, 1]$. 
Based on these parameters, we generate the \ac{GW} waveform from the IMRPhenomPv2 model \cite{Hannam:2014}.

Throughout this analysis, we consider multiple configurations for the space-borne \ac{GW} detectors as described below:  
\begin{itemize}
  \item{(i) \emph {TianQin}}: the default case where three satellites form a constellation and operate in a ``3 month on + 3 month off'' mode, with a mission life time of 5 years \cite{Luo:2015};  
  \item{(ii) \emph {TianQin I+II}}: a twin constellation of satellites are used that have perpendicular orbital planes to avoid the 3-month gaps in data \cite{Wang:2019, Liu:2020eko};  
  \item{(iii) \emph {LISA}}: we consider the mission as described in \cite{LISA:2017pwj, Robson:2019}, with a nominal mission life time of 4 years;  
  \item{(iv) \emph {TianQin+LISA}}: TianQin and LISA observing together, with 4 years of overlap in operation time;  
  \item{(v) \emph {TianQin I+II+LISA}}: similar to above but with the TianQin I+II configuration considered. 
\end{itemize}
Previous work indicates that both TianQin \cite{Wang:2019} and LISA \cite{Klein:2016} have good detection capabilities for \ac{MBHB} mergers. 
In Table \ref{detection_rate} we list the anticipated total detection rates for mergers at redshift $z < 3$ under different detector configurations. 
It is worth noting that the actual merger rate of \ac{MBHB} is likely to increase by about twice as much as the three population models predict \cite{Klein:2016}, and our actual detection rate would therefore also likely increase by about twice as much.

\begin{table}[]
  \caption{Expected total detection rate of \ac{GW} events with $z < 3$ and \ac{SNR} $\rho > 8$, detected over the entire observation time based on the \ac{MBHB} population models for five different detector configurations: TianQin, TianQin I+II, LISA, TianQin+LISA, and TianQin I+II+LISA. }
    \vspace{12pt}
    \renewcommand\arraystretch{1.1}
    \centering
    \begin{tabular}{ccccccc}
        \hline
        \hline
        ~ & \multicolumn{2}{c}{popIII} & \multicolumn{2}{c}{Q3d} & \multicolumn{2}{c}{Q3nod}  \\
        \cline{2-7}
        \thead[c]{Detectors \\ configuration} & \thead[c]{Detection \\ rate} & \thead[c]{Detection \\ percentage} & \thead[c]{Detection \\ rate} & \thead[c]{Detection \\ percentage} & \thead[c]{Detection \\ rate} & \thead[c]{Detection \\ percentage}  \\
        \hline
        TianQin            & 7.7    & 33.0\%   &  4.1    & 32.6\%   & 25.5   & 26.8\%   \\
        TianQin I+II       & 12.0   & 51.4\%   &  7.2    & 56.6\%   & 41.8   & 44.0\%   \\
        LISA               & 11.1   & 61.0\%   &  6.5    & 64.2\%   & 37.8   & 50.8\%   \\          
        TianQin+LISA       & 14.0   & 68.1\%   &  8.0    & 71.0\%   & 46.9   & 56.4\%   \\          
        TianQin I+II+LISA  & 15.6   & 72.3\%   &  9.1    & 75.9\%   & 53.0   & 60.2\%   \\          
        \hline  
        \hline     
    \end{tabular}
    \label{detection_rate}
\end{table}

\begin{figure}[htbp] 
\centering
\includegraphics[width=12.cm, height=12.cm]{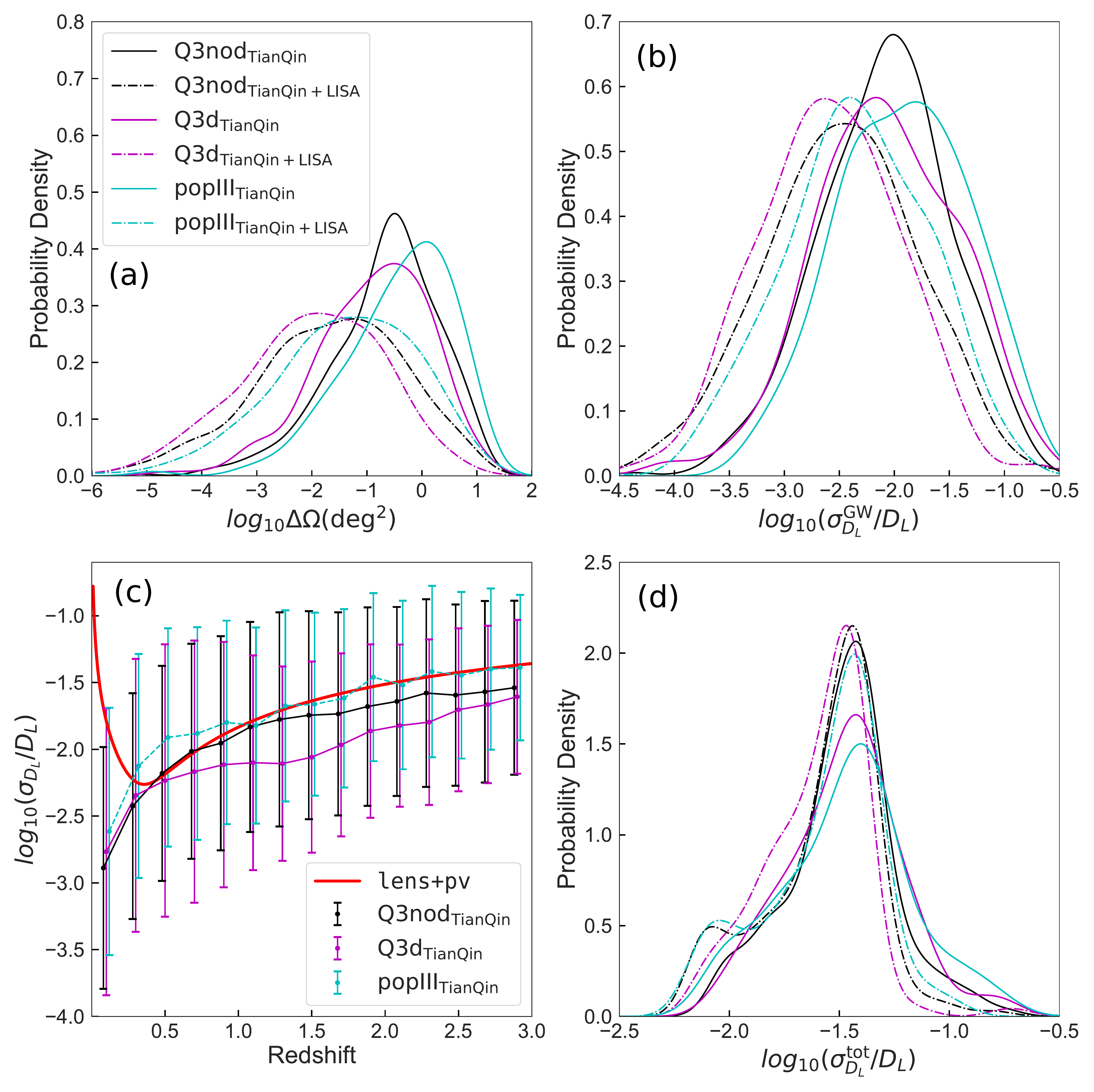}
  \caption{Spatial localisation error distribution for TianQin (solid line) and the LISA-TianQin network (dot-dashed line). 
  The top panels show the error due to \ac{GW} observations, with the panel (a) showing the sky localisation error, and the panel (b) showing the relative uncertainty on luminosity distance. 
  In the bottom panels, we illustrate uncertainties of luminosity distances, also considering uncertainties due to weak lensing and peculiar velocity.
  The panel (c) shows the uncertainty evolution over redshift summarised from 1000 simulations, and the panel (d) shows similar information as the panel (b) but with all uncertainty sources considered. 
  The results with TianQin I+II are quite similar to those of TianQin so for clarity we do not include them here. }
\label{paras_err_3model}
\end{figure}

In order to identify the host galaxy, the most important information that we gather from the \ac{GW} observation would be the spatial localisation, equivalently $\alpha$, $\delta$, and $D_L$.
In Fig. \ref{paras_err_3model}, we illustrate the marginalised distribution on the sky localisation error $\Delta \Omega$, as well as the relative error on the luminosity distance $\sigma_{D_L}/D_L$.

As shown in Fig. \ref{paras_err_3model}, a typical \ac{MBHB} merger can be localised to better than $1 ~{\rm deg}^2$ with TianQin alone, while a combination of both TianQin and LISA can improve the localisation precision by a factor of $1 \sim 2$ orders of magnitude, where a small fraction of sources can even be localised to within $10^{-4} ~{\rm deg}^2$.
However, this is generally not sufficient to pinpoint the host galaxy uniquely, especially considering the relatively large uncertainty in luminosity distance as well as the large distance range to which the detectors can reach.

Thanks to the relatively high \ac{SNR}, the typical $D_L$ uncertainty arising from the \ac{GW} observations is around the level of 1\% across the three models, while the combination of both TianQin and LISA can further improve the precision by a factor of about three.
However, as indicated in the bottom left panel of Fig. \ref{paras_err_3model}, often the \ac{GW} measurement error is the subdominant contribution. 
Therefore the overall error in luminosity distance, after considering the effects of weak lensing and peculiar velocity, is of order 10\%.

\subsection{$M_{\rm MBH}-L_{\rm bulge}$ relation}  \label{M-L-relation}
Once the \ac{MBHB} mergers and galaxy catalogs are ready, the next step is to associate the appropriate galaxy as the host galaxy for each merger.
Throughout this paper, we link the \ac{MBHB} mergers and their host galaxies through the $M_{\rm MBH}-L_{\rm bulge}$ relation.
We adopt the form of this relation as \cite{Graham:2007uq, Bentz:2008rt, Kormendy:2013dxa}
\begin{equation} \label{M-L_bulge}
  \log_{10} \left(\frac{M_{\rm MBH}}{M_\odot}\right) = \log_{10} \left(\frac{M_0 }{M_\odot}\right)+ k_c \Big[\log_{10} \left(\frac{L_{\textrm{bulge}}}{L_{\odot}}\right) + \log_{10} \left(\frac{L_0}{L_\odot}\right)\Big],
\end{equation}
where $M_0$, $k_c$ and $L_0$ are fitting parameters from astronomical observations. 
In the $K$-band, the values of the fitting parameters are $\log_{10} \left(\frac{M_0 }{M_\odot}\right)= 8.735$, $k_c = 1.22$, $\log_{10} \left(\frac{L_0}{L_\odot}\right)= -11.604$, respectively, and with an intrinsic scatter $\sigma^{\textrm{int}}_{M-L} \approx 0.3$ in the logarithm of mass \cite{Kormendy:2013dxa}. 

This $M_{\rm MBH}-L_{\rm bulge}$ relation is fairly well supported by observations for \acp{MBH} in the mass range of $\sim 10^7 - 10^{10} ~M_{\odot}$ \cite{Kormendy:2013dxa, Jiang:2011bt}. 
For the low-mass \ac{GW} sources with $M < 10^6 ~M_{\odot}$, although one can still use Eq. (\ref{M-L_bulge}) to describe the $M_{\rm MBH}-L_{\rm bulge}$ relation, it is accompanied with significantly larger scatters \cite{Jiang:2011bt}. 
Therefore, in the low-mass end, we adopt a different intrinsic scatter $\sigma^{\textrm{int}}_{M-L} \approx 0.5$ \cite{Jiang:2011bt}. 
This means that the intrinsic scatter for \ac{MBHB} mass from the $M_{\rm MBH}-L_{\rm bulge}$ relation is a step function  
\begin{equation} \label{M-L-scatter}
\sigma^{\textrm{int}}_{M-L} (M_{\rm MBHB}) = \left\{  \begin{array}{ll}
0.3,   \ \   &  M_{\rm MBHB} > 5 \times 10^6 ~M_\odot;  \\
0.5,   \ \   &  M_{\rm MBHB} \leq 5 \times 10^6 ~M_\odot.
\end{array} \right.
\end{equation}
And in the following, we will refer to the luminosity in terms of solar luminosity and black hole mass in terms of solar masses, denoted by $L_{\odot}$ and $M_\odot$, respectively.

\subsection{Simulation of the statistical redshift information}  \label{host-selection}
The dark standard siren method relies on the measurement of luminosity distance from \ac{GW} observations, and the inference of redshift information from galaxy catalogs.
So the appropriate identification, or at least the association, of the host galaxy with the merging \ac{MBHB} is of the utmost importance.

We first obtain a conservative range of luminosity distance for the possible host galaxy, $[D_L^{-}, D_L^{+}] = [(\bar D_L - 3\sigma_{D_L}^{\textrm{tot}}), (\bar D_L + 3\sigma_{D_L}^{\textrm{tot}})]$, where $\bar D_L$ is mean estimated value.
Notice that we shall not use the galaxy luminosity distance information directly, otherwise it is pointless to introduce \ac{GW} observations for constraining cosmology.
Instead, we convert the luminosity distance $D_L$ into redshift $z$, under a wide range of priors on the cosmological parameters, to obtain an appropriate boundary on redshift $[z_{\min}, z_{\max}]$.
We choose the boundary of $H_0 \in [60, 80] ~\textrm{km/s/Mpc}$, $\Omega_M \in [0.04, 0.6]$, and $\Omega_{\Lambda} \in [0.4, 1]$ for $\Lambda$CDM model, and the boundary of $w_0 \in [-2, -0.5]$ and $w_a \in [-1, 1]$ for CPL dark energy model, and therefore $D_L^{-} = \min \{D_L(z_{\min}, H_0, \Omega_M, \Omega_{\Lambda})\}$ and $D_L^{+} = \max \{ D_L(z_{\max}, H_0, \Omega_M, \Omega_{\Lambda})\}$.
This choice of parameters ensures that mainstream estimates of Hubble constant, albeit in disagreement with each other, are nonetheless well encapsulated by our prior \cite{Riess:2019cxk, Aghanim:2018eyx, Verde:2019ivm, Riess:2020sih, Riess:2020fzl}.

\begin{figure}[htbp] 
\centering
\includegraphics[width=12.cm, height=11.cm]{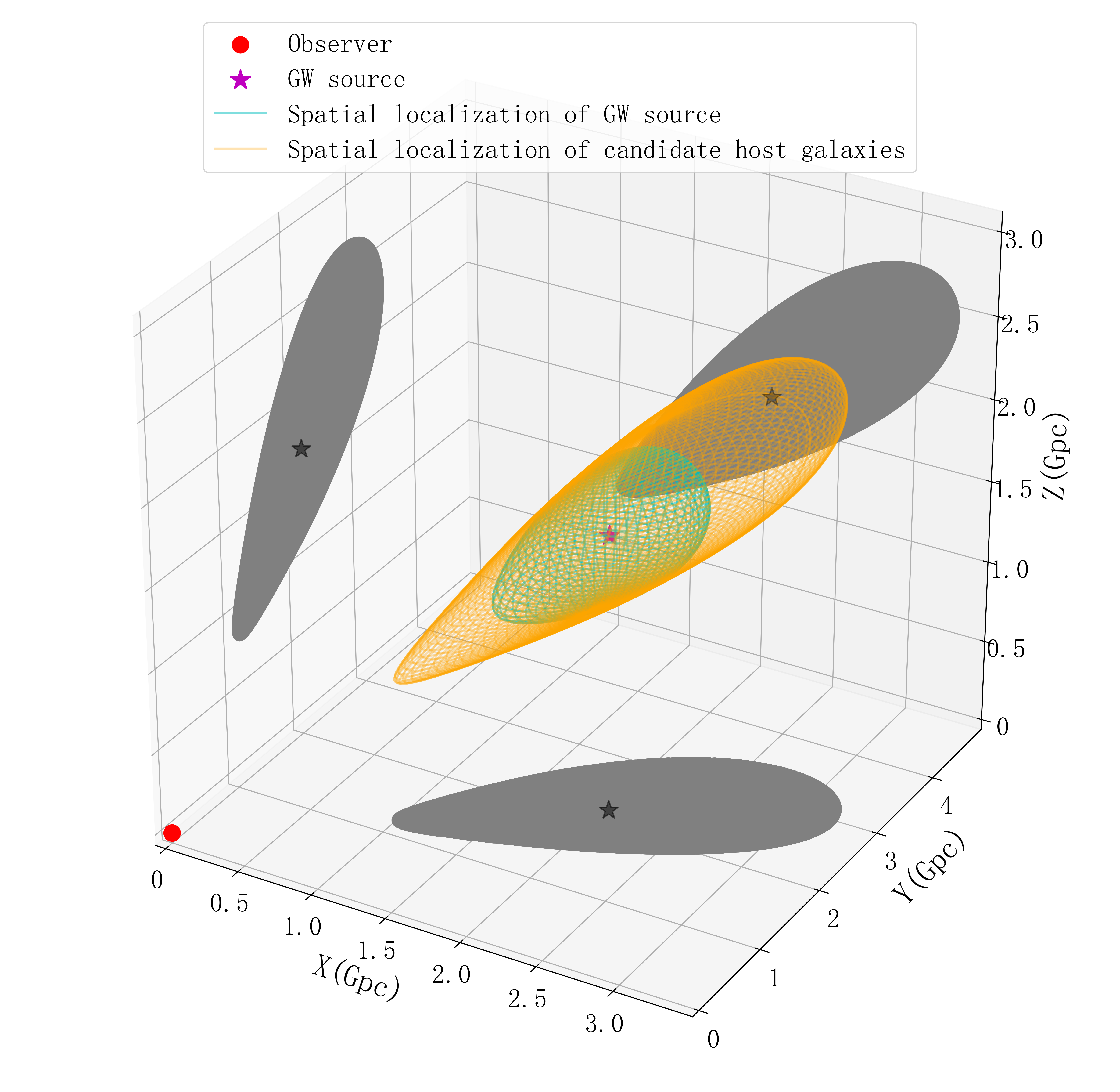}
  \caption{An example spatial localization error for a \ac{GW} source. 
  The cyan volume represents the spatial error due to the \ac{GW} observation, while for the orange volume, we convert the $[z_{\min}, z_{\max}]$ back into luminosity distance assuming the correct cosmology. The gray shadows are the projections of the spatial localization error, the red dot is the observer, and the purple star is the true input position of the \ac{GW} source. }
\label{location_error_box}
\end{figure}

It is worth noting that the spatial localisation errors of the \ac{GW} sources are indeed of irregular shape, instead of 3D ellipsoids or cylinders as some previous analyses have assumed for simplicity. We illustrate the localisation error for a typical event in Fig. \ref{location_error_box}. 
Notice also that in order to further simplify the ensuing calculation, we replace the elliptic sky localisation error with a circular shape but of the same area, which shall not alter the statistical conclusions of our paper. 

We then artificially assign a random galaxy to be the actual host galaxy. 
For a galaxy with bulge luminosity $L_{\rm bulge}$, we assign a weight according to a log-normal distribution with the $M_{\rm MBH}-L_{\rm bulge}$ relation. 
Next, we aim to simulate the real observation, to obtain a catalog of galaxies and to assign a probability of each galaxy hosting the merging black hole binary. 
Furthermore, we simulate the Malmquist bias to be more realistic.
For the $j$th galaxy with luminosity $L_j$ and redshift $z_j$, the probability of it being recorded is
\begin{equation} \label{selection_function}
{\rm erfc} (\log_{10} L_j) = \frac{1}{\sqrt{2\pi \sigma_{\log_{10} L_j}}} \int_{\log_{10} L_{\rm limit}(z_j, \vec \Omega)}^{\infty} \exp \bigg( -\frac{1}{2} \frac{(\log_{10} L - \log_{10} L_j)^2}{\sigma_{\log_{10} L_j}^2} \bigg) d (\log_{10}L) 
\end{equation}
where $ \log_{10} L_{\rm limit}(z_j, \vec \Omega) = \frac{(+4.83 - m_{\rm limit} - 5)}{2.5} + 2\log_{10}(\frac{D_L({z_j, \vec \Omega})}{1 \rm pc}) $, dependent on a given set of cosmological parameters $\vec \Omega$, and we adopt a limiting magnitude of $m_{\rm limit} = +24 \ \rm mag$ and equal measurement error of luminosity $\sigma_{\log_{10} L} =0.04$ (correspond to an uncertainty in magnitude of $0.1 \ \textrm{mag}$) \cite{York:2000gk, DES:2017myt, Gong:2019yxt, Laureijs:2011gra}.

In the analysis stage, we try to eliminate the Malmquist bias by introducing a redshift-dependent correction factor using the luminosity function of both the galaxies \cite{MonteroDorta:2008ib, Cohen:2001ag, Marchesini:2006vg, Faber:2005fp} and the bulge.
We need to manually augment sample sizes for further galaxies. 
We first divide the error box into  multiple small regions both in sky location as well as in redshift.
For each region, we can calculate the number of supplementary galaxies $\hat N_{\textrm{sup}}$ using a luminosity function $\Phi(L)$, 
\begin{equation} \label{N_sup}
\hat N_{\textrm{sup}} = N_{\textrm{obs}} \frac{\int_{0}^{L_{\rm limit}} \Phi(L) d L}{\int_{L_{\rm limit}}^{\infty} \Phi(L) d L}, 
\end{equation}
where we derive the luminosity function $\Phi(L)$ from the \ac{MDPL} simulated catalog \cite{Conroy:2009gw}. 
For a given \ac{GW} source, the debiased prior of the location is determined by the possible host galaxies, which can be expressed as 
\begin{align} \label{catalog_debias}
p_0^{\textrm{debiased}}(z,\alpha,\delta,L_{\rm bulge}|\vec{\Omega},I) \simeq \frac{1}{N_{\rm tot}} \sum  \bigg( & \sum_{m=1}^{N_{\textrm{obs}}} \mathcal{N}(z | {\bar z^m}, \sigma_z^m) \delta(\alpha - \alpha^m) \delta(\delta - \delta^m) \mathcal{N}(L_{\rm bulge} | \bar{L}_{\rm bulge}^m, \sigma_{L_{\rm bulge}}^m)  \nonumber  \\
+ & \hat N_{\textrm{sup}} \mathcal{N}(z | {\bar z_{\rm obs}}, \sigma_{z,{\rm obs}}) \delta(\alpha - \bar \alpha_{\textrm{obs}}) \delta(\delta - \bar \delta_{\textrm{obs}}) \Phi'(L_{\rm bulge})  \bigg),
\end{align}
where $N_{\textrm{tot}} = \sum(N_{\textrm{obs}} + \hat N_{\textrm{sup}})$, and $\bar z_{\textrm{obs}}$, $\bar \alpha_{\textrm{obs}}$, and $\bar \delta_{\textrm{obs}}$ are the mean values of redshift, longitude and latitude for the observed galaxies in the small region. 
For nearby galaxies ($z < 1$), we assume that spectroscopic redshift information is available and therefore the error can be neglected \cite{DESI:2016fyo}. 
However, for further galaxies the redshift if more likely obtained through photometric measurement, which is assumed to be associated with an error $\Delta z = 0.03(1 + z)$ \cite{Ilbert:2013bf, Dahlen:2013fea}.
Therefore, we adopt
\begin{equation}  \label{photo_z-error}
\sigma_z(z) = \left\{  \begin{array}{ll}
0,             \ \      &  z < 1;  \\
0.03(1 + z),   \ \      &  z \geq 1.
\end{array} \right.
\end{equation}
The bulge luminosity function with an apostrophe, $\Phi'(L_{\rm bulge})$, is defined as
\begin{equation}  \label{LF_phi-primed}
\Phi'(L_{\rm bulge}) = \left\{  \begin{array}{ll}
\frac{\Phi(L_{\rm bulge})}{\int_{0}^{L_{\rm bulge}^{\min}} \Phi(L_{\rm bulge}) d L_{\rm bulge}},   \ \   &  0\leq L_{\rm bulge} \leq L_{\rm bulge}^{\min};  \\
  \\
0,   \  \      &  L_{\rm bulge} \geq L_{\rm bulge}^{\min} ;
\end{array} \right.
\end{equation}
where $L_{\rm bulge}^{\min}$ is the minimum bulge luminosity of observed galaxies in the small region, and we derive the bulge luminosity function $\Phi(L_{\rm bulge})$ from the \ac{MDPL} simulated catalog. 
We remark that the second term in Eq. (\ref{catalog_debias}) does not represent a new batch of galaxies, but rather an adjustment to the weights of existing galaxies.

Next we assign different weights for galaxies with different positions and bulge luminosities. 
We consider the two following methods:  
\begin{itemize}
  \item{(i) \emph{fiducial method}}: each galaxy in the spatial localisation error box of the GW source has equal weight regardless of its position and luminosity information;  
  \item{(ii) \emph{weighted method}}: the weight of a galaxy is the product of both its positional weight and bulge luminosity weight.  
\end{itemize}
The positional weight is simply determined by the 3D space localisation from the \ac{GW} parameter estimation.
The luminosity-related weight, on the other hand, is more complicated.
For an observed galaxy, the weight is assigned through a log-normal distribution, with the expected redshifted mass value being the mean value derived from the \ac{GW} parameter estimation, and standard deviation $\sigma_{\log_{10} M}$.
And for a manually supplemented galaxy, a further integration of this log-normal distribution over luminosity up to $L_{\rm bulge}^{\min}$ is needed. 
A detailed expression for both the positional weight and the luminosity-related weight for a given galaxy is provided in Appendix \ref{expression_weight}. 

The luminosity-related weight of galaxies can be affected by the intrinsic scatters of the $M_{\rm MBH}-L_{\rm bulge}$ relation as well as the uncertainties related to the measurements. 
We define $\sigma_{\log_{10} M}$ as the root sum squared of the intrinsic scatter of the relation between the central \ac{MBH} mass and the galactic bulge luminosity $\sigma_{M-L}^{\textrm{int}}$, the measurement error on the total mass of \ac{GW} sources $\sigma_{\log_{10} M}^{\rm GW}$, and the error related to the bulge luminosity $k_c \sigma_{\log_{10} L_{\rm bulge}}^{\rm EM}$. 
We adopt $\sigma_{\log_{10} M}^{\rm GW} \equiv 0.05$ as a conservative choice since the mass parameter can usually be accurately determined through \ac{GW} observation.
For the measurement error on the bulge luminosity, since we do not have complete information on the bulge luminosity distribution for high redshift galaxies, we adopt the galaxy luminosity error as a proxy through $\sigma_{\log_{10} L_{\rm bulge}}^{\rm EM} \simeq 5(1+z) \sigma_{\log_{10} L} = 0.2(1+z)$.

\begin{figure}[htbp] 
\centering
\includegraphics[height=12.cm, width=12.cm]{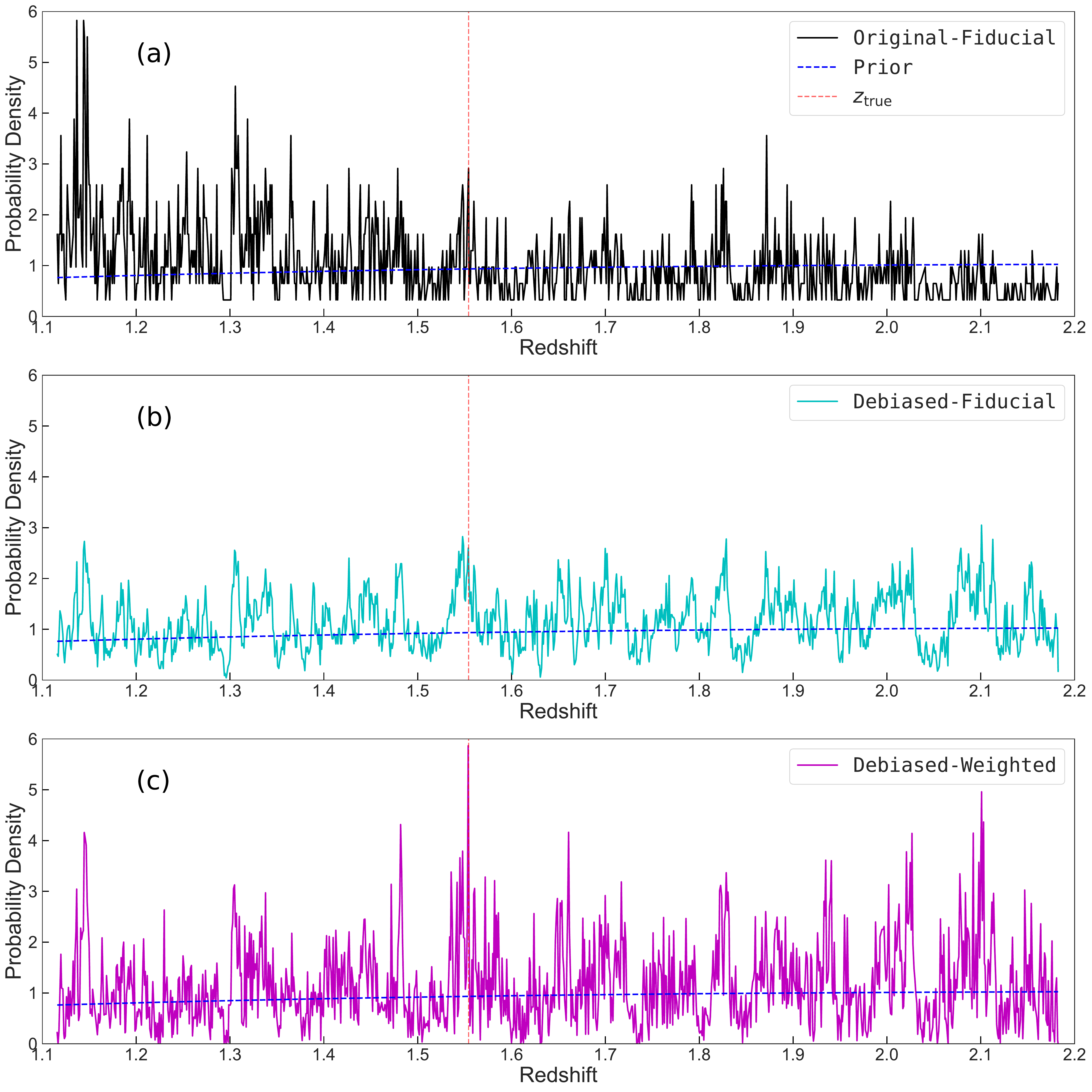}
\caption{Example: Redshift distribution of galaxies for an example \ac{GW} source. 
Top panel: Considering no debias correction, and assuming all galaxies within the error box as equally likely to be the host galaxy. 
Middle panel: Same as the top panel but including a correction for Malmquist bias. 
Bottom panel: Corrected for Malmquist bias, and assigning galaxy weights according to their position as well as luminosity. 
The vertical orange dashed line represents the real redshift of the \ac{GW} event, and the horizontal blue dashed line represents the prior distribution where galaxies have constant density over comoving volume.}
\label{z_distribution}
\end{figure}

We demonstrate the effect of the debias and different weighting in Fig. \ref{z_distribution}.
For the top panel, we do not correct for Malmquist bias, and there is an apparent excess of galaxies at low redshift. 
In the middle and bottom panel, the Malmquist bias is corrected by manually introducing supplementary galaxies; in these cases the distribution roughly follows the prior.
For the top and middle panel, all galaxies within the error box are assumed equally likely to be the host galaxy of the \ac{MBHB} merger, while for the bottom panel, we assign weights to the galaxies using the weighted method.
One can observe that the bottom panel is less smooth, and the injected source stands out from the contaminating galaxies.

\section{Cosmological Constraints}    \label{cosmo_constraint-results}

In this section, we explore quantitatively the prospects for parameter estimation with \ac{GW} cosmology.
We perform a series of \ac{MCMC} studies using the widely adopted library \textsf{emcee}, a \textsf{Python} package that implements an affine-invariant \ac{MCMC} ensemble sampler \cite{ForemanMackey:2012ig, ForemanMackey:2019ig}.
For the detector(s), we consider various scenarios, including: TianQin, TianQin I+II, LISA, TianQin+LISA, and TianQin I+II+LISA. 
We investigate both the dark standard siren case, where no \ac{EM} counterpart is available, and the bright standard siren case where we can use an \ac{EM} counterpart to identify the host galaxy and therefore to obtain explicitly the redshift from \ac{EM} observations.
For the popIII and Q3d models we generate 200 mock \ac{GW} events from \ac{MBHB} mergers, while for the Q3nod model, the event rate of which is expected to be higher than the other two models, we generate 600 mock events. 
These events form three sample pools, and the \ac{GW} events on which each realization of the cosmological parameter estimation relies are chosen randomly from these three pools. 

In order to comprehensively probe the systematic and random errors in the inferred cosmological parameters, we repeat the cosmological analysis multiple times with independent runs for each detector configuration/\ac{MBHB} population model/weighting method.

\subsection{TianQin and TianQin I+II}    \label{TQ-result}

We first consider the most pessimistic scenario, where TianQin is operating alone and no \ac{EM} counterpart is expected to be observed.
For convenience, in this analysis we simply select all galaxies that fall into the range of $9\Delta \Omega \times [z_{\min}, z_{\max}]$ (although the actual error box would correspond to a shape type characterised by Fig. \ref{location_error_box}).

\begin{figure}[htbp]
\centering
\includegraphics[width=12.cm, height=11.cm]{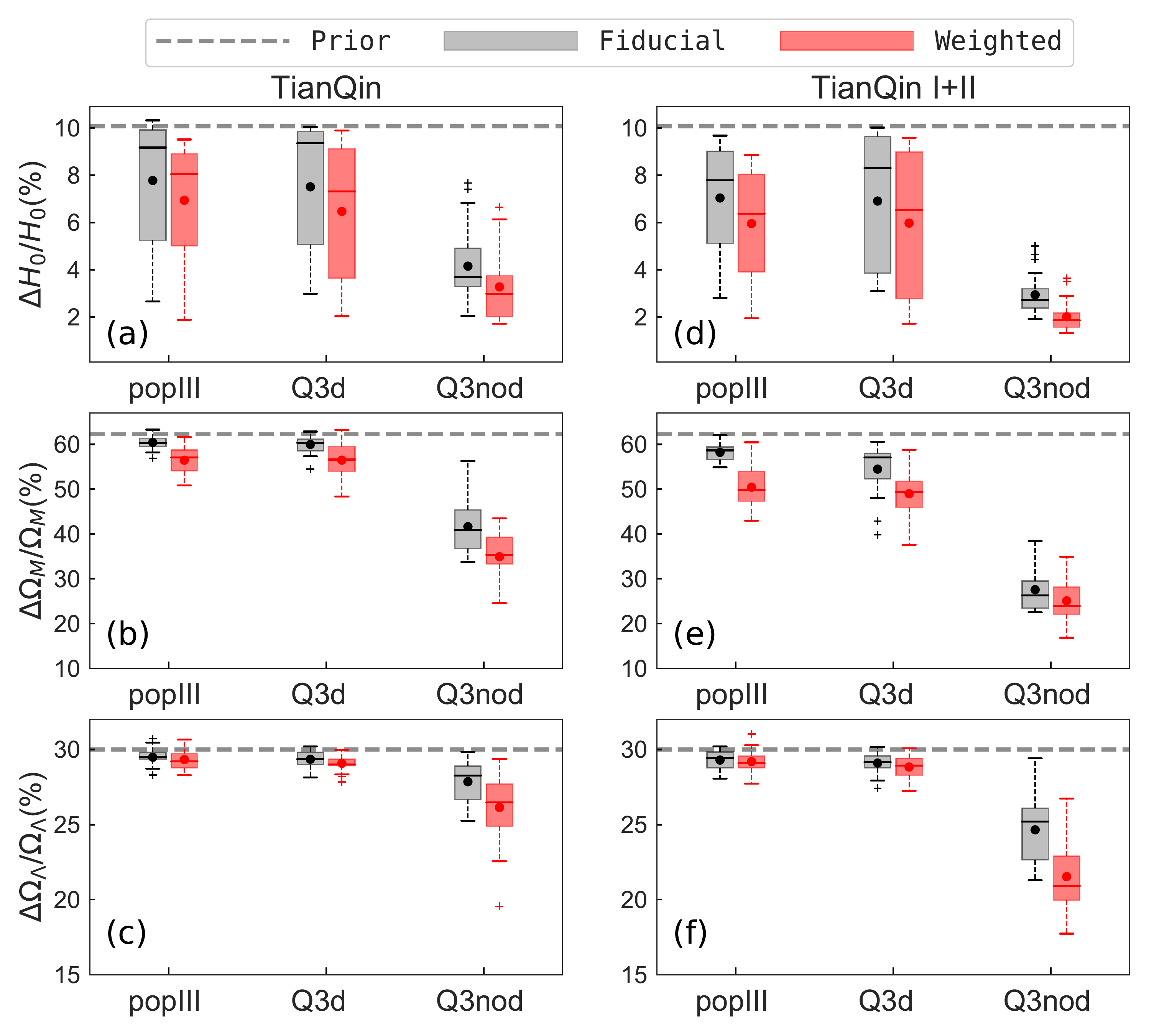}
\caption{Boxplot of the precision of the estimated cosmological parameters for the three \ac{MBHB} models, assuming two detector configurations, i.e., TianQin (left column) and TianQin I+II (right column).  
The top, middle, and bottom rows illustrate results for three cosmological parameters, $H_0$, $\Omega_M$, and $\Omega_\Lambda$, respectively. 
The horizontal gray dashed line represents a fiducial $68.27\%$ statistical interval from the prior. 
For each result, the box represents the range of $25\%-75\%$ of the data distribution, the upper limit of the whisker length is 1.5 times the box length, and the crosses are outliers. 
In each box, the dot represents the mean value, the short horizontal line represents the median value, using the fiducial method (gray) and the weighted method (red), respectively.} 
\label{H0ML_accuracy_TQ}
\end{figure}

\begin{figure}[htbp]
\centering
\includegraphics[width=13.cm, height=7.5cm]{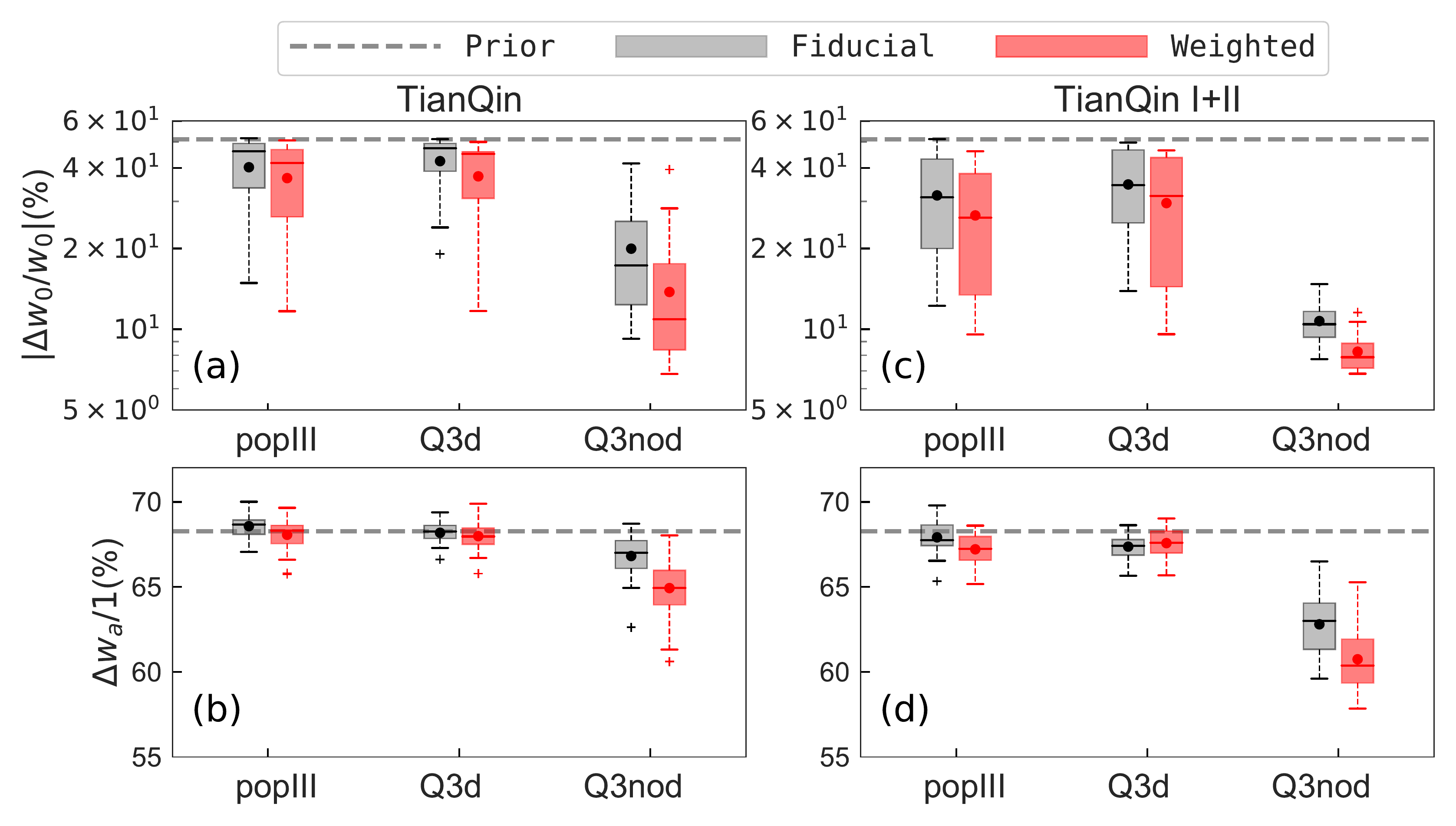}
\caption{Boxplot of the precision of the estimated dark energy \ac{EoS} parameters for the three \ac{MBHB} models, assuming two detector configurations, i.e., TianQin (left column) and TianQin I+II (right column), and keeping non-\ac{CPL} parameters fixed.  
The top and bottom illustrate results for three cosmological parameters, $w_0$ and $w_a$, respectively. 
The horizontal gray dashed line represents a fiducial $68.27\%$ statistical interval from the prior.
The gray and red boxplots represent the distribution of estimation precisions of the parameters using the fiducial method and the weighted method, respectively. }
\label{w0_accuracy_TQ}
\end{figure}

As illustrated by \cite{2019PhRvD..99l3002F}, the \ac{SNR} accumulates rather quickly just before the final merger.
Therefore, the difference in the precision of cosmological parameter estimation between TianQin and TianQin I+II is rooted in their different detection numbers.
The distributions of the constraint precisions of various cosmological parameters for TianQin (left column) and TianQin I+II (right column) under the three \ac{MBHB} population models are shown in Fig. \ref{H0ML_accuracy_TQ}. 
In order to eliminate the random fluctuations caused by the specific choice of any event, we repeat this process 24 times, and the constraint results are presented in the form of boxplots. 

The top, middle, and bottom rows represent the estimation precision of $H_0$, $\Omega_M$, and $\Omega_\Lambda$, respectively. 
The constraining ability of these parameters are decreasing as with this order.
In each panel, from left to right, we show the results adopting popIII, Q3d, and Q3nod as the underlying astrophysical model, respectively.
In each model, we generate a number of \ac{GW} event catalogs, assuming an event rate that follows a Poisson distribution with the rate parameter determined by Table \ref{detection_rate}.
It can be observed that more events leads to better constraints on the cosmological parameters.
Furthermore, the weighted method (red box) shows better constraining ability than the fiducial method (gray box).

In general, under the models with a lower MBHB event rate, such as popIII and Q3d models, we can only constrain the Hubble constant, which is almost entirely determined by low redshift \ac{GW} events. 
The constraints of other cosmological parameters require more GW events, which is feasible under the Q3nod model. 
For TianQin, the relative precision on $H_0$ is estimated to be $7.8 \%$, $7.5 \%$ and $4.2 \%$ via the fiducial method; by using the weighted method, the relative precision can reach $6.9\%$, $6.5\%$ and $3.3\%$, for the popIII, Q3d and Q3nod models, respectively. 
For TianQin I+II, the number of \ac{MBHB} mergers would be boosted by a factor of about 2, and the relative error of $H_0$ reduces to $6.0\%$, $6.0\%$, and $2.0\%$ using the weighted method, respectively. 
Under the Q3nod model, $\Omega_M$ and $\Omega_\Lambda$ are expected to be constrained to a relative precision of $34.9 \%$ ($25.1 \%$) and $26.1 \%$ ($21.5 \%$) for TianQin (TianQin I+II), respectively. 

Finally, we study the constraining power of the \ac{GW} observations on the \ac{EoS} of dark energy, assuming the \ac{CPL} model for its redshift evolution. 
Here we fix $H_0=67.8$ km/s/Mpc, $\Omega_M = 0.307$, and $\Omega_{\Lambda} = 0.693$.
We find that, for all astrophysical models and detector configurations, $w_a$ can hardly be constrained; however, meaningful constraints can be obtained on $w_0$, as shown in Fig. \ref{w0_accuracy_TQ}. 
And similarly, the weighted method also leads to more precise constraints on the parameters of \ac{EoS} of dark energy compared to the fiducial method. 
If using the weighted method, the relative precision of $w_0$ can be reach a level of $36.7\%$, $37.2\%$, and $13.8\%$ for TianQin, and $26.6 \%$, $29.6 \%$, and $8.1 \%$ for TianQin I+II --- for popIII, Q3d, and Q3nod models respectively. 

Besides, in order to more graphically represent the advantages of the weighted method over the fiducial method, we show in Appendix \ref{cosmoexample-z_survey}, the typical posterior probability distributions of the cosmological parameters and the parameters of \ac{EoS} of the dark energy constrained by TianQin.

\subsection{Network of TianQin and LISA}    \label{TQ_LISA-result}

For \ac{GW} observations, the power of the cosmological constraints is crucially related to how well the sky positions of the sources may be determined.
With a network of multiple \ac{GW} detectors working simultaneously, one can greatly improve this sky localisation error. 
The orbital planes of the TianQin constellation and LISA constellation are pointing in different directions; thus, joint detections can break the degeneracy between longitude and latitude of GW source, and the difference in the arrival time between the two detectors can also narrow the margin of the localisation error. 
However, the measurement of $D_L$ suffers from the systematics arising from weak lensing and peculiar velocities, and thus can hardly be improved even in the network observation case. 

\begin{figure}[htbp]
\centering
\includegraphics[width=12.cm, height=11.cm]{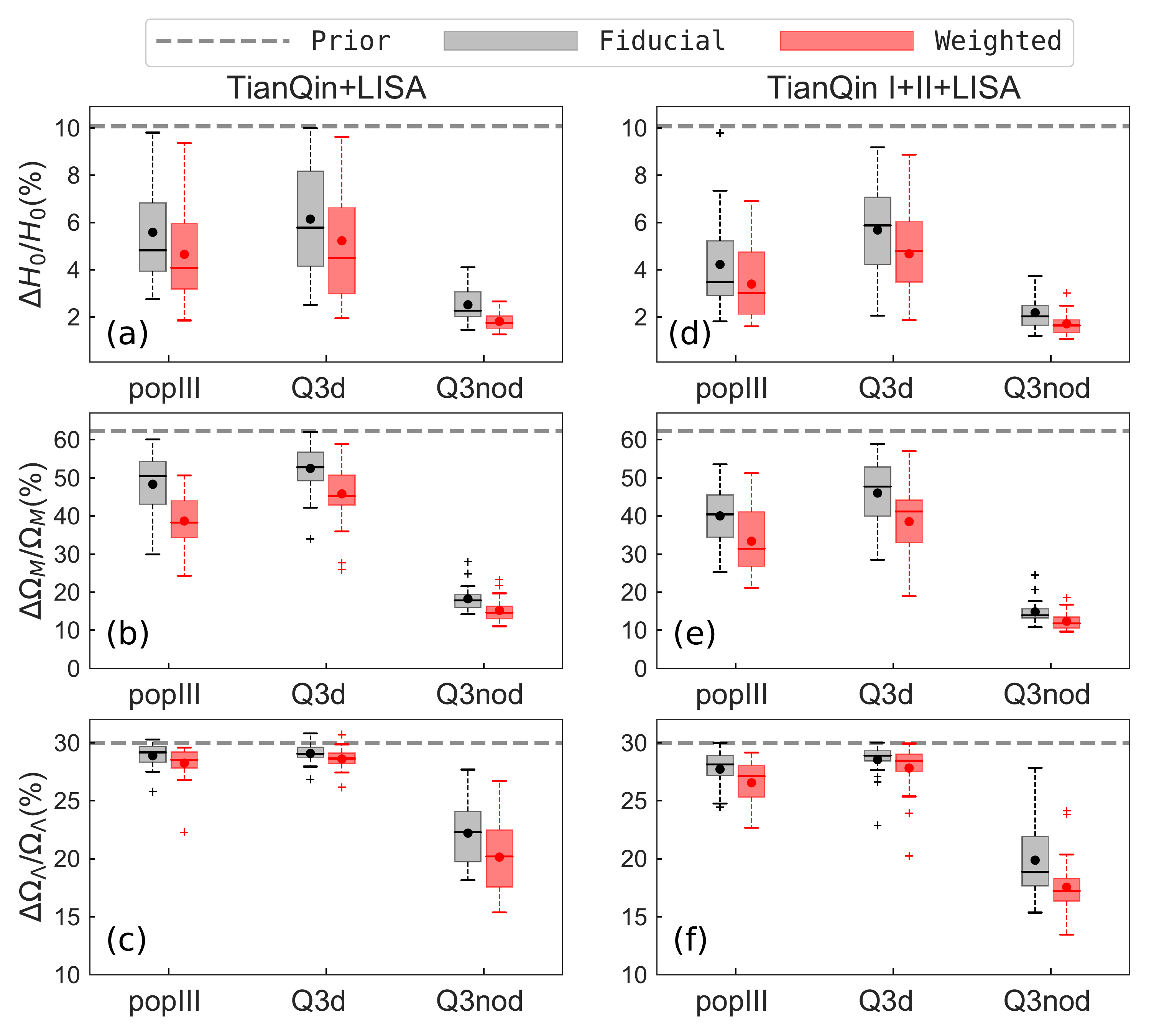}
\caption{Same as Fig. \ref{H0ML_accuracy_TQ}, but for TianQin+LISA (left column) and TianQin I+II+LISA (right column), respectively.} 
\label{H0ML_accuracy_TQLISA}
\end{figure}

\begin{figure}[htbp]
\centering
\includegraphics[width=13.cm, height=7.5cm]{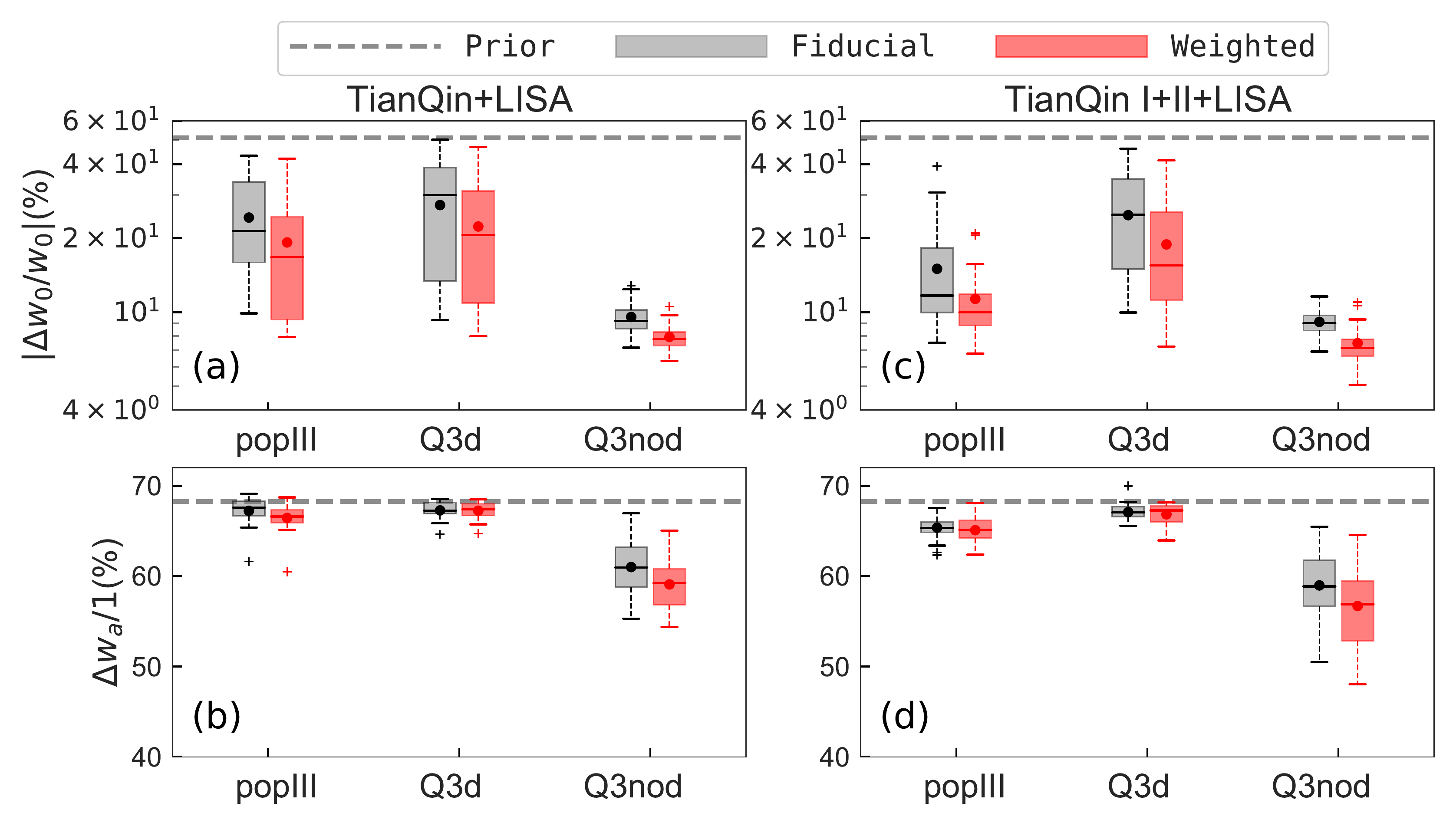}
\caption{Same as Fig. \ref{w0_accuracy_TQ}, but for TianQin+LISA (left column) and TianQin I+II+LISA (right column), respectively. }
\label{w0_accuracy_TQLISA}
\end{figure}

In what follows, we consider the cosmological constraints from a network of TianQin and LISA.
As shown in Figs. \ref{H0ML_accuracy_TQLISA} and \ref{w0_accuracy_TQLISA}, compared with the TianQin alone scenario, the network of TianQin and LISA can consistently improve the constraints on the cosmological parameters. 
This improvement benefits from both the increased detection numbers with more detectors (as illustrated in Table \ref{detection_rate}) and the better localisation capability of the network (as illustrated in Fig. \ref{paras_err_3model}).

Figure \ref{H0ML_accuracy_TQLISA} illustrates how the relative precision of constrained cosmological parameters corresponding to different detector network configurations, MBHB population models, and weighting methods, with a similar setup as in Fig. \ref{H0ML_accuracy_TQ}. 
Comparing these two figures, it is easy to see that the joint detections can effectively improve the constraint precision on the Hubble constant $H_0$. 
Similarly, the fractional dark energy density $\Omega_\Lambda$ is still poorly constrained, except under the Q3nod model. 
However, different from the TianQin alone case, here we can obtain a meaningful constraints on the fractional total matter density $\Omega_M$ in all cases. 
Again,  compared with the fiducial method, the weighted method can significantly reduce the uncertainties on the Hubble constant. 

In particular, using the weighted method, TianQin+LISA \ac{GW} detections lead to a precision of $4.7\%$, $5.2\%$, and $1.8\%$ on measurements of $H_0$ for the popIII, Q3d, and Q3nod models, respectively; while the TianQin I+II+LISA detections further improve the precision of $H_0$ measurements to $3.4\%$, $4.7\%$, and $1.7\%$, respectively for these three astrophysical models.
These kinds of measurements would be interesting, since they could be helpful for resolving (or confirming) the Hubble tension \cite{Freedman:2017yms, Verde:2019ivm, Aghanim:2018eyx, Riess:2020sih, Riess:2020fzl}. 

When assuming a dynamical dark energy model, the joint detection of TianQin and LISA can improve the constraints on $w_0$ (see Fig. \ref{w0_accuracy_TQLISA}), while $w_a$ is still poorly constrained. 
With the weighted method scheme, for TianQin+LISA, $w_0$ can be constrained to a level of  $19.2\%$, $22.3\%$ and $7.9\%$ in popIII, Q3d and Q3nod model, respectively; 
assuming a joint detection of TianQin I+II+LISA, one can better constrain the dark energy \ac{EoS} parameters, $w_0$ can be constrained to a level of $11.3\%$, $18.9\%$ and $7.5\%$, respectively.
Especially, in the case of adopting the Q3nod model, the constraint precisions on the parameters of \ac{EoS} of dark energy become comparable to those from current \ac{EM} observations \cite{Aghanim:2018eyx, Abbott:2018wzc}. 
Thus, the \ac{GW} observations can provide an independent verification of our current understanding of the Universe \cite{Freedman:2017yms, Riess:2020sih, Riess:2020fzl, Benevento:2020fev, Alestas:2020mvb, DiValentino:2017iww, Cai:2021wgv}.

\begin{table}[]
    \caption{Expected relative precision on ($H_0$, $\Omega_M$, $\Omega_\Lambda$) and ($w_0$, $w_a$) constraints for the three \ac{MBHB} models, assuming different configurations of detectors or networks. 
    When constraining $H_0$, $\Omega_M$, and $\Omega_\Lambda$, no evolution of dark energy is assumed, while all other cosmological parameters are fixed when studying the \ac{CPL} parameters. 
    Numbers are shown for mean values (median values in brackets). 
    The symbol ``$-$'' indicates that the corresponding parameter is not effectively constrained under the corresponding condition.  }
    \vspace{12pt}
    \renewcommand\arraystretch{1.1}
    \centering
    \begin{tabular}{|c|c|c|c||c|c||c|c|}
        \hline
        \multirow{3}*{Cosmological}  &  \multirow{3}*{Detector }  & \multicolumn{6}{c|}{Expected relative precision(\%) }  \\
        \cline{3-8}
        ~  &  ~  & \multicolumn{2}{c||}{popIII} & \multicolumn{2}{c||}{Q3d} & \multicolumn{2}{c|}{Q3nod}   \\
        \cline{3-8}
        \thead[c]{parameter} & \thead[c]{configuration}  &  \thead[c]{~~~Fiducial~~~~}  & \thead[c]{~~Weighted~~~~}  &  \thead[c]{~~~Fiducial~~~~}  & \thead[c]{~~Weighted~~~~}  &  \thead[c]{~~~Fiducial~~~~}  & \thead[c]{~~Weighted~~~~} \\
        \hline
        ~                       &  TianQin            & $7.8$ (9.2)  & $6.9$ (8.1)    & $7.5$ (9.4)  & $6.5$ (7.3)    & $4.2$ (3.7)  & $3.3$ (3.0)     \\
        \cline{2-8}
        ~                       &  TianQin I+II       & $7.0$ (7.8)  & $6.0$ (6.4)    & $6.9$ (8.3)  & $6.0$ (6.5)    & $2.9$ (2.7)  & $2.0$ (1.9)     \\          
        \cline{2-8}
        $\Delta H_0 / H_0$      &  LISA               & $7.2$ (7.2)  & $5.6$ (5.8)    & $7.7$ (8.2)  & $6.8$ (7.8)    & $2.8$ (2.9)  & $2.1$ (1.9)     \\
        \cline{2-8}
        ~                       &  TianQin+LISA       & $5.6$ (4.8)  & $4.7$ (4.1)    & $6.1$ (5.8)  & $5.2$ (4.5)    & $2.5$ (2.3)  & $1.8$ (1.8)     \\          
        \cline{2-8}
        ~                       &  TianQin I+II+LISA  & $4.2$ (3.5)  & $3.4$ (3.0)    & $5.7$ (5.9)  & $4.7$ (4.8)    & $2.2$ (2.0)  & $1.7$ (1.7)     \\          
        \hline
        
        ~                       &  TianQin            & $-$  & $56.5$ (57.1)    & $-$  & $56.5$ (56.6)    & $41.6$ (40.9)  & $34.9$ (35.3)     \\
        \cline{2-8}
        ~                       &  TianQin I+II       & $58.2$ (58.7)  & $50.5$ (49.8)    & $54.5$ (57.1)  & $49.0$ (49.4)    & $28.1$ (27.7)  & $25.1$ (23.9)     \\          
        \cline{2-8}
        $\Delta \Omega_M / \Omega_M$   &  LISA        & $56.2$ (56.5)  & $49.6$ (49.8)    & $55.4$ (55.7)  & $49.7$ (49.9)    & $21.7$ (20.0)  & $17.5$ (16.4)     \\
        \cline{2-8}
        ~                       &  TianQin+LISA       & $48.3$ (50.4)  & $38.7$ (38.3)    & $52.5$ (52.8)  & $45.8$ (45.2)    & $18.3$ (17.9)  & $15.2$ (14.6)     \\          
        \cline{2-8}
        ~                       &  TianQin I+II+LISA  & $40.0$ (40.4)  & $33.4$ (31.4)    & $46.0$ (47.7)  & $38.5$ (41.2)    & $14.8$ (13.9)  & $12.4$ (11.8)     \\          
        \hline
        
        ~                       &  TianQin            & $-$  & $-$    & $-$  & $-$    & $27.9$ (28.3)  & $26.1$ (26.5)     \\
        \cline{2-8}
        ~                       &  TianQin I+II       & $-$  & $-$    & $-$  & $-$    & $24.6$ (25.0)  & $21.5$ (20.9)     \\          
        \cline{2-8}
        $\Delta \Omega_{\Lambda} / \Omega_{\Lambda}$   &  LISA   & $-$  & $-$    & $-$  & $-$    & $23.6$ (24.2)  & $21.5$ (22.5)     \\
        \cline{2-8}
        ~                       &  TianQin+LISA       & $-$  & $-$    & $-$  & $-$    & $22.2$ (22.3)  & $20.1$ (20.2)     \\          
        \cline{2-8}
        ~                       &  TianQin I+II+LISA  & $-$  & $26.6$ (27.1)    & $-$  & $-$    & $19.9$ (18.9)  & $17.6$ (17.2)     \\          
        \hline
        \hline
        ~                       &  TianQin            & $40.3$ (46.1)  & $36.7$ (41.7)   & $42.4$ (47.4)  & $37.2$ (45.1)   & $20.0$ (17.3)  & $13.8$ (10.9)   \\
        \cline{2-8}
        ~                       &  TianQin I+II       & $31.6$ (31.1)  & $26.6$ (26.1)   & $34.7$ (34.5)  & $29.6$ (31.5)   & $10.5$ (9.8)  & $8.1$ (7.6)   \\          
        \cline{2-8}
        $|\Delta w_0 / w_0|$    &  LISA               & $35.1$ (35.6)  & $23.7$ (23.8)   & $36.0$ (41.0)  & $29.7$ (33.7)   & $11.4$ (11.0)  & $8.1$ (7.9)   \\
        \cline{2-8}
        ~                       &  TianQin+LISA       & $24.3$ (21.4)  & $19.2$ (16.7)   & $27.3$ (29.9)  & $22.3$ (20.6)   & $9.6$ (9.2)  & $7.9$ (7.8)    \\          
        \cline{2-8}
        ~                       &  TianQin I+II+LISA  & $15.0$ (11.7)  & $11.3$ (10.0)   & $24.8$ (24.9)  & $18.9$ (15.5)   & $9.1$ (9.0)   & $7.5$ (7.2)    \\          
        \hline
        
        ~                       &  TianQin            & $-$  & $-$   & $-$  & $-$   & $-$  & $-$   \\
        \cline{2-8}
        ~                       &  TianQin I+II       & $-$  & $-$   & $-$  & $-$   & $62.8$ (63.0)  & $60.7$ (60.4)   \\          
        \cline{2-8}
        $\Delta w_a / 1$    &  LISA                 & $-$  & $-$   & $-$  & $-$   & $62.8$ (62.8)  & $60.0$ (59.1)   \\
        \cline{2-8}
        ~                       &  TianQin+LISA       & $-$  & $-$   & $-$  & $-$   & $61.0$ (61.0)  & $59.1$ (59.2)    \\          
        \cline{2-8}
        ~                       &  TianQin I+II+LISA  & $-$  & $-$   & $-$  & $-$   & $59.0$ (58.9)  & $56.7$ (56.9)    \\          
        \hline
    \end{tabular}
    \label{HOw0_accuracy_z_survey}
\end{table}

Table \ref{HOw0_accuracy_z_survey} summarises the relative precision on $H_0$, $\Omega_M$, and $\Omega_\Lambda$, as well as $w_0$ and $w_a$, under various detection scenarios. 
Each result is obtained by marginalizing over the other parameters, while all three parameters of $\Lambda$CDM model are fixed when constraining $w_0$ and $w_a$. 
In order to mitigate the impact of random errors, we independently generate 24 GW event catalogs for each detector configuration/MBHB population model/weighting method, and present the mean values and median values of the relative constraint precisions of these five parameters. 
This table shows that the positional and bulge luminosity weighting of the host galaxies is very helpful for improving the precision of the parameter estimation. 
And a network of TianQin and LISA yield better constraints than both TianQin and LISA. 
In the most ideal scenario, $H_0$, $\Omega_M$, and $\Omega_\Lambda$ can be estimated with a relative precision of $1.7\%$, $12.4\%$, and $17.6\%$, while $w_0$ and $w_a$ can be constrained to $7.5\%$ and $56.7\%$, respectively.

\subsection{EM-bright scenario}    \label{GW_EM-result}

Some literature has suggested that \ac{MBHB} mergers can be accompanied by x-ray, optical, or radio activity \cite{Dotti:2011um,2016JCAP...04..002T}, although it is not yet certain about the physical properties of the \ac{EM} counterpart for \ac{GW} events. 
In most cases, the gas-rich environment near the \acp{MBHB} is considered to be responsible for such \ac{EM} transients. 
For example, in the late stage of binary evolution, the gas within the binary orbit would be driven inward by the inspiralling \ac{MBHB}. 
The accretion rate can exceed the Eddington limit, forming high-velocity outflows and emitting strong \ac{EM} radiation \cite{Armitage:2002uu}. 
If the component \acp{MBH} are highly spinning with aligned spin, the external disk can extract energy from the orbiting \ac{MBHB} until merged, forming dual jets and observable emissions in a way similar to the Blandford-Znajek mechanism\cite{Palenzuela:2010nf}. 
General relativistic magnetohydrodynamic simulations of magnetized plasma show that \acp{MBHB} can amplify magnetic fields by strong accretion and emit strong \ac{EM} signals \cite{Giacomazzo:2012iv}.
Some have argued that jets are increasingly dominated by magnetic fields, and can lead to transient emission after the merger \cite{Gold:PRD2014}.
Hydrodynamic simulations including viscosity indicate that, before merger, the \ac{MBHB} will transfer orbital energy into internal shocks and emit through x-ray radiation \cite{Farris:mnrasl2014}. 

In view of these literatures, it is interesting to consider the optimistic scenario, where an \ac{EM} counterpart of each \ac{MBHB} merger is identified, and these counterparts are further used to extract redshift information from the host galaxy observation \cite{2016JCAP...04..002T}. 
This possibility would be enhanced by a number of current or planned \ac{EM} facilities with large field of view and high sensitivity, including Einstein Probe \cite{EinsteinProbeTeam:2015bcj, Liu:2020bgc}, Chinese Space Station Telescope (CSST) \cite{Gong:2019yxt}, Euclid telescope \cite{Laureijs:2011gra}, Vera Rubin Observatory \cite{Ivezic:2008fe}, Five-hundred-meter Aperture Spherical radio Telescope (FAST) \cite{Jiang:2019rnj}, and Square Kilometre Array (SKA) \cite{Feng:2020nyw}.
We remark also that this bright standard siren analysis can serve as a lower limit for the precision of cosmological parameter estimation in the dark standard siren scenario --- at least in the absence of any additional mitigating strategy to deal with the impact of weak lensing and/or peculiar velocity errors. 
In our bright standard siren analysis, we apply the same selection criteria as \cite{2016JCAP...04..002T}, and only consider the \ac{GW} events with $\rho \geqslant 8$ and $\Delta \Omega \leqslant 10 \ \textrm{deg}^2$ (and $z < 3$, additionally).

\begin{figure}[htbp]
\centering
\includegraphics[width=8.cm, height=7.3cm]{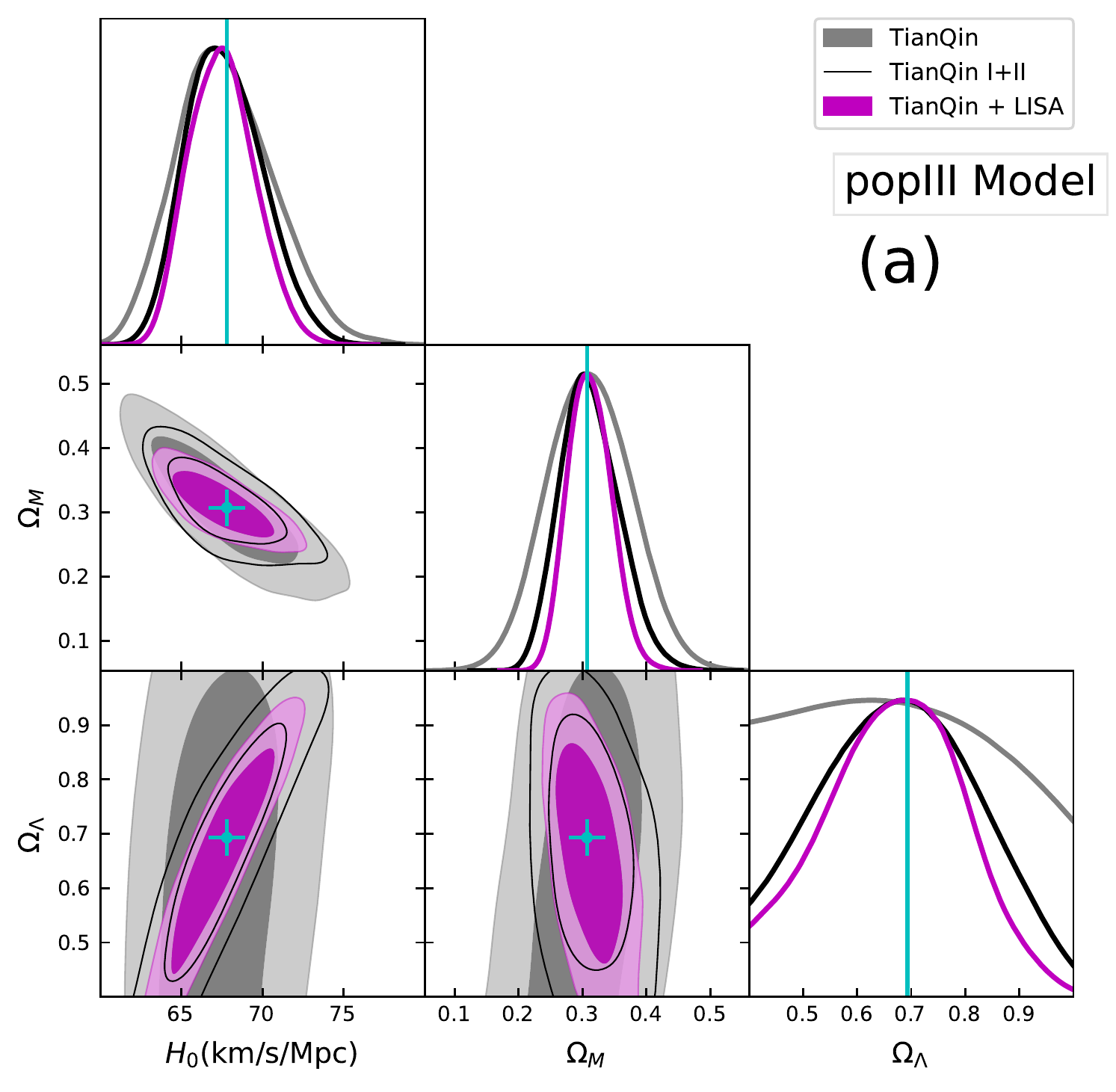} \\
\includegraphics[width=8.cm, height=7.3cm]{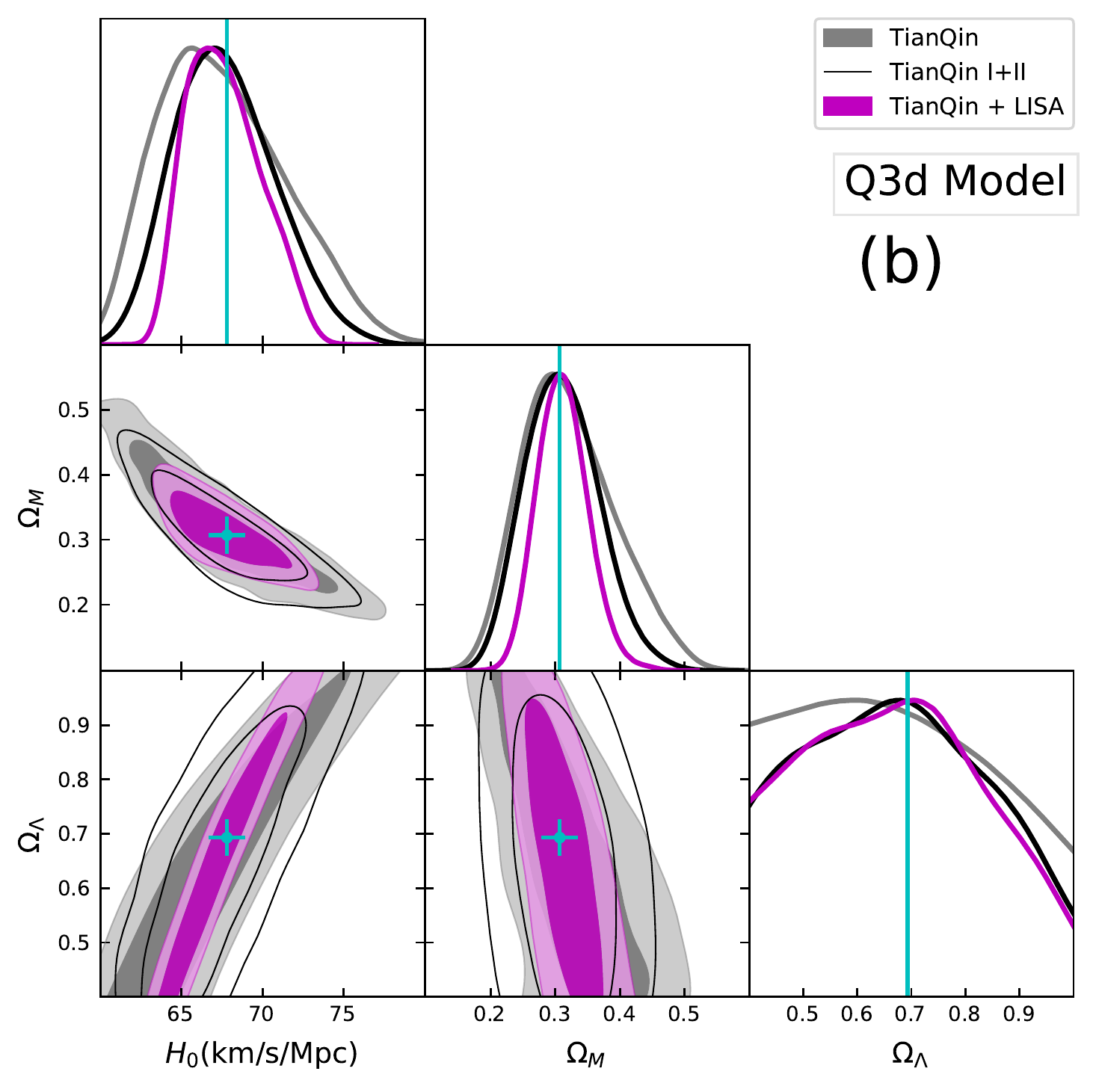} ~~~~
\includegraphics[width=8.cm, height=7.3cm]{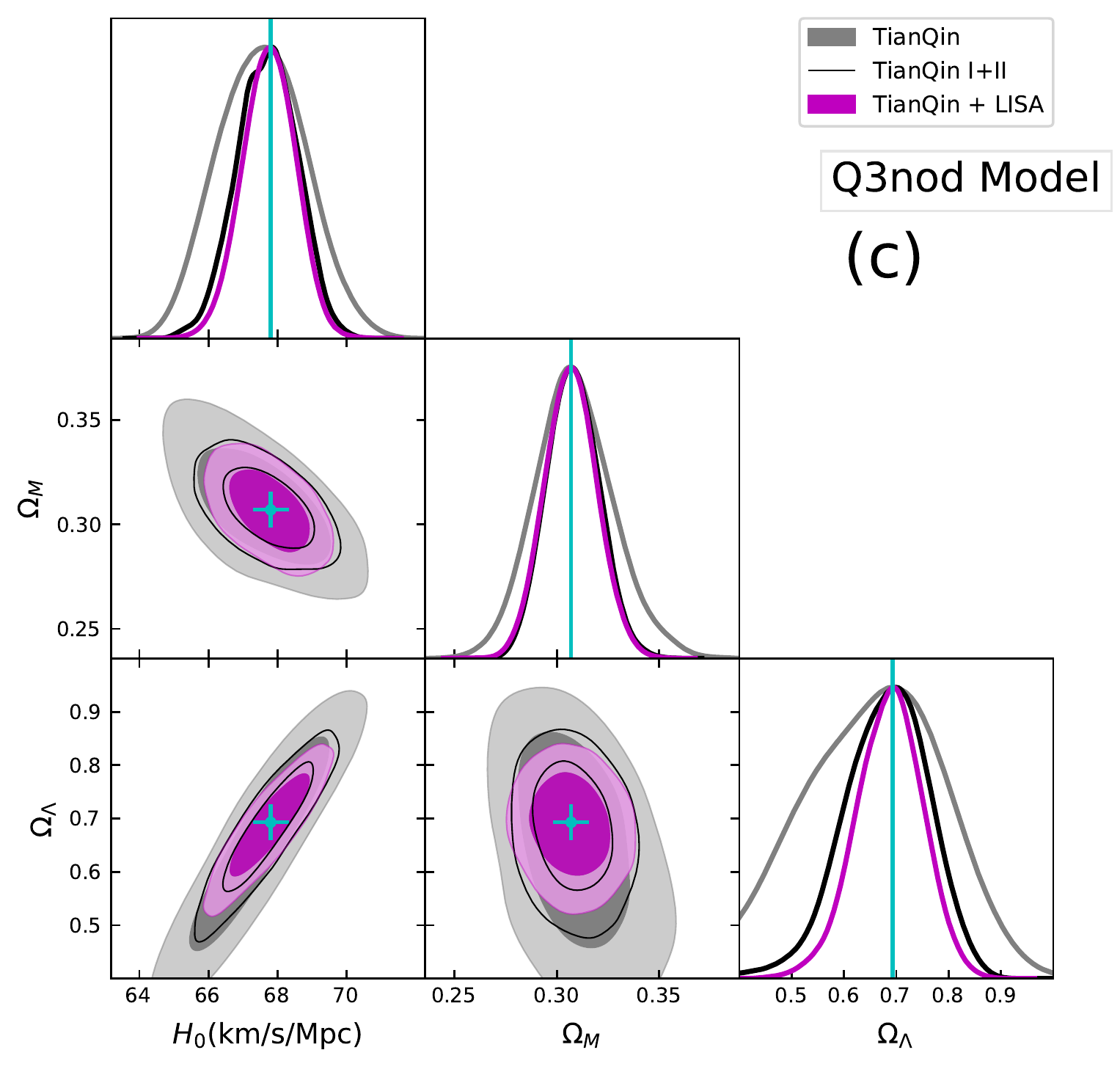}
  \caption{Typical results for the estimation of cosmological parameters under the optimistic scenario with an explicit \ac{EM} counterpart. Contours show the $68.27\%$ and $95.45\%$ confidence levels, assuming different detector configurations of TianQin (grey shadow), TianQin I+II (black line), and TianQin+LISA (magenta shade), respectively. 
The three subplots correspond to adopting popIII, Q3d, and Q3nod respectively as the underlying model for \ac{MBHB} mergers. } 
\label{H0ML_z_direct}
\end{figure}

\begin{figure}[htbp]
\centering
\includegraphics[width=5.5cm, height=4.8cm]{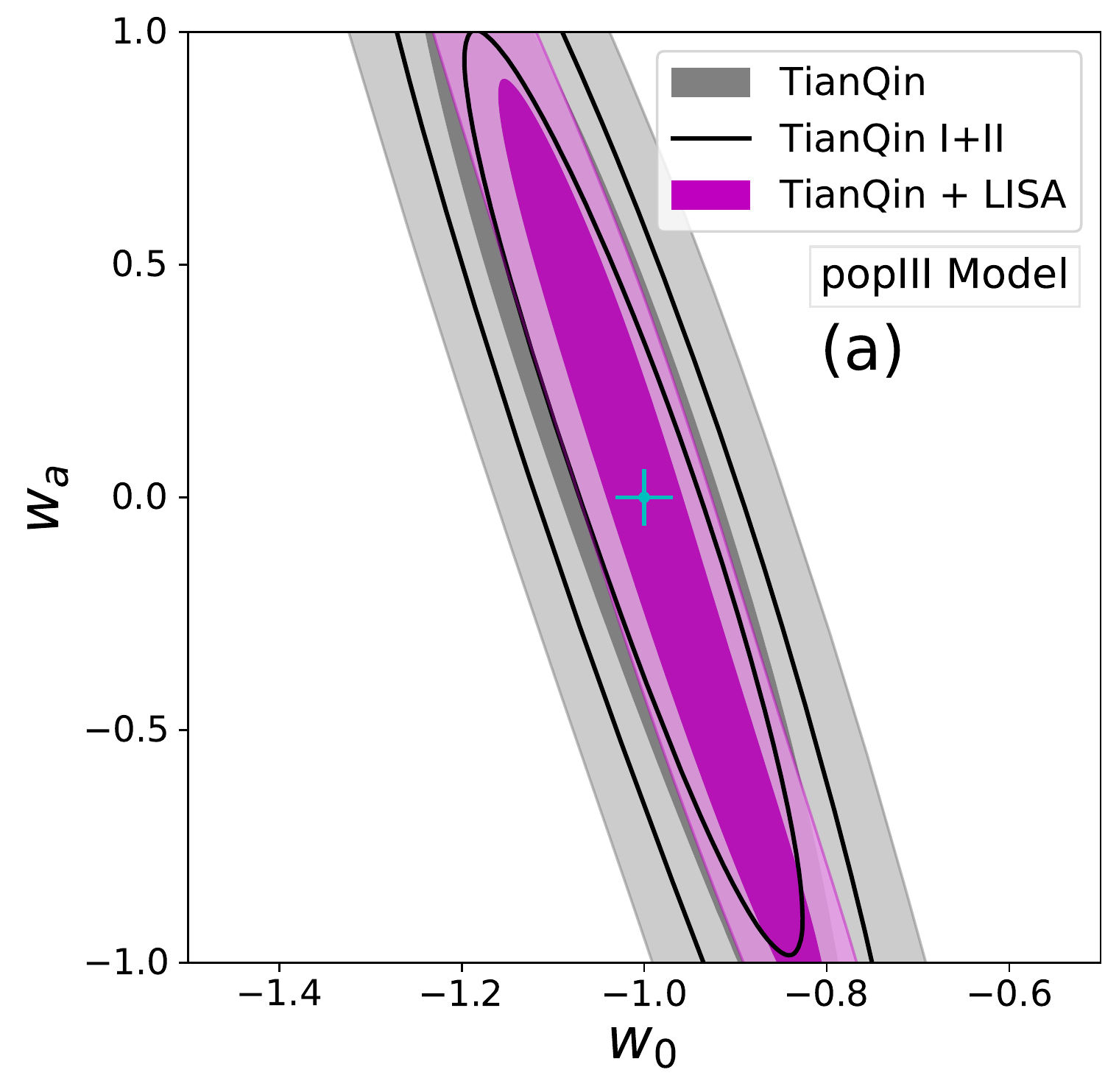}
\includegraphics[width=5.5cm, height=4.8cm]{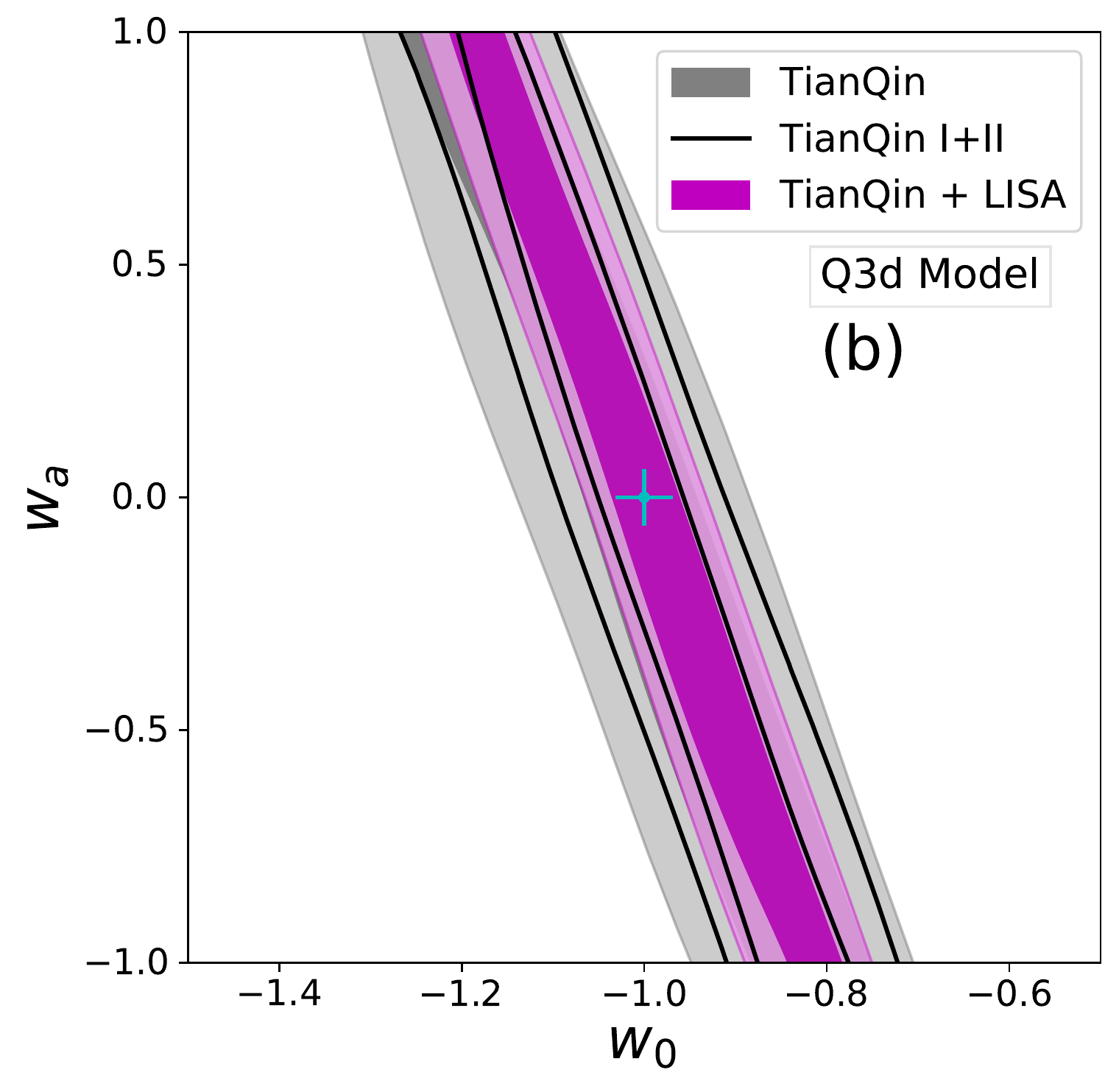}
\includegraphics[width=5.5cm, height=4.8cm]{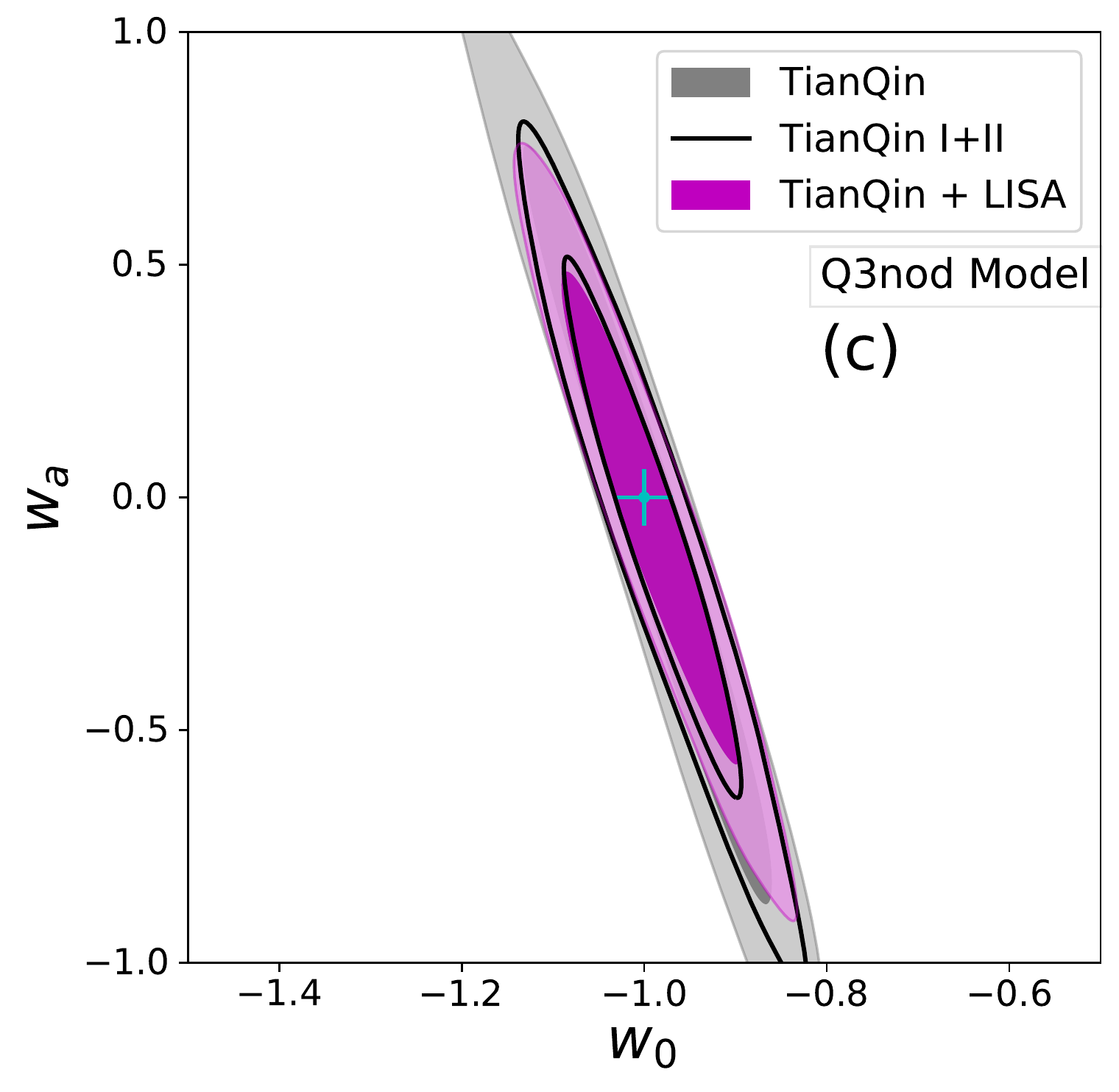}
\caption{Typical results for the estimation of dark energy \ac{EoS} parameters under the optimistic scenario with an explicit \ac{EM} counterpart. Contours show the $68.27\%$ and $95.45\%$ confidence levels, assuming different detector configurations of TianQin (grey shadow), TianQin I+II (black line), and TianQin+LISA (magenta shade), respectively. 
The three subplots correspond to adopting popIII, Q3d, and Q3nod respectively as the underlying model for \ac{MBHB} mergers.
The other cosmological parameters, $H_0$, $\Omega_M$, and $\Omega_{\Lambda}$ are fixed. }  
\label{w0wa_z_direct}
\end{figure}

In Fig. \ref{H0ML_z_direct}, we illustrate the expected probability distribution of $(H_0, \Omega_M, \Omega_{\Lambda})$, while in Fig. \ref{w0wa_z_direct} we show the expected results for the \ac{CPL} parameters $(w_0, w_a)$. 
In both cases, we consider three detector configurations, namely TianQin, TianQin I+II, and TianQin+LISA, represented by the grey shadow, black line, and magenta shades, respectively. 
We also consider the popIII, Q3d, and Q3nod models. 
Correspondingly, the marginalised one dimensional relative errors of the parameters for various detector configuration are listed in Table \ref{HOMLw0wa_accuracy_z_direct}. 

\begin{table}[]
  \caption{Constraints on the cosmological parameters and dark energy \ac{EoS} parameters under the optimistic scenario that an \ac{EM} counterpart is observed, considering different detector configurations and underlying \ac{MBHB} merger models.
Notice that the cosmological parameters ($H_0$, $\Omega_M$, $\Omega_{\Lambda}$) and \ac{CPL} parameters $(w_0,\ w_a)$ are studied separately. 
Each result is obtained based on 108 replicated and independent random realizations, the error represents $68.3 \%$ confidence interval. } 
    \vspace{12pt}
    \renewcommand\arraystretch{1.5}
    \centering
    \begin{tabular}{|c|c|c|c|c|c|c|}
        \hline
        ~ & ~ & \multicolumn{5}{c|}{Relative error(\%)} \\
        \cline{3-7}
        \thead[c]{Cosmological parameter} & \thead[c]{Population \\ model} & \thead[c]{~~~~~TianQin~~~~~} & \thead[c]{~TianQin I+II~~} &\thead[c]{~~~~~~~LISA~~~~~~~} & \thead[c]{TianQin+LISA} & \thead[c]{~TianQin I+II+LISA}  \\
        \hline
        ~                                              & popIII & $4.3_{-2.0}^{+3.0}$   & $3.4_{-1.3}^{+1.9}$   & $3.9_{-1.4}^{+2.8}$   &  $3.1_{-1.0}^{+1.3}$   & $2.5_{-0.7}^{+1.5}$     \\
        \cline{2-7}
        $\Delta {H_0}/H_0$                             & Q3d    & $6.2_{-3.4}^{+2.2}$   & $4.5_{-2.4}^{+2.2}$   & $4.9_{-2.3}^{+2.0}$   &  $3.8_{-1.7}^{+3.0}$   & $3.6_{-1.5}^{+2.7}$     \\
        \cline{2-7}
        ~                                              & Q3nod  & $1.9_{-0.5}^{+0.7}$   & $1.4_{-0.2}^{+0.3}$   & $1.4_{-0.3}^{+0.5}$   &  $1.3_{-0.2}^{+0.4}$   & $1.2_{-0.2}^{+0.3}$     \\          
        \hline
        ~                                              & popIII & $19.9_{-6.7}^{+11.9}$   & $13.1_{-3.3}^{+7.8}$   & $14.8_{-3.8}^{+10.2}$   &  $9.9_{-1.9}^{+4.5}$   & $7.9_{-1.2}^{+4.5}$    \\
        \cline{2-7}
        $\Delta {\Omega_M} /\Omega_M$                  & Q3d    & $27.3_{-11.9}^{+8.8}$   & $16.3_{-5.5}^{+9.6}$   & $19.6_{-7.1}^{+9.3}$   &  $14.6_{-4.7}^{+10.1}$   & $12.9_{-3.5}^{+9.3}$    \\
        \cline{2-7}
        ~                                              & Q3nod  & $6.5_{-0.7}^{+1.3}$   & $4.7_{-0.4}^{+0.6}$   & $5.0_{-0.4}^{+0.8}$   &  $4.4_{-0.4}^{+0.7}$   & $3.9_{-0.3}^{+0.7}$     \\          
        \hline
        ~                                              & popIII & $27.7_{-2.8}^{+0.8}$   & $25.0_{-4.1}^{+2.9}$   & $26.5_{-3.6}^{+1.4}$  &  $21.8_{-3.0}^{+4.1}$  & $20.2_{-4.1}^{+4.1}$    \\
        \cline{2-7}
        $\Delta {\Omega_{\Lambda}} / \Omega_{\Lambda}$ & Q3d    & $27.8_{-1.4}^{+0.9}$   & $27.0_{-6.2}^{+1.1}$   & $27.4_{-4.3}^{+1.0}$  &  $25.6_{-6.4}^{+2.3}$  & $24.9_{-6.5}^{+2.4}$    \\
        \cline{2-7}
        ~                                              & Q3nod  & $16.0_{-2.4}^{+3.5}$   & $11.3_{-1.3}^{+1.8}$   & $12.3_{-1.7}^{+3.0}$  &  $10.7_{-0.9}^{+2.0}$  & $9.5_{-1.0}^{+1.5}$    \\          
        \hline
        \hline
        ~                       &  popIII   & $11.6_{-1.9}^{+4.2}$  & $10.6_{-1.8}^{+2.5}$   & $11.4_{-2.0}^{+3.2}$   &  $9.8_{-1.5}^{+2.5}$   & $9.4_{-1.6}^{+1.9}$    \\
        \cline{2-7}
        $|\Delta {w_0} / w_0|$  &  Q3d      & $13.8_{-3.3}^{+2.9}$  & $12.3_{-2.4}^{+2.0}$   & $13.1_{-2.3}^{+2.2}$   &  $12.2_{-2.4}^{+2.2}$  & $11.2_{-1.7}^{+2.6}$    \\
        \cline{2-7}
        ~                       &  Q3nod    & $8.5_{-1.3}^{+1.7}$   & $6.7_{-0.6}^{+1.1}$    & $6.8_{-0.9}^{+1.5}$    &  $6.6_{-0.9}^{+1.0}$   & $5.7_{-0.8}^{+1.1}$    \\
        \hline
        ~                       &  popIII   & $66.1_{-3.8}^{+2.0}$  & $61.7_{-3.5}^{+4.0}$   & $64.0_{-4.5}^{+2.9}$   &  $57.8_{-4.9}^{+5.7}$   & $53.5_{-3.9}^{+6.0}$    \\
        \cline{2-7}
        $\Delta {w_a} / 1$      &  Q3d      & $67.0_{-3.9}^{+1.1}$  & $64.9_{-4.2}^{+2.4}$   & $66.3_{-4.4}^{+1.2}$   &  $63.7_{-5.7}^{+3.2}$   & $62.6_{-4.9}^{+3.5}$    \\
        \cline{2-7}
        ~                       &  Q3nod    & $49.3_{-3.9}^{+5.3}$  & $39.0_{-2.8}^{+5.2}$   & $40.8_{-3.2}^{+5.4}$   &  $37.3_{-3.9}^{+4.2}$   & $33.0_{-2.7}^{+4.0}$    \\          
        \hline
    \end{tabular}
    \label{HOMLw0wa_accuracy_z_direct}
\end{table}

As expected, with the identified \ac{EM} counterparts, one can significantly improve the ability of the standard sirens to constrain the cosmological parameters. 
For TianQin, the relative precision of $H_0$ can be as small as $1.9\%$, which translates into an absolute precision of about $\Delta H_0 = 1.3 ~\rm{km/s/Mpc}$. 
For TianQin I+II, the relative precision of $H_0$ can be as small as $1.4\%$, which exceeds the constraint accuracy for TianQin+LISA in the dark standard siren scenario. 
If TianQin I+II+LISA is implemented, the relative precision of $H_0$ can be as small as $1.2\%$, which translates into an absolute precision of about $\Delta H_0 = 0.8 ~\rm km/s/Mpc$. 
Furthermore, the fractional density parameter $\Omega_{\Lambda}$ and the \ac{EoS} parameter $w_a$ for the dark energy, which in the dark standard siren scenario are hardly constrained, can also be constrained to a relative precision of as small as $9.5\%$ and $33.0\%$ when the \ac{EM} counterpart available. 
These constraints are comparable to those from analysis of other cosmological \ac{EM} survey data \cite{Abbott:2018wzc, Abdullah:2020qmm}.

\section{Discussion}    \label{discussion-part}

\subsection{Constraining ability from different redshifts}    \label{constraint_ability}

For a given \ac{GW} event, its ability to constrain the cosmological parameters depends on two factors: its spatial localization accuracy, and the cosmological evolution between the source and the observer. 

\begin{figure}[htbp]
\centering
\includegraphics[width=10.cm, height=6.cm]{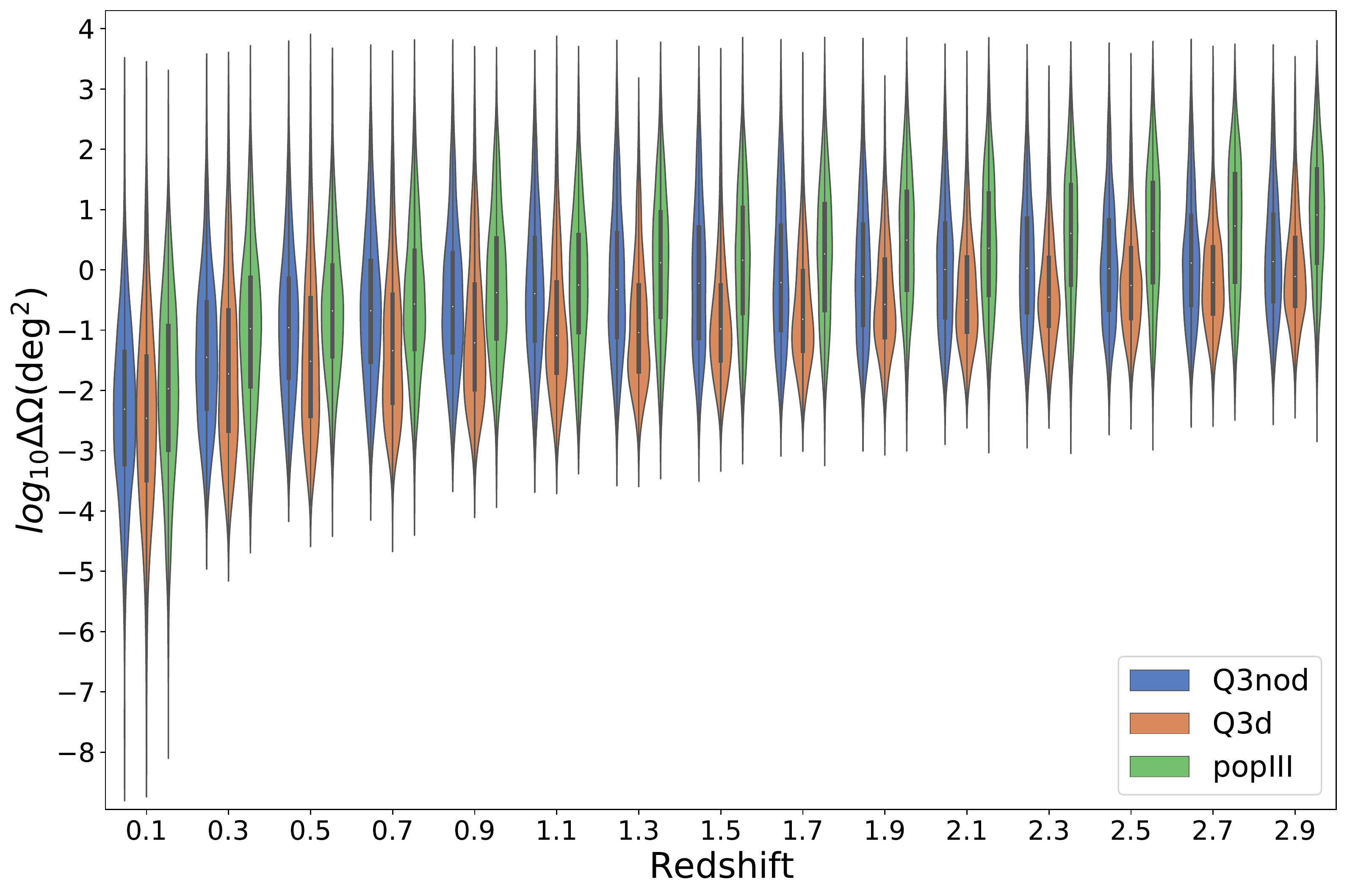}
\caption{Distribution of TianQin localization areas. Each violin plot stacks 1000 \ac{GW} events around the given redshifts $[(z-0.1),(z+0.1)]$.}\label{dOmega_z}
\end{figure}

\begin{figure}[htbp]
\centering
\includegraphics[width=10.cm, height=6.cm]{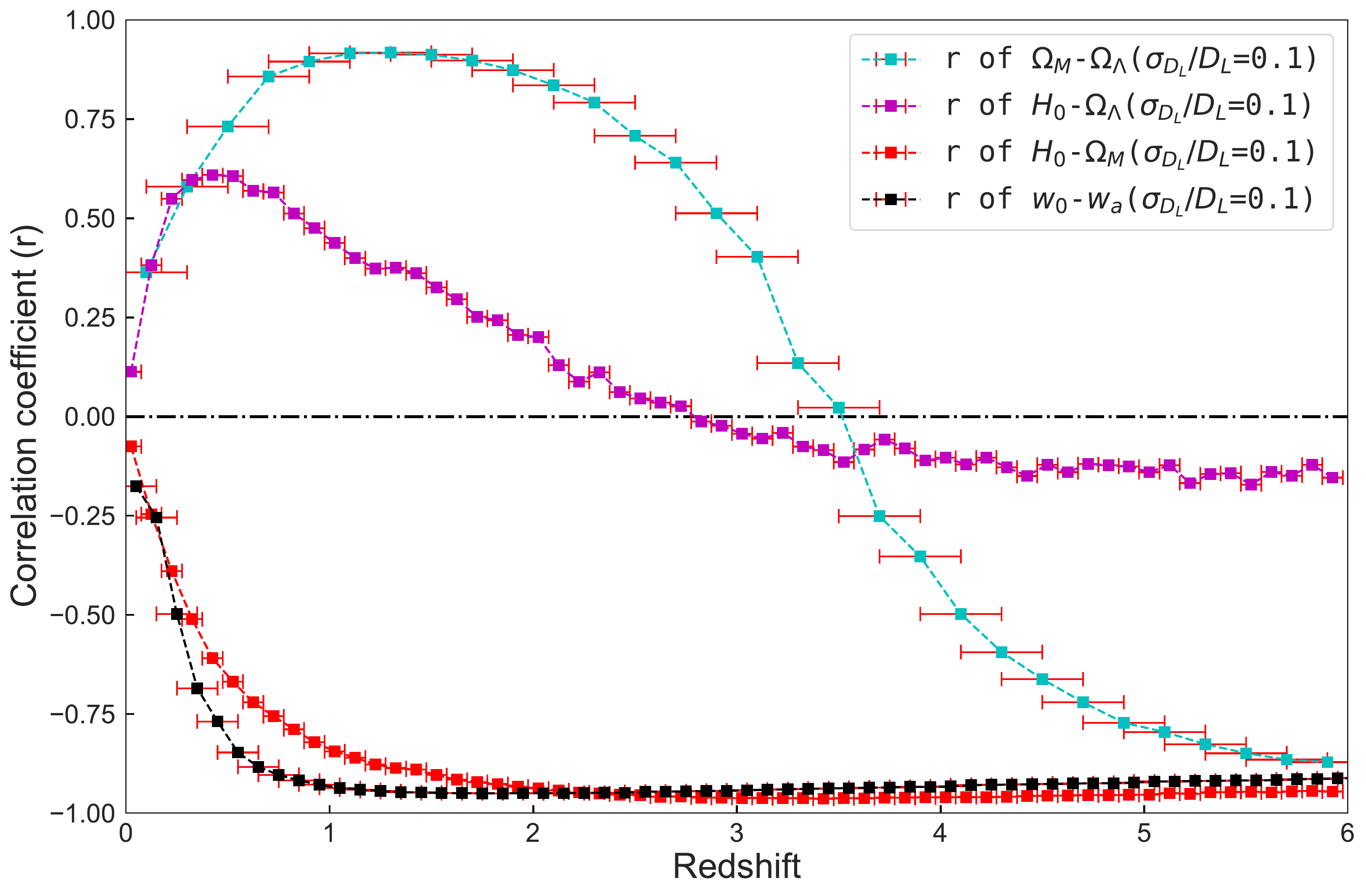}
  \caption{Evolution of the correlation coefficient (indicated by $r$) with redshift. 
  For a given redshift bin of width 0.05/0.1/0.2, we randomly generate a number of \ac{EM}-bright standard sirens, obtain the \ac{MCMC} samples and compute correlation coefficients on $H_0-\Omega_M$/$H_0-\Omega_{\Lambda}$/$\Omega_M-\Omega_{\Lambda}$/$w_0-w_a$, respectively. Due to the large uncertainties for some parameters, we choose a slightly different prior for $h \in \textrm{U}[0.4, 1]$, $\Omega_M \in \textrm{U}[0, 0.6]$, $\Omega_{\Lambda} \in \textrm{U}[0.4, 1]$, and $w_0 \in \textrm{U}[-2.5, 0.5]$, $w_a \in \textrm{U}[-3, 3]$. }   
\label{r_z}
\end{figure}

In Fig. \ref{dOmega_z}, we show the distribution of sky localization errors for \acp{MBHB} mergers at different redshifts. We see the sky localization area $\Delta \Omega$ tends to increase as the redshift increases.
Meanwhile, nearby events are usually accompanied by larger \ac{SNR}, which leads to better determination of the luminosity distance.
To sum up, for nearby events, the smaller localization area together with the better distance estimation leads to a smaller error box, and therefore a smaller number of candidate host galaxies. 

On the other hand, given the same $\Delta D_L$ and $\Delta \Omega$, the \ac{GW} signals at different redshifts lead to different constraints on the cosmological parameters. 
In Fig. \ref{r_z}, we illustrate the degeneracy of the cosmological parameters by showing the evolution of correlation coefficients between different pairs of parameters. 
To do this, we estimate cosmological parameters with a sample of \ac{EM}-bright \ac{GW} events, assuming a relative error on luminosity distance $\sigma_{D_L}/D_L=0.1$. 

We find that, the $H_0$ constraint from high redshifts suffers from strong degeneracy with $\Omega_M$ and $\Omega_{\Lambda}$, and this degeneracy becomes less significant only at very low redshifts. 
Notice that the correlation coefficient between $\Omega_M$ and $\Omega_{\Lambda}$ switches its sign at $z \sim  3.5$, suggesting that observations of both higher and lower redshift events are needed to alleviate the degeneracy. 
Also, the correlation coefficient between $w_0$ and $w_a$ is approaching total anticorrelation at $z \gtrsim 1$. 
These two parameters cannot be precisely measured, partly because of the strong anticorrelation between them. 

In conclusion, a small number of low redshift \ac{GW} sources is sufficient to provide a good constraint on the $H_0$, thanks to the precise position estimation, as well as the weaker parameter degeneracies. 
Meanwhile, high redshift events are useful to better constrain other cosmological parameters like $\Omega_M$ and $\Omega_{\Lambda}$.
For cases where the \ac{EM} counterpart can be uniquely identified, the \ac{GW} events with large redshift can give tight constraints on the cosmological parameters; otherwise the large positional error would weaken such constraints.

\subsection{Effects of redshift limit}    \label{z-limit}

Throughout this study we have applied a cutoff for events beyond redshift $z=3$. 
To some extent, this cutoff reflects the incompleteness of galaxy catalogs at higher redshifts. 
It is very challenging to pursue relatively complete galaxy catalog at high redshift. 
Such as, sky surveys like the Sloan Digital Sky Survey (SDSS) \cite{Alam:2015mbd, Abolfathi:2017vfu, SDSS-IV:2019txh} and Dark Energy Survey (DES) \cite{DES:2017myt} can reliably map galaxies as far as $z \simeq 1.2$. 

Our paper highlights the need for more detailed galaxy surveys with redshift limit of $z>3$. 
Actually, it is not beyond the imagination that, once \ac{MBHB} mergers are routinely detected, intensive observations would be triggered to map the higher redshift Universe --- at least within the sky localization regions of the observed mergers --- and thus provide a more complete catalog of galaxies therein \cite{2015ApJ...801L...1B}. 

Moreover, the redshift limit for quasar observations is much larger than that for galaxies.
For example, the redshift limit of the quasar catalog mapped by SDSS can reach $z \simeq 4$ \cite{Abolfathi:2017vfu, SDSS-IV:2019txh}. 
Meanwhile, some have argued that quasars can host the \acp{MBHB} \cite{Sobacchi:2016yez, Connor:2019gun, Penil:2020uqg}. 
In addition to the quasars, one can also explore the potential of using Lyman-$\alpha$ forest effect to obtain redshift information \cite{Hess:2018qit, Ravoux:2020bpg, Qin:2021gkn, Porqueres:2019dpn}. 
It is still possible to obtain statistical redshift for the high-redshift GW sources.

\subsection{Importance of bulge luminosity information}    \label{position_weight-role}

\begin{figure}[htbp]
\centering
\includegraphics[width=15.cm, height=6.5cm]{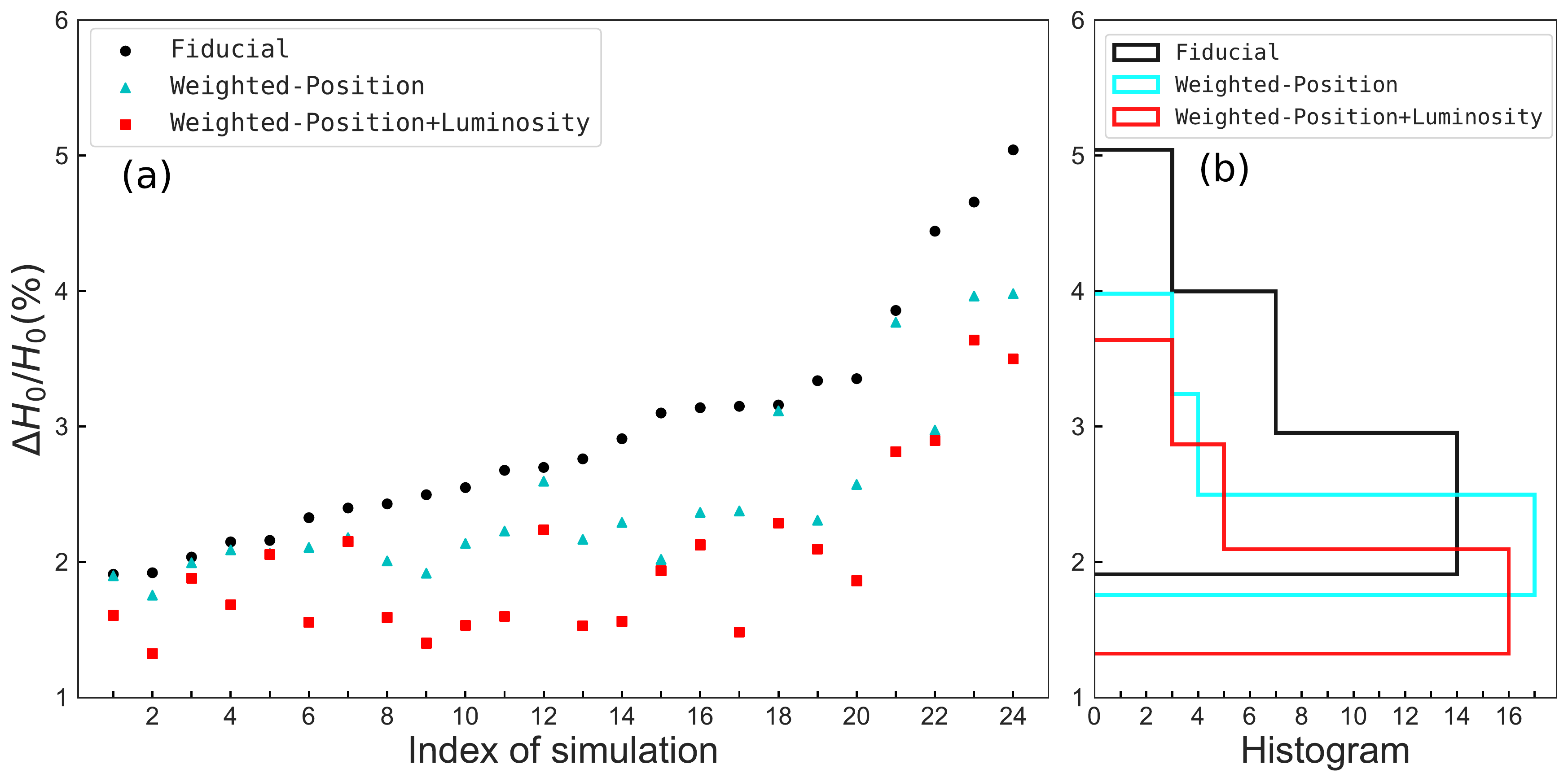}
\caption{Comparison of the relative errors on the Hubble constant $H_0$ under different weighting methods, assuming \ac{GW} detections with TianQin I+II under the Q3nod \ac{MBHB} model and no \ac{EM} counterpart. 
Left panel: Each point represents the relative error on $H_0$ for an individual simulation corresponding to the fiducial (black dot), the position-only weighted (cyan triangle), and the position+luminosity weighted (red square) method, respectively. 
Right panel: Histogram of the relative error on $H_0$ for the three different weighting methods.  } 
\label{compare-PositionalWeight}
\end{figure}

Throughout this paper, we have incorporated both position and bulge luminosity information for the calculation of weights that we assign to candidate host galaxies. 
Here we demonstrate the importance of the bulge luminosity alone, by weighting only on spatial location information.
As shown in Fig. \ref{compare-PositionalWeight}, compared with the fiducial method, the relative error on $H_0$ shrinks when sky location information is included, but the bulge luminosity weight consistently improves the estimation precision.
For mergers with higher redshift, the importance of the bulge luminosity is weakened, due to the large localization error volume and the incompleteness of the galaxy catalog. 
The large number of galaxies within this volume, combined with large uncertainty associated with the bulge luminosity,  decreases the effectiveness of the weighting scheme. 
On the other hand, the galaxy catalog is incomplete at high redshift, and the introduction of the luminosity function and the bulge luminosity function only compensates the weights of high redshift bright galaxies to some extent [see Eqs. (\ref{N_sup}) and (\ref{catalog_debias})], which does not help to weight the ``correct'' host galaxies for the high redshift \ac{GW} sources with lower mass. 

\begin{figure}[htbp]
\centering
\includegraphics[width=9.cm, height=6.cm]{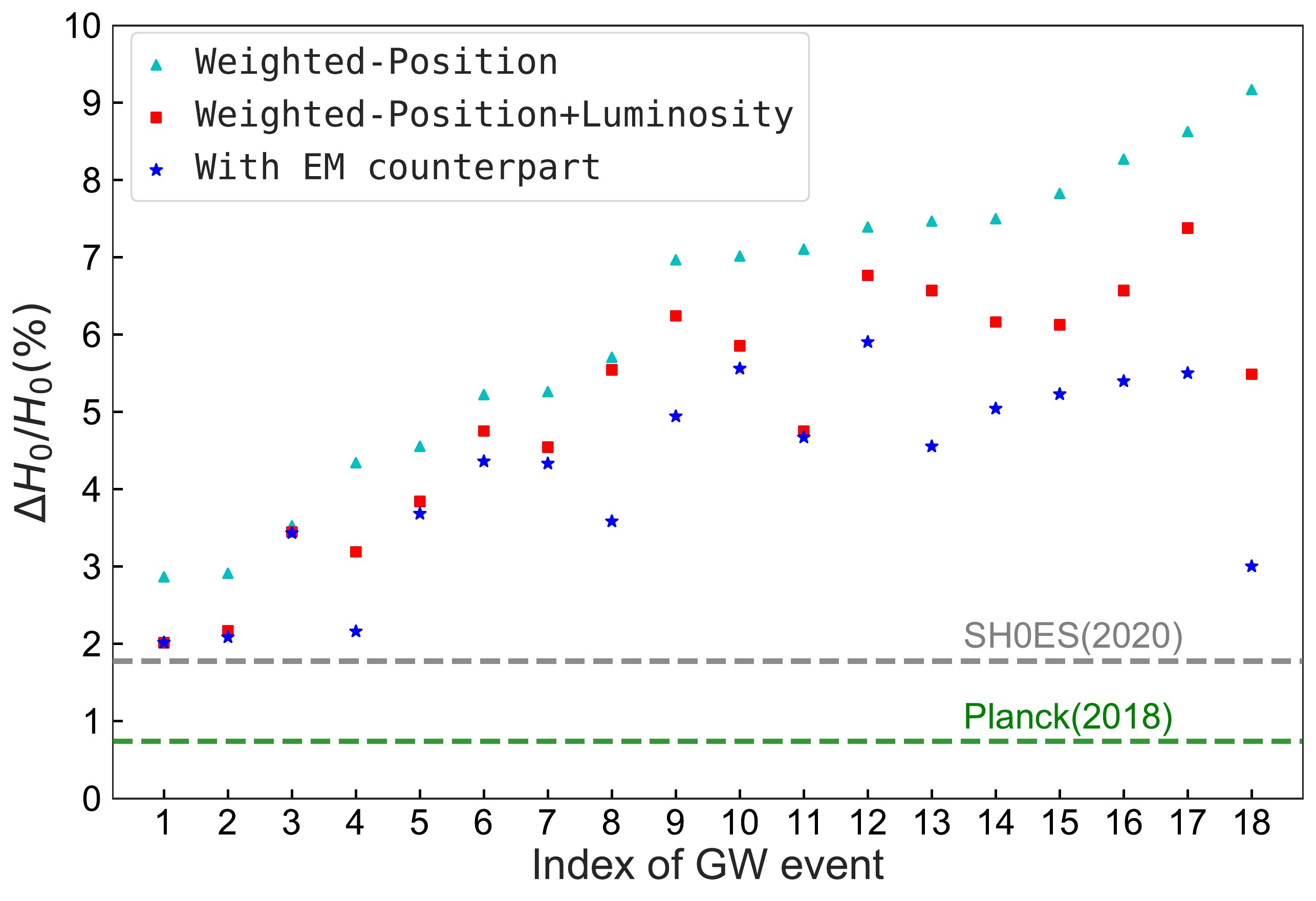}
\caption{Comparison of the constraints on the Hubble constant $H_0$ by an individual nearby event for the following scenarios: with \ac{EM} counterpart (blue star), using the position+luminosity weighted method (red square), and using position-only weighted method (cyan triangle). Each index represents one nearby GW event. 
The green line represents the $H_0$ error measured by $Planck$ using \ac{CMB} anisotropies \cite{Aghanim:2018eyx}, and the gray line represents the $H_0$ error measured by the SH0ES project using type Ia supernovae data \cite{Riess:2020fzl}. 
These events are selected from the sample pool of the Q3nod model with the conditions of both $z \lesssim 0.5$ and $\Delta \Omega \lesssim 5 \times 10^{-3} {\rm deg}^2$.  }  
\label{H0_constraint-goldenEvents}
\end{figure}

However, the bulge luminosity weight plays a crucial role especially when nearby events are considered, since only a relatively small number of galaxies is located within the error box. In this case, therefore one host galaxy can be largely identified through the $M_{\rm MBH}-L_{\rm bulge}$ relation. 
Fig. \ref{H0_constraint-goldenEvents} illustrates the constraints on the Hubble constant for the nearby events with both $z \lesssim 0.5$ and $\Delta \Omega \lesssim 5 \times 10^{-3} ~{\rm deg}^2$ in three scenarios, i.e., with \ac{EM} counterpart, using the position plus luminosity weighted method, and using position-only weighted method. 
We can observe that, by using the $M_{\rm MBH}-L_{\rm bulge}$ relation, the constraint precision on the Hubble constant in the dark siren scenario can be greatly improved, sometimes even comparable to scenarios with \ac{EM} counterpart. 
Since we also expect that it would be easier to obtain more complete information on the bulge luminosity for nearby galaxies, this fact highlights the importance and potential efficacy of the bulge luminosity information.

\subsection{Scope of the $M_{\rm MBH}-L_{\rm bulge}$ relation and the luminosity function}    \label{M-L-PhiL-evolution}

We have demonstrated that performing a weighting process of candidate host galaxies according to the $M_{\rm MBH}-L_{\rm bulge}$ relation is vital to improve the precision of the measured cosmological parameters. 
In this paper, a default assumption is adopted in our calculation, which is that the $M_{\rm MBH}-L_{\rm bulge}$ relation applies to all \acp{MBHB} independent of the mass of the GW source. 
This relation behaves differently at the low-mass end than at the high-mass end \cite{Jiang:2011bt, Kormendy:2013dxa}, and we have used different values of intrinsic scatter at the high and low-mass ends, see Eq. (\ref{M-L-scatter}).
In this part we discuss the scope of application of this relation. 

For TianQin, the \ac{MBHB} events with $M > 5 \times 10^6 M_{\odot}$ account for about $18\%$, $16\%$ and $21\%$ of the total detectable events under popIII, Q3d and Q3nod models, respectively; 
for LISA, these percentages are about $44\%$, $36\%$ and $40\%$, respectively. 
The high-mass \ac{MBHB} \ac{GW} sources account for quite a large proportion of the total detectable events. 
For the low-mass \acp{MBH} with $M < 10^6 M_{\odot}$, their host galaxies show no co-evolution with the central \acp{MBH} \cite{Jiang:2011bt, Jiang:2011bu}. 
The low-mass \acp{MBH} are more likely to be remnants of black hole seeds, 
for a given \ac{MBH} mass, the possible bulge luminosity spans a larger range \cite{Jiang:2011bt}. 

Bulges at the centers of galaxies can be classified as either classical bulges or pseudo-bulges, depending on whether they contain disk structures (pseudo-bulges) or not (classical bulges). 
Based on numerical simulations of galaxy collisions, it is generally accepted that classical bulges are made in major mergers of galaxies, whereas pseudobulges are the result of internal evolution of galaxy disks since their formation \cite{Kormendy:2013dxa}. 
It is noteworthy that the host galaxies of the low-mass \acp{MBH} that deviate from the $M_{\rm MBH}-L_{\rm bulge}$ relation almost all have pseudobulges in their centers \cite{Jiang:2011bt}. 
This implies that the deviations from the $M_{\rm MBH}-L_{\rm bulge}$ relation at the low-mass end do not unduly affect the scope of the weighted method proposed in this paper, and it is sufficient to account for this effect by adopting a larger intrinsic scatter, see Eq. (\ref{M-L-scatter}). 

In addition to the bulge luminosity information of galaxies, other observable information about galaxies such as stellar velocity dispersion and galaxy morphology may provide additional constraints on \ac{MBH} mass for the lighter ones. 
The relation between \ac{MBH} mass and stellar velocity dispersion (the $M_{\rm MBH}-\sigma$ relation) at the low-mass end is consistent with that at the high-mass end \cite{Xiao:2011xg}. 
Even such information is not available \textit{a prior}, telescopes would be motivated to perform deep observations after the \ac{GW} detection, especially when the sky location can be very precisely determined.

We also examine other aspects of this relation. 
Throughout this paper we have assumed a universal $M_{\rm MBH}-L_{\rm bulge}$ relation, obtained from fitting the observed data at $z \approx 0$ \cite{Bentz:2008rt, Kormendy:2013dxa}. 
However, due to the lack of reliable models, we ignore the possible redshift evolution of this $M_{\rm MBH}-L_{\rm bulge}$ relation, which might cause some bias in our analysis.
In the weighting process, we choose a moderately large $\sigma_{\log_{10} M}$ so that such bias can be partially absorbed. 

Another possible source of bias is introduced by ignoring the redshift evolution of the luminosity function. 
We manually augmented our galaxy samples, according to the modelled luminosity function, to remove the Malmquist bias. 
However, the luminosity function is only complete for nearby redshifts, and becomes more and more incomplete as the redshift increases \cite{Marchesini:2006vg, Faber:2005fp, Shen:2020obl}. 
We therefore highlight also the need for a more thorough and precise luminosity function evolution model, which can be obtained by existing \cite{McLure:2012fk, Bouwens:2014fua} and future ultradeep-field observations \cite{Gong:2019yxt,Laureijs:2011gra}. 

Notice that most galaxy catalogs constructed from cosmological surveys do not list the galactic bulge luminosity of each source.
In order to use bulge luminosity to improve our weighting scheme, we need to extract this information from data. 
Since the luminosity of a galaxy decreases with its radius, which can be described empirically using the S$\acute{e}$rsic function (also known as the $r^{1/n}$ law) \cite{Boroson1981ApJS, Simien:1986vt, Caon:1993wb, Hopkins:2008ev, Kormendy:2008np}, we can approximately estimate the bulge luminosity from the total luminosity and the morphology of a galaxy based on the S$\acute{e}$rsic function, but this conversion would bring additional large uncertainties.

\subsection{Consistency check}    \label{consistency-check}

The possibility exists that there will be only a small number of \ac{GW} events available throughout the observation time, and fluctuations resulting from small number statistics could potentially bias the estimated cosmological parameters. 
This bias can happen especially when the true host galaxy is relatively dim, so that the galaxy survey could mistakenly associate the \ac{GW} event with some other galaxy. 
Following \cite{Petiteau:2011we}, we discuss the impact of this bias, as well as how to perform a consistency check, which can identify and remove it.

Without loss of generality, in the following analysis we focus on the results for $H_0$ and $w_0$. For a given simulated catalog, we first perform a cosmological parameter estimation using all \ac{GW} events.
Then, event by event, we remove one event (hereafter the $k$th event), and perform the cosmological analysis using the events, which remain, with $p(\vec{\Omega}|I) \prod_{i \neq k} p(D_i,S|\vec{\Omega},I)$.
We find that, the mis-association of the host galaxy could occur, and could severely bias our estimation. 

\begin{figure}[htbp]
\centering
\includegraphics[width=8.cm, height=8.cm]{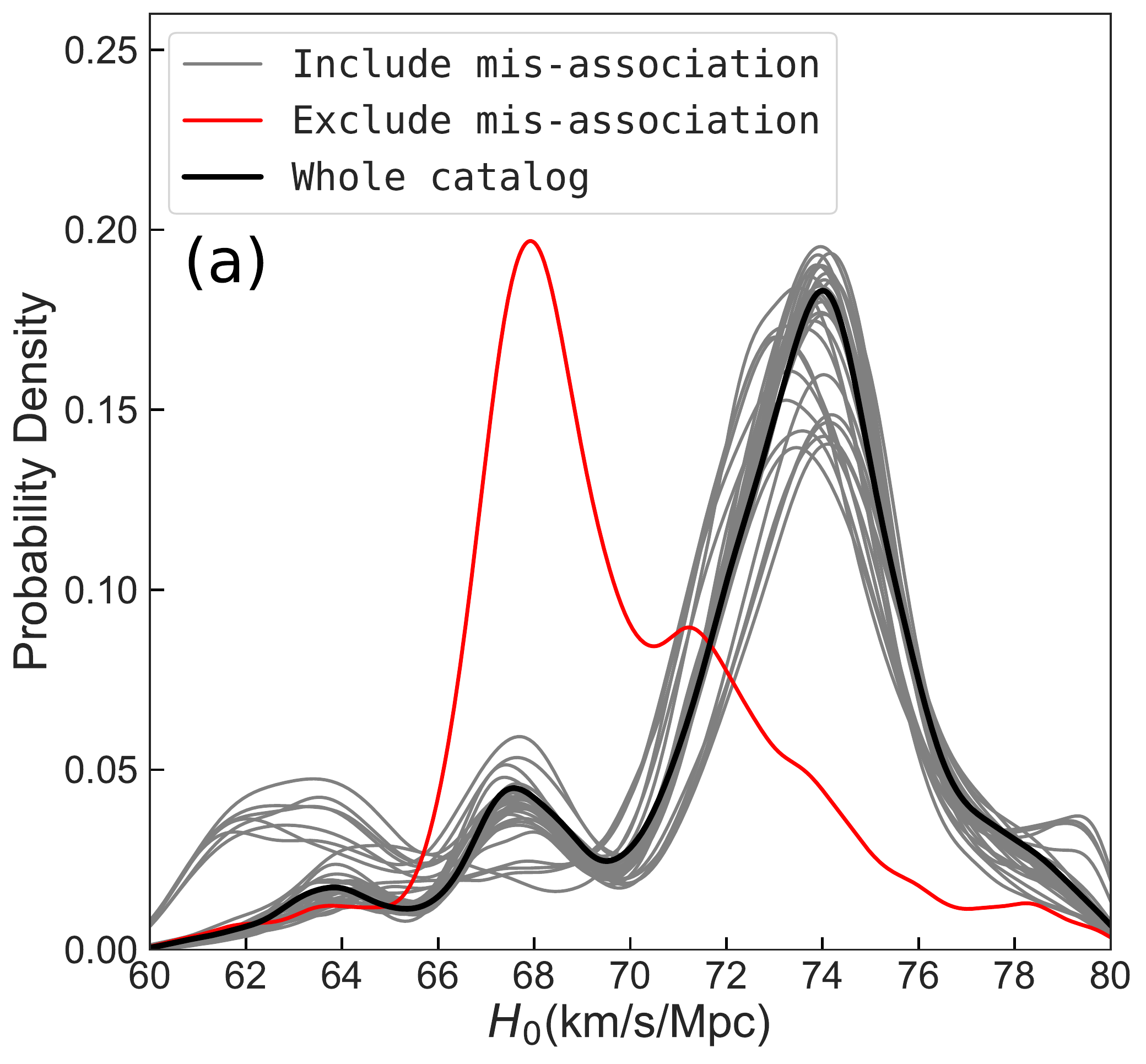} ~
\includegraphics[width=8.cm, height=8.cm]{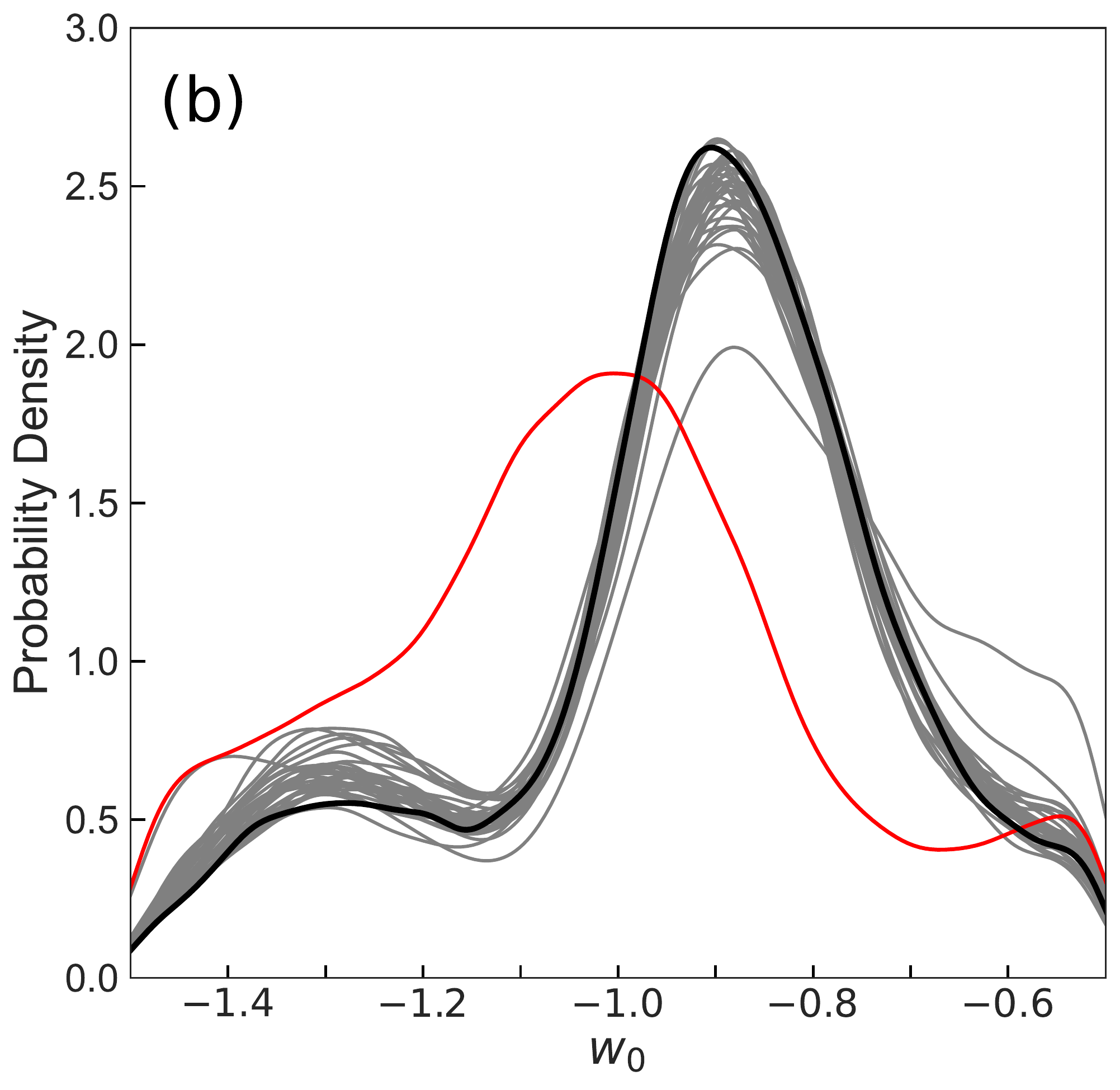}
\caption{A consistency check on an example catalog, showing marginalized posterior distributions of $H_0$ [panel (a)] and $w_0$ [panel (b)]. 
In each panel, the black line represents the result for the whole catalog, the grey lines represent the results for subsets that include the mis-association, and the red line represents the result for the subset that exclude the mis-association.
The obvious deviation of posterior modes between the red line and the grey lines indicates the possible occurrence of a mis-association bias. } 
\label{consistency_check_sample}
\end{figure}

An example of this bias is illustrated in Fig. \ref{consistency_check_sample}, where the mis-association of one event occurs, so that posteriors on $H_0$ and $w_0$ obtained from all subcatalogs that contain this event (indicated by the grey lines) yield estimates of the parameters that deviate significantly from their true values. 
Notice, however, that the posteriors obtained from the subset that {\em excludes} this particular mis-associated event (indicated by the red line) show in each case an obvious deviation from the remaining posteriors, and thus provide a clear diagnostic for the potential mis-association bias. 

The above analysis therefore provides a consistency check for the mis-association.
If the number of events is too large to carry out such a consistency check, one can remove more events in every iteration in order to enhance its efficiency. 
(Moreover, we also note that when the total number of events is larger, then the impact of mis-association of a single event will also be less severe. ) 

\section{Conclusions and Outlook}    \label{conclusion-outlook}

In this paper, we develop a Bayesian analysis framework for constraining the cosmological parameters, using simulated observations of the \acp{MBHB} mergers from space-borne \ac{GW} observatories like TianQin and LISA. 
We obtain the luminosity distance information directly from the \ac{GW} observations, while a statistical analysis of simulated galaxy catalogs from \ac{EM} surveys is used to obtain the corresponding redshift information. 
With the identification of an explicit \ac{EM} counterpart, one can indeed perform very precise cosmological measurement. However, we also show that one can still obtain useful cosmological constraints even in the dark standard siren scenario where no \ac{EM} counterpart exists --- provided that Malmquist bias is properly accounted for. 
Furthermore, if we include the localization and mass information from the \ac{GW} observation to inform the weighting of candidate host galaxies, making use of the relation between the central \ac{MBH} mass and the bulge luminosity of the host galaxy, we can significantly improve the cosmological constraints. 

For the dark standard siren scenario, we consider two weighting schemes, namely the fiducial method, one with uniform weights for all galaxies within the error box, and the weighted method, the weights related to its location and bulge luminosity. 
With the weighted method scheme, the precision of cosmological parameter estimates can be greatly improved.  
In this scheme, TianQin can constrain the Hubble constant $H_0$ to a precision of $6.9\%$, $6.5\%$, and $3.3\%$, and the \ac{CPL} parameter $w_0$ to a precision of $36.7\%$, $27.2\%$, and $13.8\%$, for the popIII, Q3d and Q3nod \ac{MBHB} population models respectively. 
For TianQin I+II, the $H_0$ precision can be improved to $6.0\%$, $6.0\%$, and $2.0\%$, and $w_0$ precision can reach $26.6\%$, $29.6\%$, and $8.1\%$ --- again for popIII, Q3d, and Q3nod models, respectively. 
The other parameters like $\Omega_M$ and $\Omega_{\Lambda}$ need a larger number of events to be significantly constrained. 
Under the Q3nod model, using the weighted method, $\Omega_M$ and $\Omega_{\Lambda}$ can be constrained to an accuracy of $34.9\%$ and $26.1\%$ for TianQin, and $25.1\%$ and $21.5\%$ for TianQin I+II, respectively. 
However, $w_a$ is always difficult to be constrained in all cases.

LISA can perform similarly to TianQin I+II, but the joint detection of TianQin and LISA can significantly improve the precision of the cosmological parameter estimates relative to the results obtained from an individual detector. 
Using the weighted method, for TianQin+LISA, the precision of $H_0$ improves to $4.7\%$, $5.2\%$, and $1.8\%$, and the precision of $w_0$ improves to $19.2\%$, $22.3\%$, and $7.9\%$ for popIII, Q3d, and Q3nod models, respectively; 
for TianQin I+II+LISA, the precision of $H_0$ improves to $3.4\%$, $4.7\%$, and $1.7\%$, and the precision of $w_0$ improves to $11.3\%$, $18.9\%$, and $7.5\%$, respectively. 

For the \ac{EM}-bright standard siren scenario, the identification of the \ac{EM} counterpart can help to pinpoint the redshift, and one can not only gain tighter constraints on $H_0$ and $w_0$, but also obtain meaningful constraints on the other cosmological parameters, including $\Omega_M$, $\Omega_{\Lambda}$, and $w_a$. 

The constraints on the Hubble constant $H_0$ are mainly derived from a few events at low redshift, but high redshift \ac{GW} events play an important role of smoothing out fluctuations in the $H_0$ posterior, as well as helping to constrain the values of $\Omega_M$ and $\Omega_{\Lambda}$. 
The CPL model describes the evolution of the dark energy \ac{EoS} with redshift, so a combination of the \ac{GW} events at different redshift is needed to constrain $w_0$ and $w_a$. 
We also discuss the application of consistency checks on the estimated cosmological parameters, so that the potential bias caused by a single mis-association of a \ac{GW} event with its host galaxy could be identified. 

There are a number of assumptions, which we adopt that could affect the applicability of our method. 
For example, we depend strongly on the availability of a relatively complete galaxy catalog, reaching out to a high redshift. 
While in reality, obtaining a relatively complete galaxy catalog at $z > 1.5$ is very challenging, when one considers the small localization area that will be provided by TianQin and/or LISA it seems likely that deep-drilled, a more complete galaxy surveys triggered by \ac{GW} detections could be carried out in order to alleviate incompleteness issues. 
We also assume a simple relationship between the \acp{MBH} mass and the bulge luminosity. 
Knowledge of the redshift evolution of this $M_{\rm MBH}-L_{\rm bulge}$ relation could also improve the cosmological constraints by reducing any potential bias arising from adopting this simple relationship. 
Extra information like stellar velocity dispersion can be used to improve the estimation of \ac{MBH} mass at the low-mass end through the $M_{\rm MBH}-\sigma$ relation.

In the future there is scope to extend our paper in several ways. 
For example, a new population model of \ac{MBHB} mergers can be added for the analysis \cite{Barausse:2020mdt}. 
The possible existence of strong gravitational lensing events can also help with pinpointing the redshift of \ac{GW} events \cite{Hannuksela:2020xor, Yu:2020agu}. 
We also plan to extend the \ac{GW} cosmology study to more types of \ac{GW} sources \cite{Liu:2020eko, Fan:2020zhy}.

\section*{Acknowledgements}
This work has been supported by the Guangdong Major Project of Basic and Applied Basic Research (Grant No. 2019B030302001),  the Natural Science Foundation of China (Grants  No. 11805286, 11690022, and 11803094), the Science and Technology Program of Guangzhou, China (No. 202002030360), and National Key Research and Development Program of China (No. 2020YFC2201400). 
We acknowledge the science research grants from the China Manned Space Project with No. CMS-CSST-2021-A03, No. CMS-CSST-2021-B01.
M.H. is supported by the Science and Technology Facilities Council (Ref. ST/L000946/1).
The authors would like to thank the Gravitational-wave Excellence through Alliance Training (GrEAT) network for facilitating this collaborative project. 
The GrEAT network is funded by the Science and Technology Facilities Council UK grant no. ST/R002770/1.
We acknowledge the use of the {\it Kunlun} cluster, a supercomputer owned by the School of Physics and Astronomy, Sun Yat-Sen University. 
The authors acknowledge the uses of the calculating utilities of \textsf{LALSuite} \cite{lalsuite}, \textsf{numpy} \cite{vanderWalt:2011bqk}, \textsf{scipy} \cite{Virtanen:2019joe}, and \textsf{emcee} \cite{ForemanMackey:2012ig, ForemanMackey:2019ig}, and the plotting utilities of \textsf{matplotlib} \cite{Hunter:2007ouj}, \textsf{corner} \cite{corner} , and \textsf{GetDist} \cite{Lewis:2019xzd}. 
The authors want to express great gratitude to the anonymous referee for the helpful feedback that improves the manuscript significantly.
The authors also thank Jie Gao, En-Kun Li, Zhaofeng Wu, and Bai-Tian Tang for helpful discussions.

\appendix
\renewcommand\thefigure{\Alph{section}\arabic{figure}}
\renewcommand\thetable{\Alph{section}\arabic{table}}

\section*{Appendix}

\section{Decomposition of the multimessenger likelihood function}    \label{derivation_Eq9}
In this Appendix we provide a detailed derivation of the multimessenger likelihood function, namely Eq. (\ref{likeli3}). 
We can factorize the left-hand side of Eq. (\ref{likeli3}) as 
\begin{align} \label{likeli3-detailed}
& p(D_i, S, D_L, z, \alpha, \delta, M_z, L_{\rm bulge}, \vec{\theta}', \vec{\phi}' | \vec{\Omega}, I)   \nonumber \\
 = &  p(D_i, S | D_L, z, \alpha, \delta, M_z, L_{\rm bulge}, \vec{\theta}', \vec{\phi}', \vec{\Omega}, I) p_0( D_L, z, \alpha, \delta, M_z, L_{\rm bulge}, \vec{\theta}', \vec{\phi}' | \vec{\Omega}, I)   \nonumber \\
 = & p(D_i | D_L, \alpha, \delta, M_z, \vec{\theta}', \vec{\Omega}, I) p(S | z, \alpha, \delta, L_{\rm bulge}, \vec{\phi}', \vec{\Omega}, I) p_0( D_L, z, \alpha, \delta, M_z, L_{\rm bulge}, \vec{\theta}', \vec{\phi}' | \vec{\Omega}, I)   \nonumber \\
 = & p(D_i | D_L, \alpha, \delta, M_z, \vec{\theta}', \vec{\Omega}, I) p(S | z, \alpha, \delta, L_{\rm bulge}, \vec{\phi}', \vec{\Omega}, I) p_0(D_L | z, \vec{\Omega}, I) p_0(M_z | z, L_{\rm bulge}, \vec{\Omega}, I)  \nonumber \\ 
 & \times p_0(z, \alpha, \delta, L_{\rm bulge} | \vec{\Omega}, I) p_0(\vec{\theta}'|\vec{\Omega}, I) p_0(\vec{\phi}'|\vec{\Omega}, I) \nonumber \\ 
 = & p(D_i | D_L, \alpha, \delta, M_z, \vec{\theta}', I) p(S | z, \alpha, \delta, L_{\rm bulge}, \vec{\phi}', I) p_0(D_L | z, \vec{\Omega}, I) p_0(M_z | z, L_{\rm bulge}, \vec{\Omega}, I)  \nonumber \\ 
 & \times p_0(z, \alpha, \delta, L_{\rm bulge} | \vec{\Omega}, I) p_0(\vec{\theta}'|\vec{\Omega}, I) p_0(\vec{\phi}'|\vec{\Omega}, I) . 
\end{align}
Since the distribution of the \ac{GW} sources and galaxies reflects the evolution of the Universe, we retain the cosmological parameters in all notation of the prior $p_0$.

\section{deriving the weights}    \label{expression_weight}
In this Appendix we provide a detailed expression of weighting coefficients applied to the observed and the supplementary galaxies respectively. 
For an observed galaxy at sky position and redshift of $(\alpha_j, \delta_j, z_j)$ and with bulge luminosity $L_{\rm bulge}$, the positional weight and the weight of bulge luminosity is given by
\begin{align}
W_{\textrm{pos}}^{\rm obs}(\alpha_j, \delta_j)  &\propto  \exp \bigg\{-\frac{1}{2} \Big[\big( \alpha_j - \bar \alpha,\delta_j - \bar \delta \big) \Sigma_{\alpha \delta}^{-1} \big(\alpha_j - \bar \alpha,\delta_j - \bar \delta\big)^{\rm T} \Big]\bigg\},  \label{w_angle_general}  \\ 
W_{\textrm{lum}}^{\textrm{obs}}(L_{{\rm bulge},j}, z_j)  &\propto  \exp \Bigg[ -\frac{1}{2} \frac{\left(\log_{10} \big((1+z_j) \hat M(L_{{\rm bulge},j})\big) - \log_{10} \bar M_z \right)^2}{\sigma_{\log_{10} M}^2} \Bigg],  \label{w_mag-obs} 
\end{align}
where $\bar \alpha$, $\bar \delta$ and $\bar M_z$ are the measurement mean value of the longitude, the latitude and the redshifted total mass of the \ac{GW} source, respectively, and $\Sigma_{\alpha \delta} = \Gamma_{\alpha \delta}^{-1}$ is the covariance matrix of sky localization of the \ac{GW} source. 
For the supplementary galaxy, its sky position and redshift $(\alpha_j, \delta_j, z_j)$ are replaced by the measurement mean $(\bar \alpha_{\rm obs}, \bar \delta_{\rm obs}, \bar z_{\rm obs})$ of observed galaxies within the divided small region, we defined the weight of bulge luminosity as
\begin{equation}  \label{w_mag-sup}
W_{\textrm{lum}}^{\textrm{sup}}\big(\Phi'(L_{\rm bulge}), \bar z_{\rm obs} \big)  \equiv  \int_{0}^{L_{\rm bulge}^{\rm min}} \Phi'(L_{\rm bulge}) W_{\textrm{lum}}^{\textrm{obs}}(L_{\rm bulge}, \bar z_{\rm obs}) d L_{\rm bulge} , 
\end{equation}
where $(\bar \alpha_{\rm obs}, \bar \delta_{\rm obs}, \bar z_{\rm obs})$ and $L_{\rm bulge}^{\rm min}$ have different values for each divided small region.

\section{Examples of the cosmological constraints}    \label{cosmoexample-z_survey}

\begin{figure}[htbp]
\centering
\includegraphics[height=7.1cm, width=7.1cm]{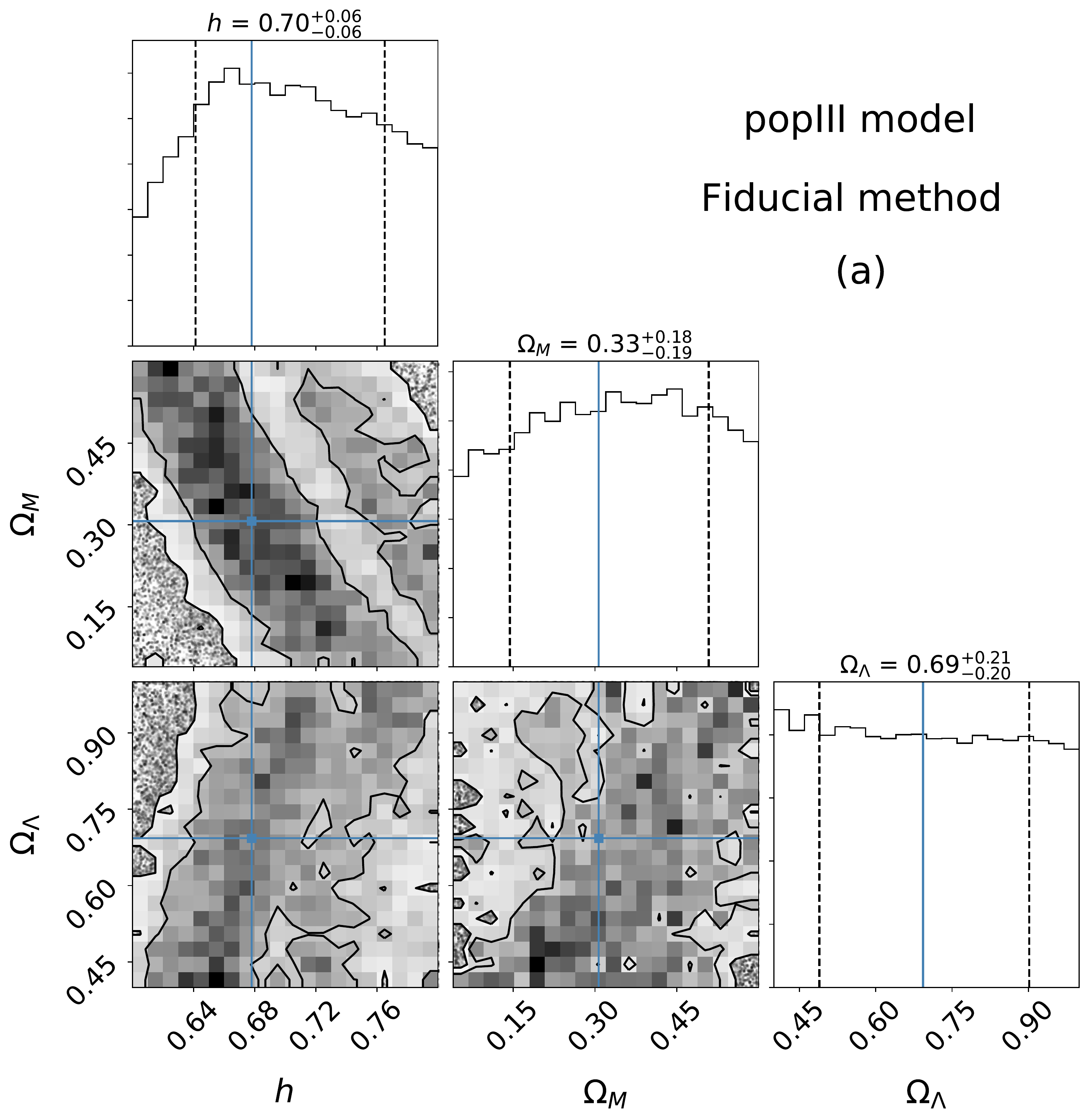} ~~~~
\includegraphics[height=7.1cm, width=7.1cm]{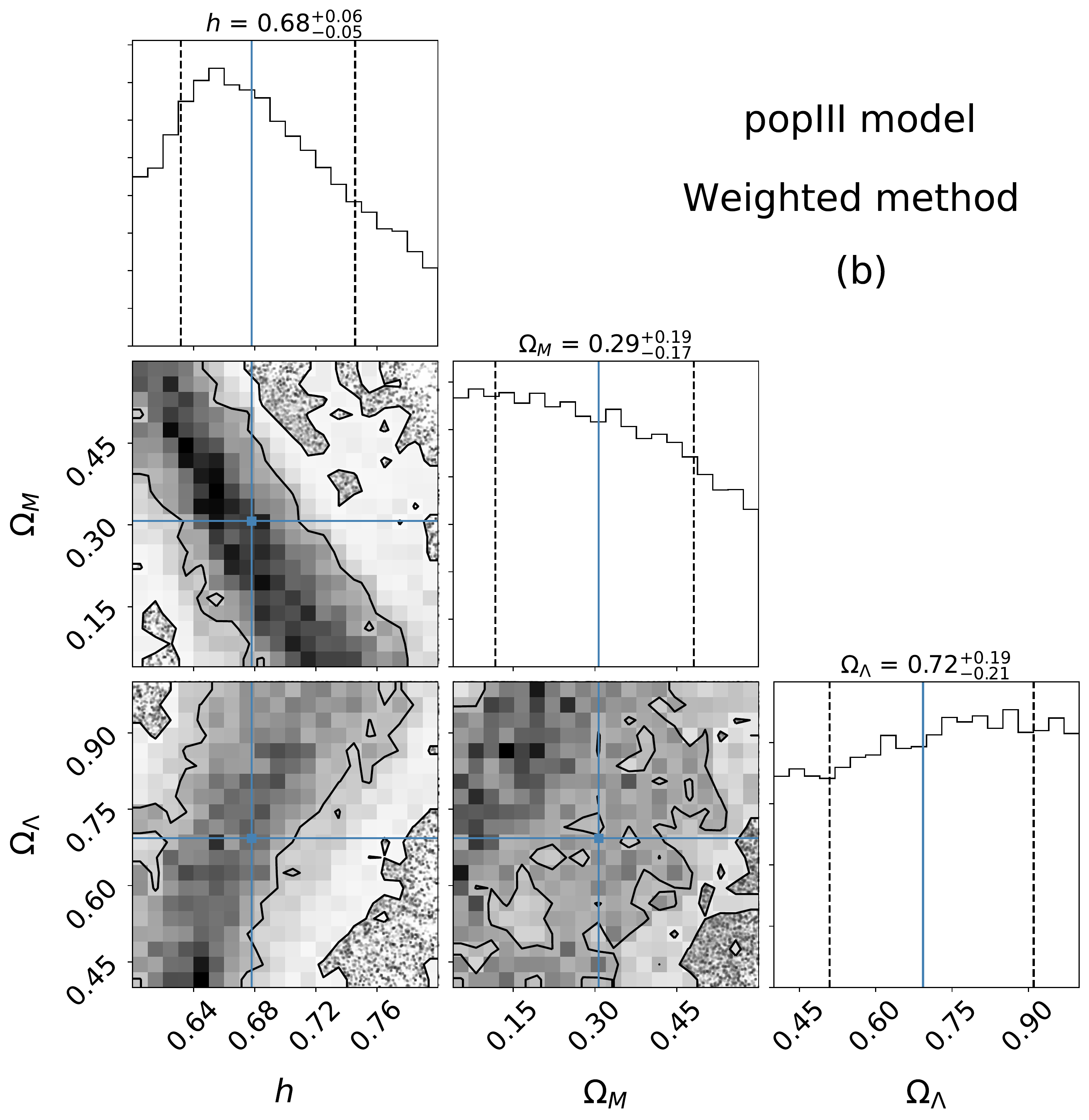} \\
\includegraphics[height=7.1cm, width=7.1cm]{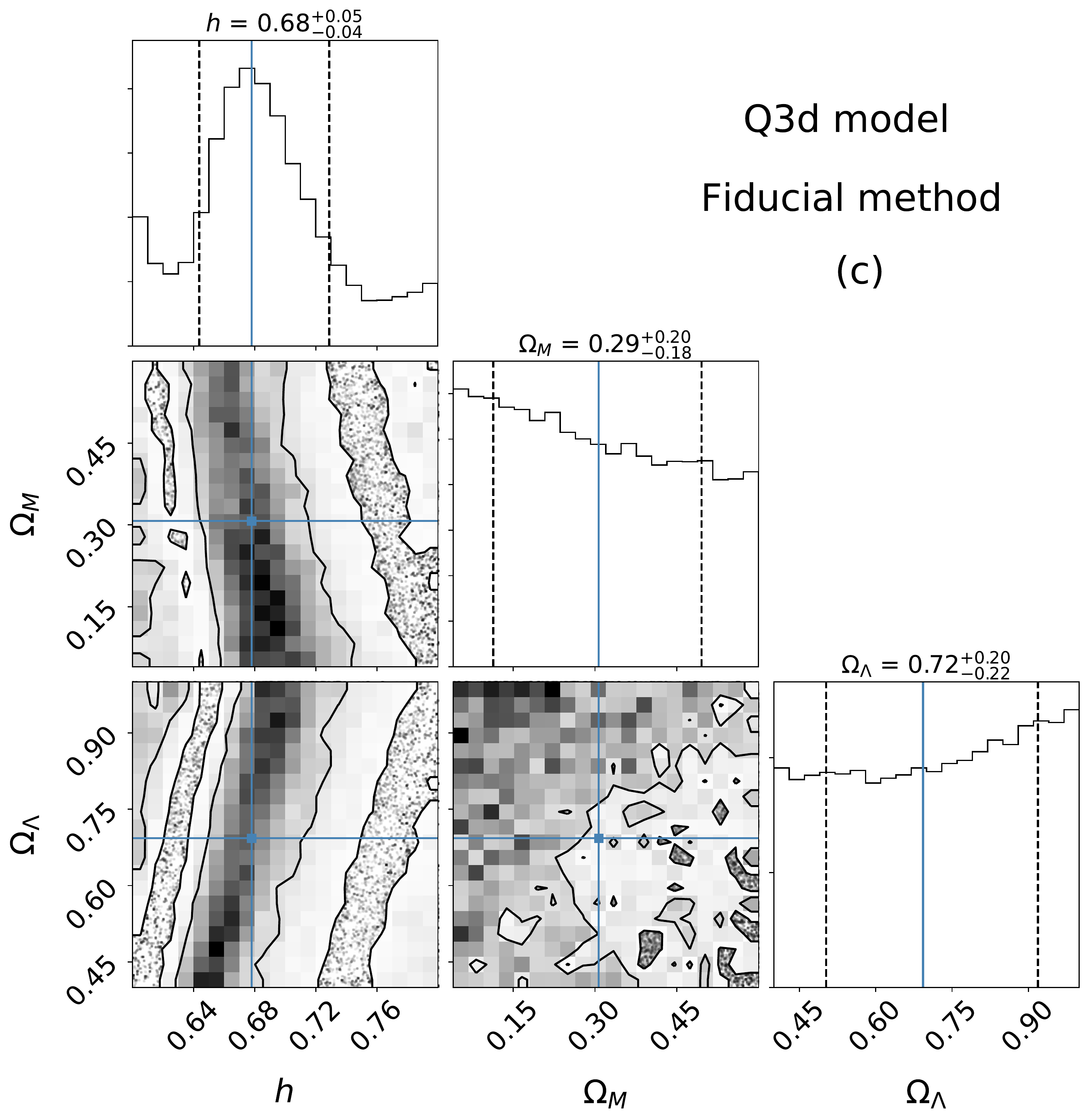} ~~~~
\includegraphics[height=7.1cm, width=7.1cm]{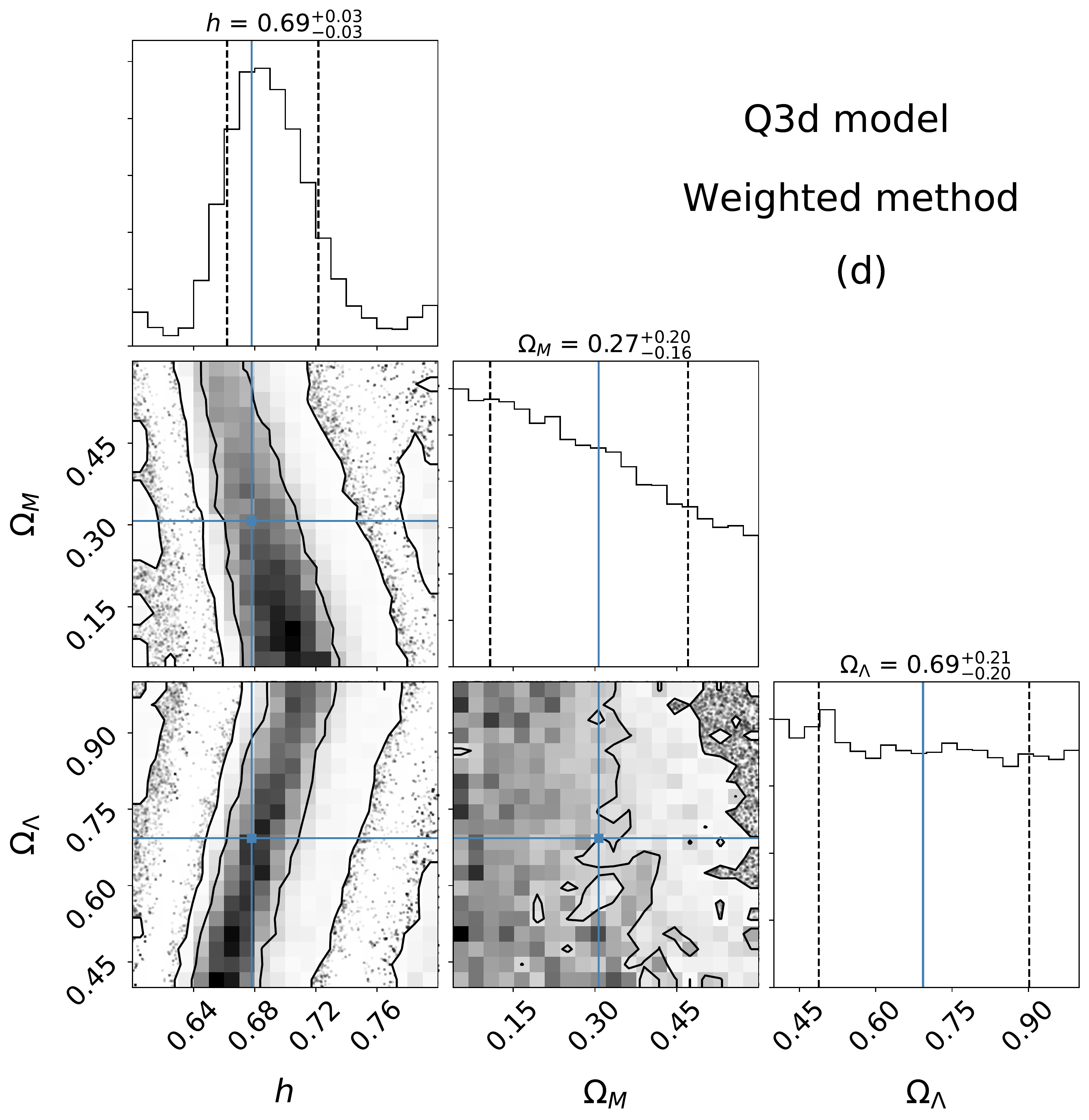} \\
\includegraphics[height=7.1cm, width=7.1cm]{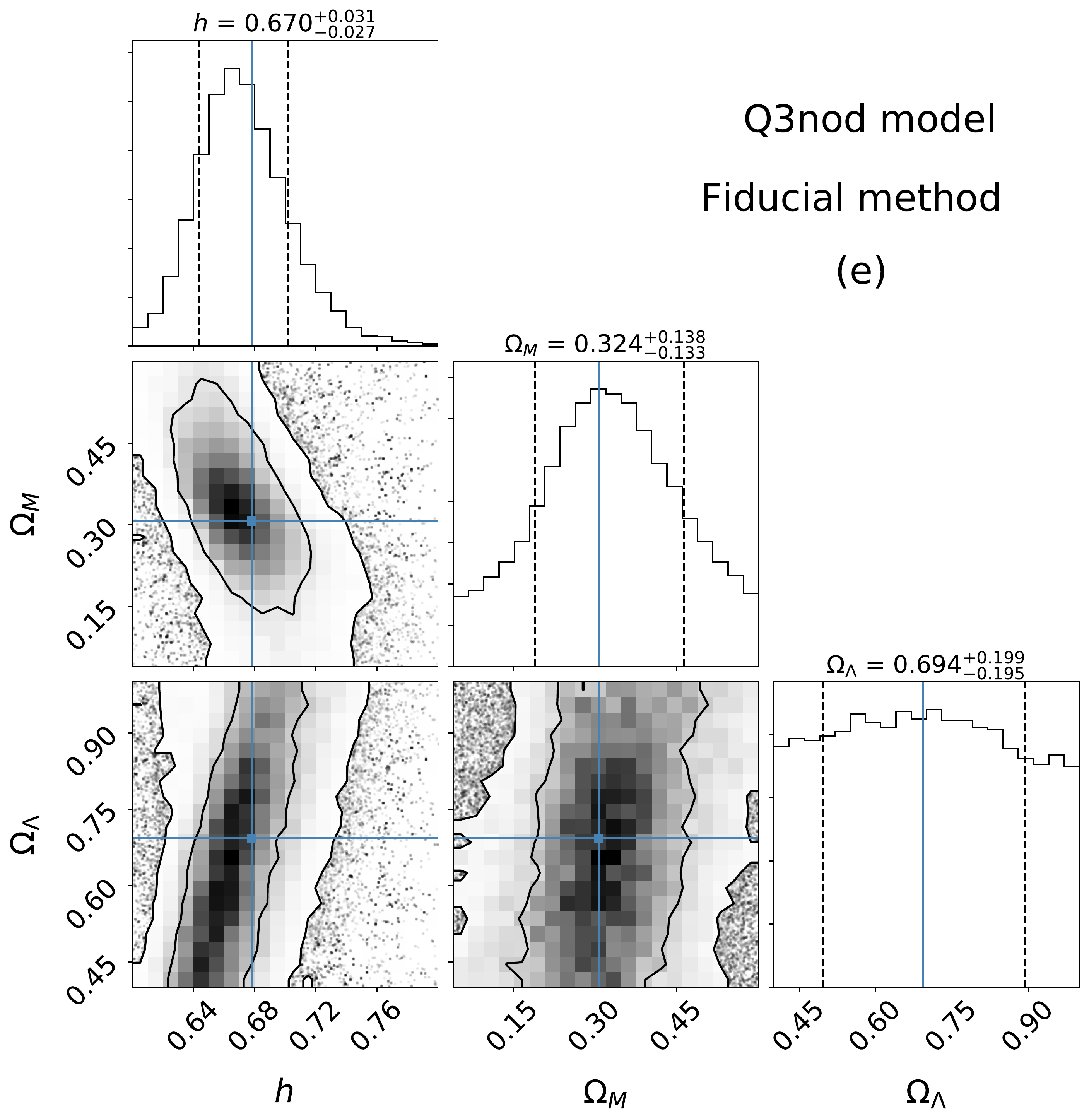} ~~~~
\includegraphics[height=7.1cm, width=7.1cm]{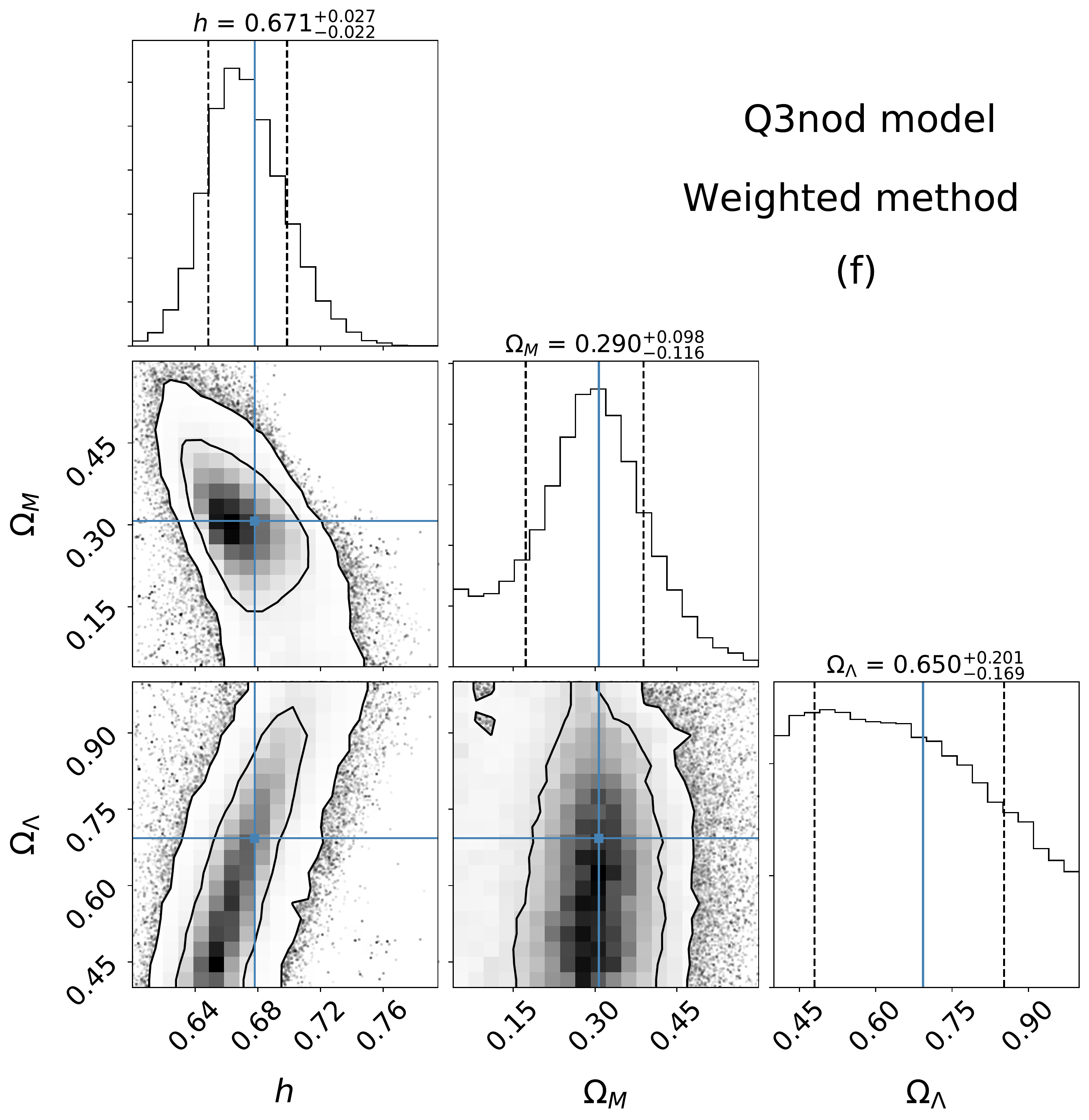}
\caption{Typical corner plots of the posteriors for the parameters $h$, $\Omega_M$, and $\Omega_{\Lambda}$ constrained by the detections of TianQin, comparing the fiducial method (left column) and the weighted method (right column). 
The top, middle, and bottom rows correspond to adopting popIII, Q3d, and Q3nod respectively as the underlying model for \ac{MBHB} mergers.
In each subplot, the three panels at the lower left show the two-dimensional joint posterior probabilities of $h-\Omega_M$, $h-\Omega_{\Lambda}$, and $\Omega_M-\Omega_{\Lambda}$, with the contours represent confidence levels of $1 \sigma (68.27\%)$ and $2 \sigma (95.45\%)$, respectively; the upper, middle and right panels show the one-dimensional posterior probabilities of the corresponding parameters, after marginalization over the other parameters, with the dashed lines indicate $1 \sigma$ credible interval. 
In each panel the solid cyan lines mark the true values of the parameters. }
\label{H0ML_examples-1TQ}
\end{figure}

\begin{figure}[htbp]
\centering
\includegraphics[height=7.1cm, width=7.1cm]{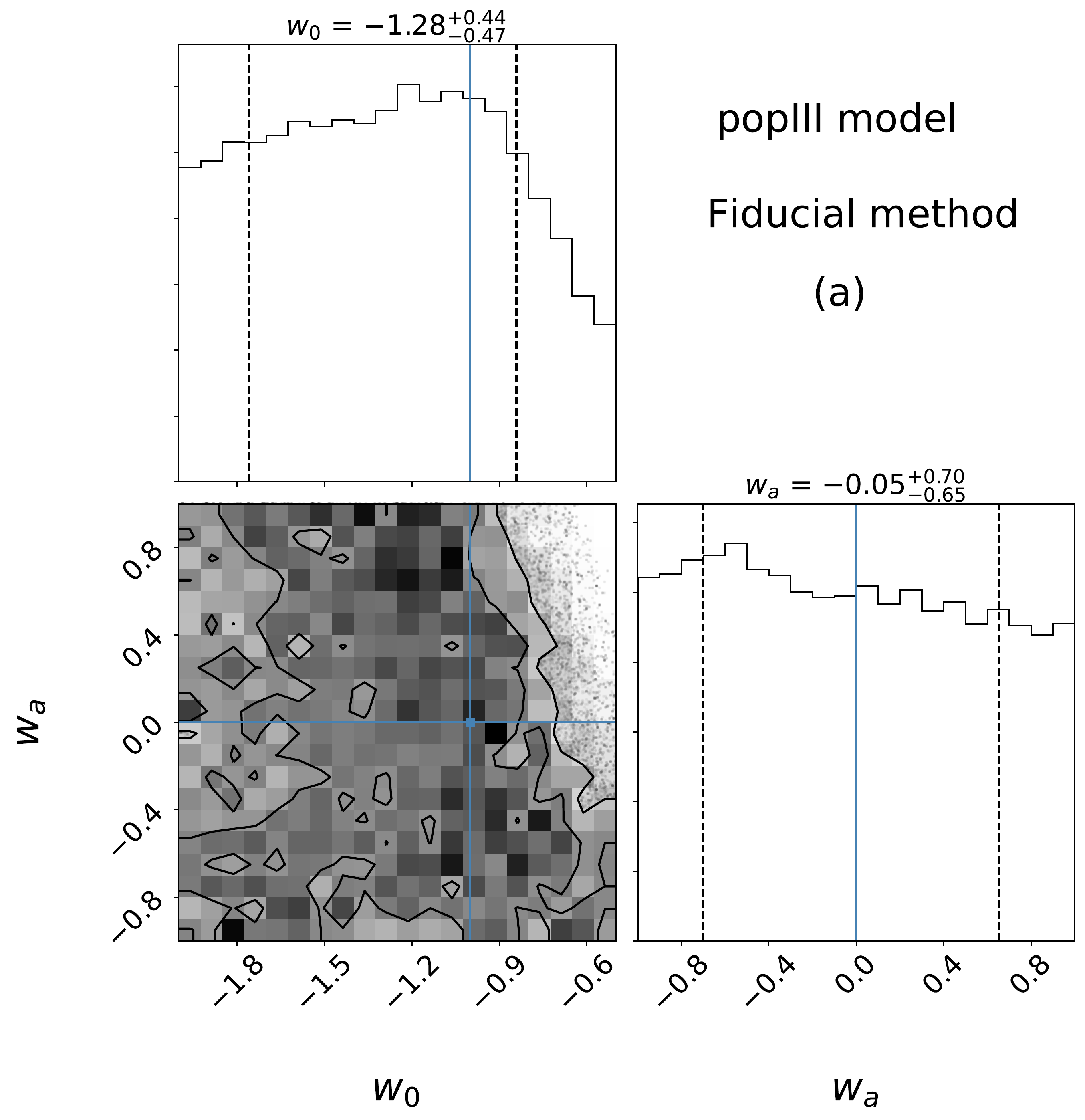} ~~~~~
\includegraphics[height=7.1cm, width=7.1cm]{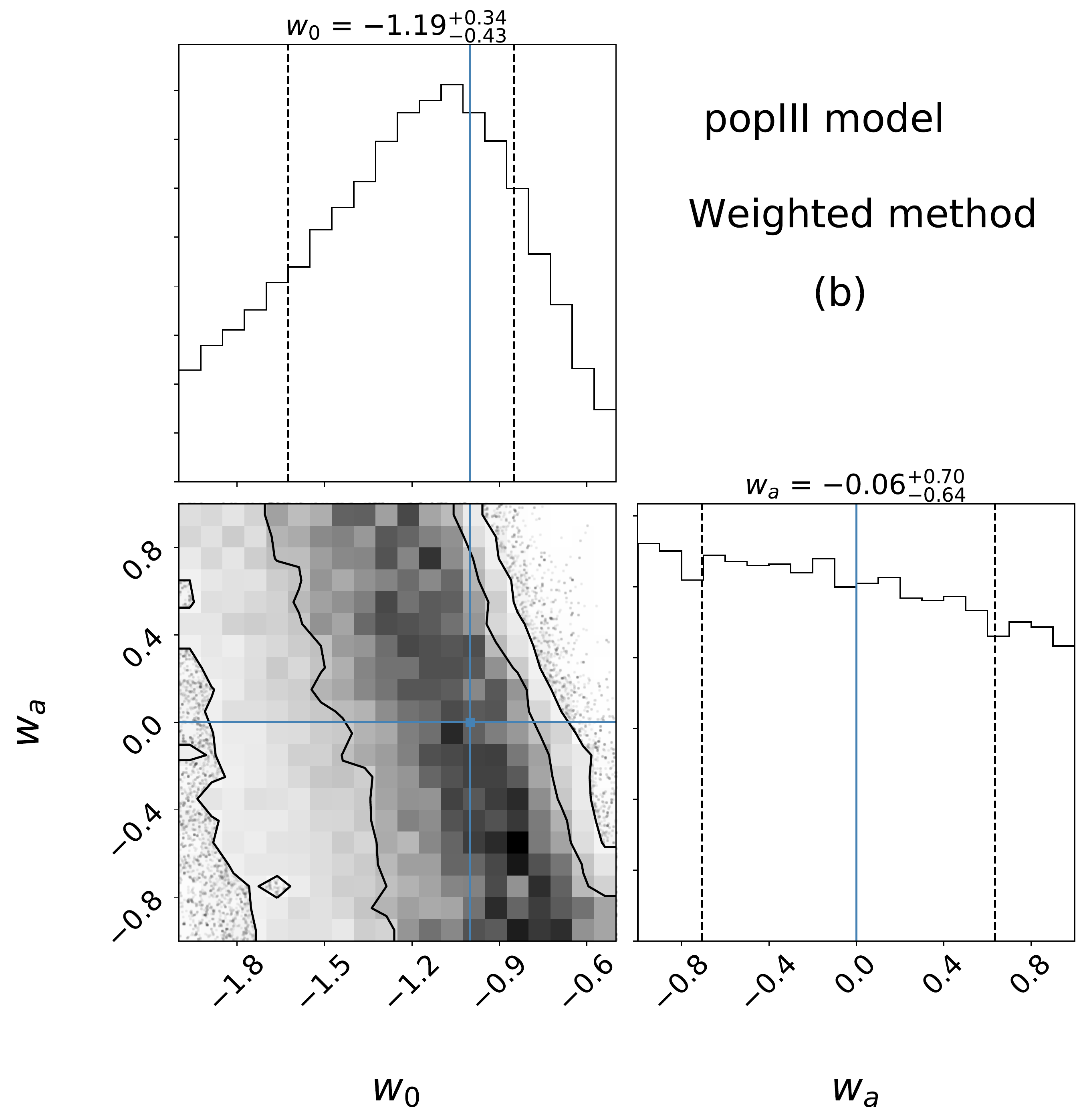} \\
\includegraphics[height=7.1cm, width=7.1cm]{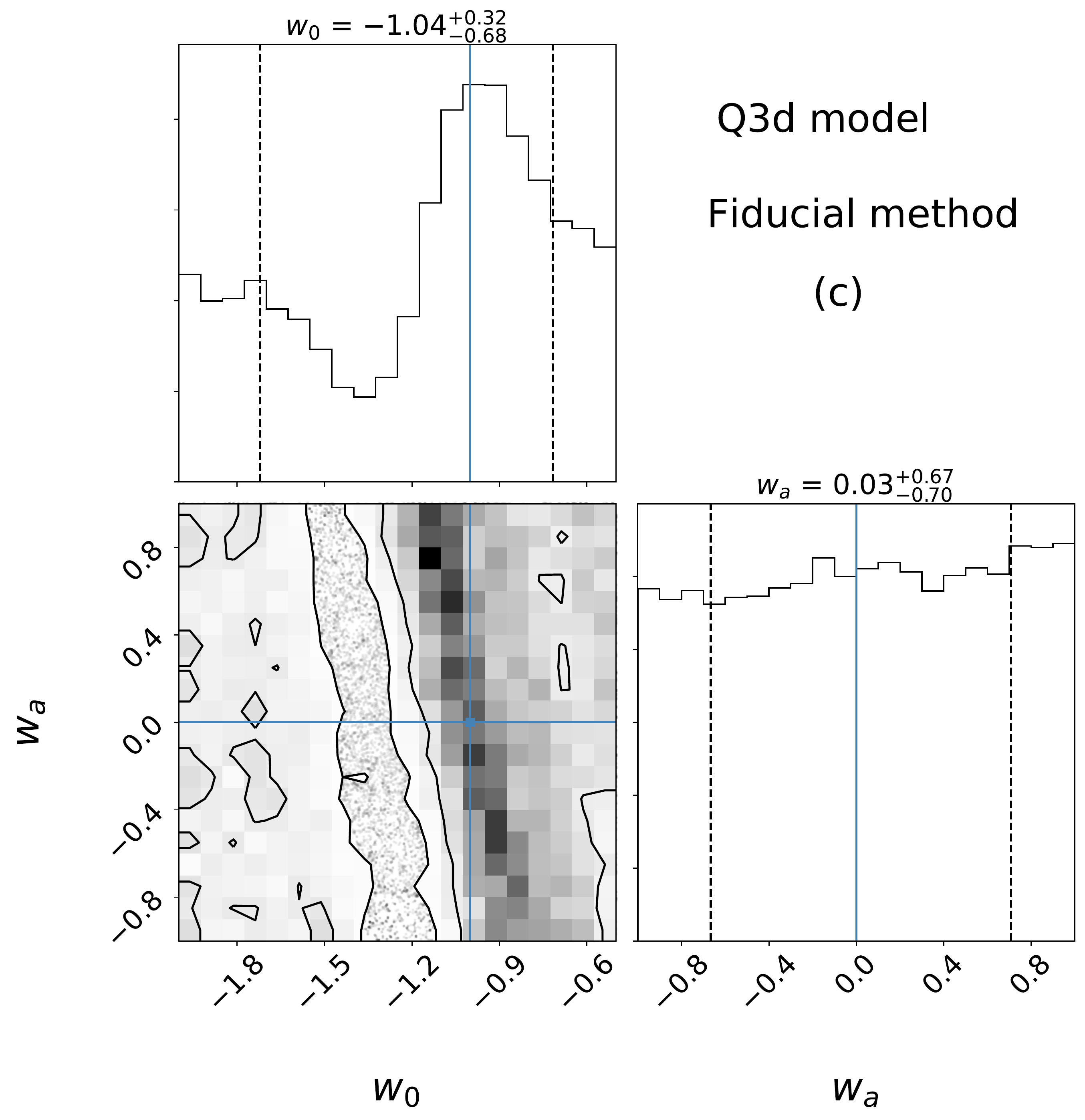} ~~~~~
\includegraphics[height=7.1cm, width=7.1cm]{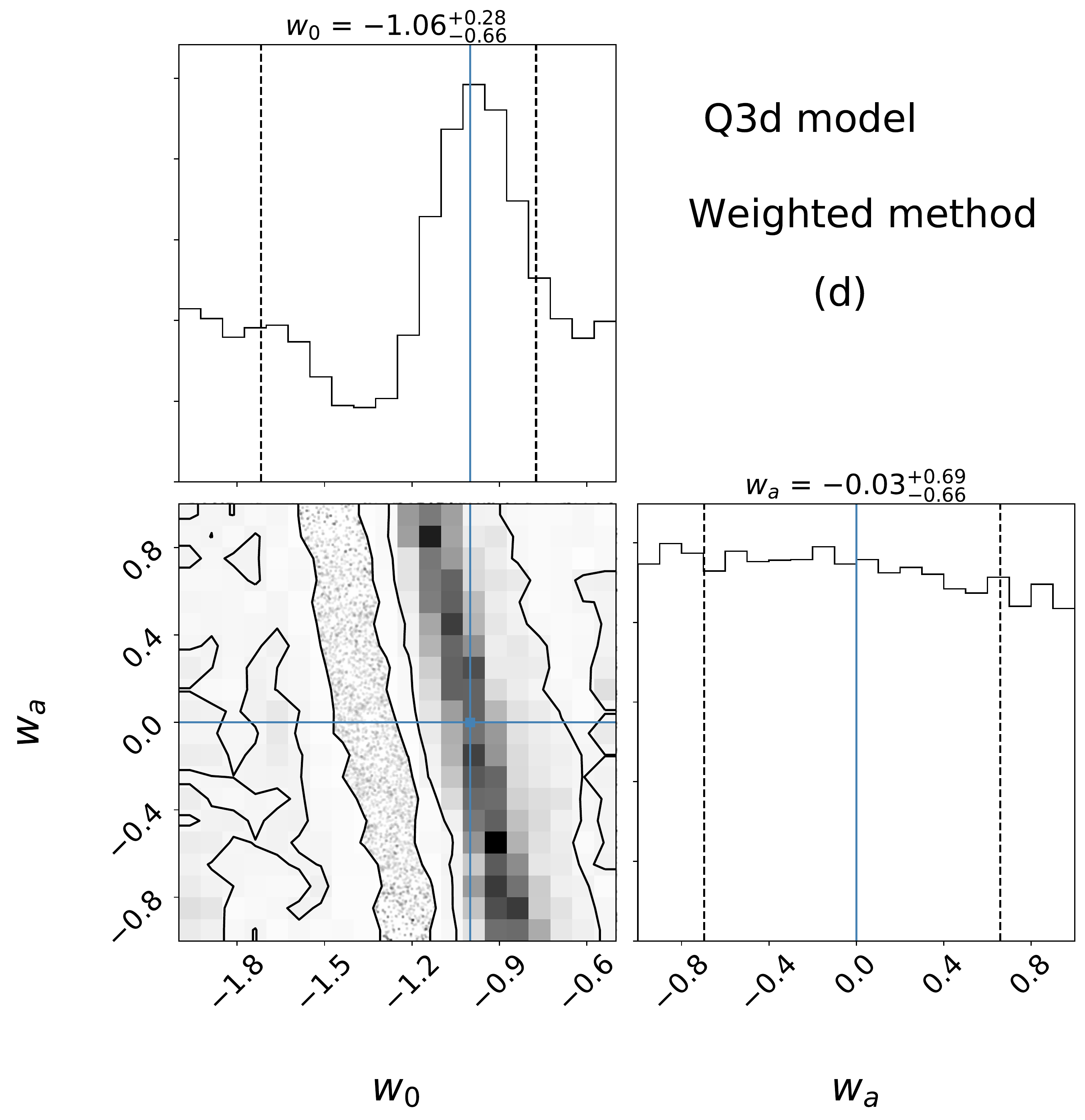} \\
\includegraphics[height=7.1cm, width=7.1cm]{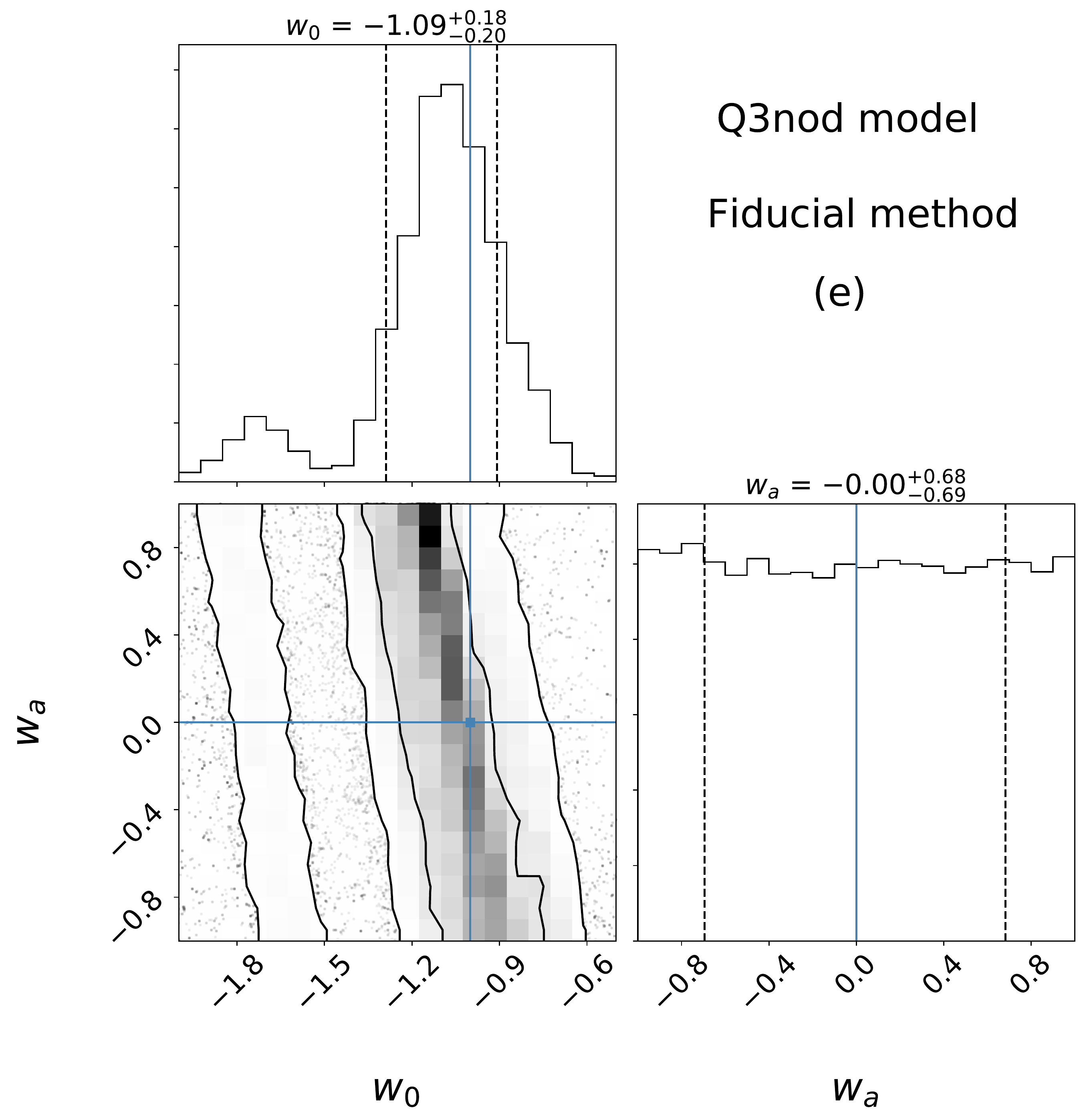} ~~~~~
\includegraphics[height=7.1cm, width=7.1cm]{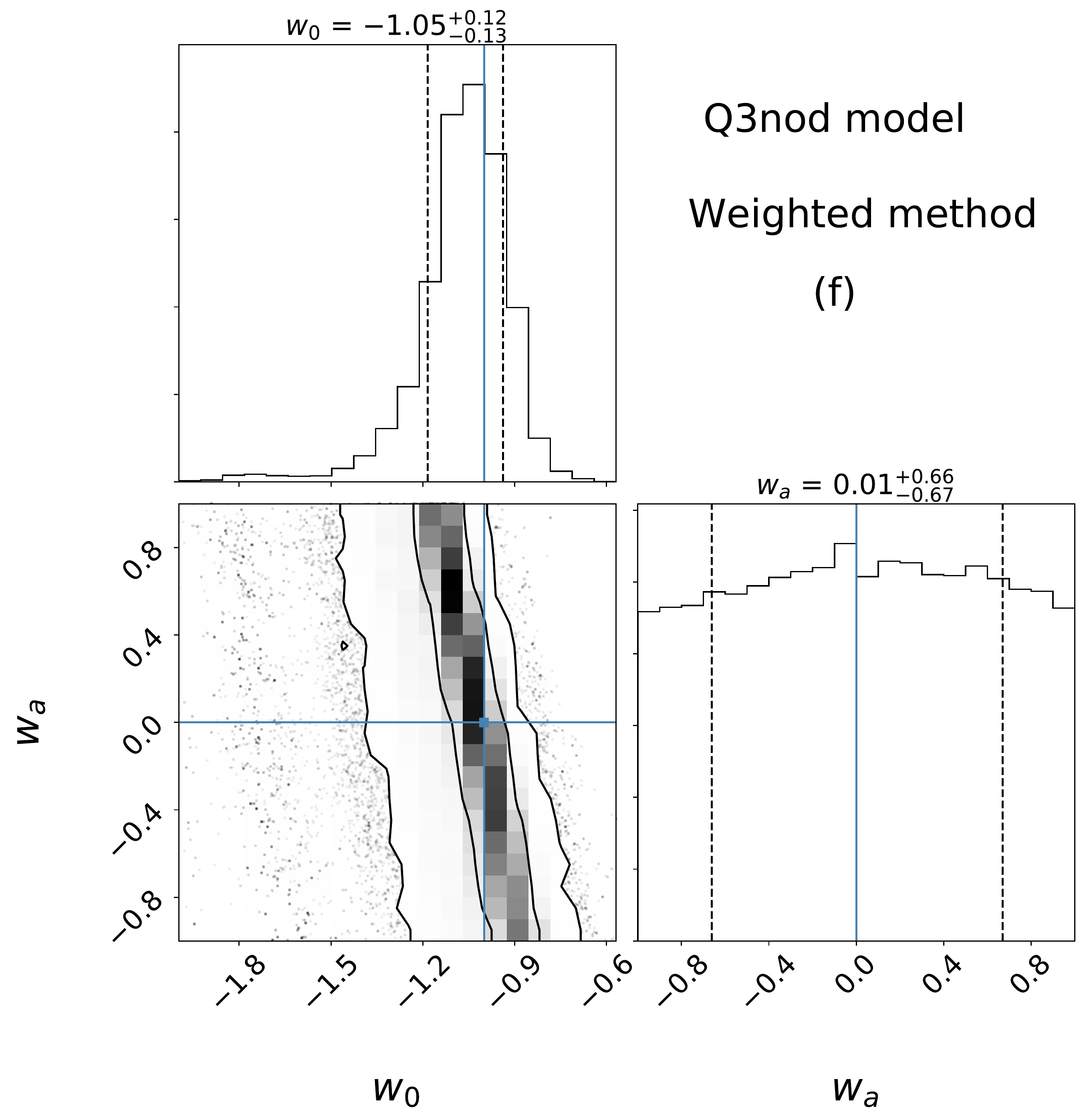}
\caption{Same as Fig. \ref{H0ML_examples-1TQ}, but for the parameters of \ac{EoS} of the dark energy $w_0$ and $w_a$. 
The other cosmological parameters, $H_0$, $\Omega_M$, and $\Omega_{\Lambda}$ are fixed.}
\label{w0wa_examples-1TQ}
\end{figure}

In this Appendix, we show the typical posterior probability distributions of the cosmological parameters and the parameters of \ac{EoS} of the dark energy for TianQin under the three population models, i.e., popIII, Q3d, and Q3nod, using the fiducial method and the weighted method, in Figs. \ref{H0ML_examples-1TQ} and \ref{w0wa_examples-1TQ}, respectively.

\newpage
\renewcommand{\refname}{Reference}
\bibliographystyle{unsrt}   
\bibliography{bibfile}   

\end{document}